\documentclass[12pt]{article}

\usepackage[utf8]{inputenc}

\usepackage{mathptmx}                 % Times-like font
\usepackage[margin=3cm]{geometry}

\usepackage{amsmath,amssymb,amsthm,mathtools}
\usepackage{bm}
\usepackage{setspace}
\usepackage{microtype}
\usepackage[longnamesfirst,sort&compress]{natbib}
\usepackage{booktabs}
\usepackage{tabularx}
\usepackage{enumitem}

\usepackage{float}
\usepackage{placeins}
\usepackage{graphicx}
\usepackage{subcaption}
\usepackage{adjustbox}
\usepackage{rotating}

\usepackage{titlesec}
\usepackage{titling}
\usepackage{etoolbox}

\usepackage[dvipsnames]{xcolor}
\usepackage[hidelinks,breaklinks=true]{hyperref}
\hypersetup{
    colorlinks=true,
    linkcolor=blue!55!black,
    citecolor=blue!55!black,
    urlcolor=blue!55!black
}

\newcommand{\VAR}{\mathrm{VAR}}
\newcommand{\LP}{\mathrm{LP}}

\theoremstyle{plain}

\newtheorem{theorem}{Theorem}
\newtheorem{corollary}{Corollary}

\theoremstyle{definition}
\newtheorem{assumption}{Assumption}[section]
\newtheorem{lemma}{Lemma}[section]
\newtheorem{remark}{Remark}

\title{When Does $N$ Buy Precision?\\
Heterogeneous Responses to Aggregate Shocks in Micro Panels}

\author{Mahmut Ablay\thanks{Central Bank of the Republic of T\"urkiye.
Email: \texttt{mahmut.ablay@tcmb.gov.tr}.}
\and Mengshan Xu\thanks{Department of Economics, University of Mannheim.
Email: \texttt{mengshan.xu@uni-mannheim.de}.}
\and Endong Wang\thanks{Corresponding author. Department of Economics,
University of Mannheim, L7 3--5, 68161 Mannheim, Germany.
Email: \texttt{endong.wang@uni-mannheim.de}.}}

\date{\today}

\begin{document}

\maketitle

\begin{abstract}
\noindent
Panel local projections (panel LPs), valued for their flexibility and robustness, are widely used to estimate heterogeneous responses to aggregate shocks from micro data. This paper, however, highlights the potential of panel vector autoregressions (panel VARs) and characterizes the trade-off between the efficiency of panel VARs and the robustness of panel LPs. In contrast to existing results in the time-series literature, we show that under correct specification, panel VAR estimators of the heterogeneous responses converge at rate $(NT)^{-1/2}$, whereas panel LP estimators converge at rate $T^{-1/2}$. When the heterogeneous response function is nonparametric, the panel VAR incurs a
nonparametric rate loss, whereas the panel LP retains its parametric
$T^{-1/2}$ rate under suitable smoothness and panel-growth conditions. We provide a Hausman-type test for choosing between them, with power against two forms of panel VAR misspecification. We apply panel LPs and panel VARs to Turkish administrative firm-level data to estimate heterogeneous firm responses to monetary policy shocks.

% Panel local projections, valued for their flexibility and
% robustness, are widely used to estimate heterogeneous responses to
% aggregate shocks from micro data. 
%  This paper, however, aims to draw attention to the potential of a vector autoregression-based approach in micro panel settings. We compare two approaches for estimating heterogeneous responses in a micro panel --  panel local projections (panel LPs) and panel vector autoregressions (panel VARs)  -- and characterize the trade-off
% between the efficiency gains of panel VARs and
% the robustness of panel LPs. We consider both parametric and nonparametric heterogeneous response functions.  In contrast to the existing results in the time-series literature, we find that under a correctly specified model, the panel VAR estimator  converges at rate $(NT)^{-1/2}$, whereas the panel LP
% estimator converges at rate $T^{-1/2}$; when the
% heterogeneous response function is nonparametric, the panel LP estimator retains the parametric
% $T^{-1/2}$ rate, whereas the panel VAR estimator suffers a
% nonparametric rate loss.
%  Finally, we provide a Hausman-type test to inform applied researchers’ choice between the two approaches. We are applying our method to Turkish administrative firm-level data to study heterogeneous firm responses to monetary policy.

 
\end{abstract}
 
\newpage
 
\section{Introduction}
\label{sec:introduction}

Aggregate shocks reach firms through heterogeneous exposures. A monetary policy
shock, for example, affects firms with different leverage and liquidity
differently, and in heterogeneous-agent models the distribution of such
differences shapes the aggregate response
\citep{kaplanmollviolante2018,auclert2019}. Firm-level micro panel data are
accordingly widely used to measure heterogeneous responses
\citep{ottonellowinberry2020,jeenas2023firm,cloyne2023,gulyas2024}. A standard
estimator is the panel local projection (panel LP). At each horizon $h$, the
researcher regresses the firm's outcome $h$ periods ahead on an identified
aggregate shock interacted with a predetermined state, such as leverage or
liquidity. The interaction coefficient measures how the response to the shock
varies with the predetermined state.
 
These micro panels typically contain observations on a large number of units $N$ over a comparatively small number of periods $T$. The aggregate shock, however, is realized only once per period, so an additional unit adds only cross-sectional variation. Whether a large $N$ improves precision depends on how effectively the chosen estimator uses cross-sectional information. In this paper, we compare the widely used  panel LPs with panel vector autoregressions (panel VARs). The latter estimates a dynamic model using the full panel and iterates it to obtain responses at longer horizons. This comparison is well developed in the time-series literature \citep{jorda2005} but has received less attention in micro panels, where conclusions drawn from time-series settings do not directly apply.  Meanwhile, empirical studies of heterogeneous responses in micro panels often treat panel LPs as the default \citep{almuzara2024}, with less attention paid to alternative approaches and their relative benefits and costs.

Our first contribution is to show that the panel VAR and the panel LP typically converge at
different rates. In time series, LPs and VARs recover the same population impulse
response \citep{plagborgmollerwolf2021} and converge at a common rate, so the choice
between them is a trade-off between  bias and variance 
\citep[e.g.,][]{li2024,montielolea2024double,montielolea2026primer}. In micro panels, this common
rate breaks down. When the exposure function is linear, the (bias-corrected) panel VAR converges at
$(NT)^{-1/2}$ under a correctly specified dynamic model, whereas the panel LP converges
at $T^{-1/2}$ at positive horizons. The rate difference arises because the two regression errors contain
different sources of uncertainty. The panel VAR error is
idiosyncratic and averages out over both units and periods. Nevertheless, the projection error of the panel LP contains
heterogeneous effects of post-impact aggregate shocks, which
time fixed effects cannot absorb. Because each shock is common
to all units at a given period, increasing $N$ cannot average away
the aggregate shocks themselves.\footnote{An exception arises at impact ($h=0$): the panel LP is only one equation of the panel VAR system.}

Our second contribution extends this comparison to the model with an unknown nonlinear (nonparametric) exposure function.
Nonparametric local projections have been developed for time series
\citep{goncalves2024nonparametric,goncalves2026semiparametric} and micro panels
\citep{paranhos2025,david2026causal}. Our paper differs from these studies by examining how nonparametric estimation interacts with the time-series limit of panel LPs, providing a more nuanced comparison of the convergence rates of panel LPs and  panel VARs when nonparametric heterogeneous response functions are allowed: (i) With an $r$-smooth exposure function of an $m$-dimensional predetermined state, an admissible choice of the series order gives integrated rates of
$(NT)^{-r/(2r+m)}$ for the panel VAR and
$T^{-1/2}+(NT)^{-r/(2r+m)}$ for the panel LP.  (ii) For the panel LP, more interestingly, when $N/T^{m/(2r)}$ diverges and $r/m>1$,
a suitable choice of the series order makes the nonparametric terms smaller
than $T^{-1/2}$. As a result, nonparametric modeling, which usually trades estimation efficiency for robustness, does not compromise the panel LP’s parametric rate $T^{-1/2}$ in this case. The nonparametric panel VAR, however, does not enjoy this “free lunch.”
This picture is summarized in
 Table~\ref{tab:intro-rates}.  
 
 As a by-product, our paper also makes a technical contribution to the semiparametric and nonparametric estimation of dynamic
micro panels. We extend the  estimator of \citet{lee2014} to a two-way within nonparametric series
estimator for both panel VARs and panel LPs. This extension accommodates both unit and time effects,
thereby accounting for the persistent unit heterogeneity and common
period-specific variation, which are central to dynamic micro panels
exposed to aggregate shocks. 
 
\begin{table}[t]
\centering
\caption{Convergence rates of the panel VAR and the panel LP}
\label{tab:intro-rates}
\small
\begin{tabularx}{0.78\textwidth}{@{}p{0.18\textwidth}XX@{}}
\toprule
Estimator
& Linear exposure
& Nonparametric exposure (*) \\
\midrule
Panel VAR
& $(NT)^{-1/2}$
& $(NT)^{-r/(2r+m)}$
\\[0.55em]
Panel LP
& $T^{-1/2}$
& $T^{-1/2}+o_p(T^{-1/2}) (**)$
\\
\bottomrule
\end{tabularx}
 
\begin{minipage}{0.75\textwidth}
\vspace{0.4em}
\footnotesize
*  Nonparametric rates are integrated rates under the assumptions of
Theorem~\ref{prop:semiparametric-rates}.
The displayed panel VAR rate additionally requires that
$K\asymp(NT)^{m/(2r+m)}$ satisfies the series-growth conditions.
 
** The panel LP rate uses a series order making the nonparametric
terms $o(T^{-1/2})$.
See Appendix~\ref{np:sec:integrated-rates} for the respective
smoothness and panel-growth restrictions.
\end{minipage}
\end{table}
 
Third, we develop a Hausman-type test to guide the choice between
panel VAR and panel LP. Under the null hypothesis that the dynamic model is correctly specified, the panel VAR estimator converges faster,
making its estimation error asymptotically negligible in the difference
between the two estimators. The test therefore requires only the
panel LP asymptotic variance. Our test extends the comparison of
SVAR-IV and LP-IV impulse response estimates in
\citet{stockwatson2018} to the panel setting. We establish
asymptotically correct size under the null and consistency against
two departures from the dynamic model: a panel VAR with omitted variables,
in which the propagation law omits lags or variables, and a panel VARMA, in
which the shock interaction has a moving-average component. A rejection provides evidence of bias in the panel VAR
estimates\footnote{With some exceptions; see Remark~\ref{rem:hausman-scope}
for an example.}; under these alternatives, the panel LP remains consistent.

This paper relates to three literatures. The first studies heterogeneous
responses to aggregate shocks using micro data, at the firm level
\citep{ottonellowinberry2020,cloyne2023,jeenas2023firm,gulyas2024} and at the
household and regional levels
\citep{mianraosufi2013,cloyne2020,nakamura2014fiscal,chodorowreich2019}. With
time effects, these designs identify differences in responses across units but not their levels \citep{guren2021crossregional,wolf2023missing}. When the errors contain
a component common to all units at a date, a larger cross section does not
average it away \citep{moulton1990,andrews2005,abadie2023}, and
\citet{almuzara2024} develop inference for panel LPs in this setting. 
%Our contribution to this literature is the comparison with the panel VAR. 
In contrast to these studies, we examine this problem through the lens of the robustness–efficiency trade-off between panel LPs and panel VARs.

The second literature compares LPs and VARs in time series. The two recover the
same population impulse response \citep{plagborgmollerwolf2021} and converge at
the same rate, so the comparison concerns bias and variance
\citep{jorda2005,li2024,montielolea2024double,montielolea2026primer}.
Nonparametric and semiparametric LPs extend the analysis to state-dependent
responses \citep{goncalves2024nonparametric,goncalves2026semiparametric}. We
extend this comparison to micro panels, where the panel VAR and the panel LP
converge at different rates under both linear and nonparametric exposure.

 The third literature studies semiparametric and nonparametric estimation in panels. Our nonparametric analysis extends \citet{lee2014}; see also related but different approaches to nonparametric dynamic panels by \cite{su2013nonparametric} and \cite{su2016semiparametric},  nonparametric
LPs for time series \citep{goncalves2024nonparametric,goncalves2026semiparametric}, and nonparametric
LPs 
 for micro panels \citep{paranhos2025,david2026causal}. 
 
\textit{Outline}--Section~\ref{sec:motivation} illustrates the convergence rates in a scalar model. Sections~\ref{sec:parametric} and~\ref{sec:semiparametric} establish the results for linear and nonparametric exposure, respectively. Section~\ref{sec:misspecification} studies misspecification of the panel VAR
and proposes a Hausman-type test. Section~\ref{sec:conclusion} concludes.

\section{A scalar example}
\label{sec:motivation}
 
In this section, we use a scalar dynamic model as a toy example to illustrate how the choice between a panel VAR and a panel LP can affect the estimation of heterogeneous responses.
 
\subsection{A scalar dynamic panel}
 
Consider
\begin{align}
y_{it}
&=
\alpha_i
+
\rho y_{i,t-1}
+
(\beta_0+\beta_1y_{i,t-1})\varepsilon_t
+
u_{it},
\label{eq:scalar-dgp}
\end{align}
where $\alpha_i$ is a unit effect, $\varepsilon_t$ is an observed aggregate
shock, and $u_{it}$ is the one-step error. The exposure of unit $i$ to the
aggregate shock at date $t$ is $\beta_0+\beta_1y_{i,t-1}$. The coefficient
$\beta_1$ governs how exposure varies with the predetermined state $y_{i,t-1}$,
and $\rho$ propagates the resulting heterogeneity over time.
 
The aggregate shock $\varepsilon_t$ has mean zero and is independent of the unit
effects and of all variables dated $t-1$ or earlier, so that
$\mathbb E[y_{i,t+j-1}\varepsilon_{t+j}]=0$ for every $j\ge0$. The one-step
error $u_{it}$ is independent and identically distributed (iid) across units
and periods, and is therefore idiosyncratic.
 
Fix the predetermined state at $y_{i,t-1}=y$. The level response at horizon
$h$ is
$b_h(y)\equiv \mathbb E[y_{i,t+h}\mid \varepsilon_t=1,\ y_{i,t-1}=y]
-\mathbb E[y_{i,t+h}\mid \varepsilon_t=0,\ y_{i,t-1}=y]
=\rho^h(\beta_0+\beta_1y)$.
The parameter of interest is the \emph{heterogeneous response} at horizon
$h$, the difference between the level responses of two units in
predetermined states $y$ and $y'$:
\begin{align}
\theta_h(y,y')
&\equiv
b_h(y)-b_h(y')
=
\beta_1\rho^h(y-y')
=
\beta_{1,h}(y-y'),
\label{eq:scalar-contrast}
\end{align}
where $\beta_{1,h}\equiv\beta_1\rho^h$. We therefore compare estimators of
$\beta_{1,h}$.
 
Throughout the paper, we focus on $\theta_h$. Such cross-sectional
differences in responses are the primary
estimand in the empirical
literature on the heterogeneous transmission of aggregate
shocks,\footnote{For example, \citet{kashyap2000banks} study how the effect
of monetary policy on bank lending varies with bank liquidity, and
\citet{ottonellowinberry2020} estimate how the semi-elasticity of firm
investment to monetary shocks varies with firm default risk, absorbing the
average response with sector-by-quarter fixed effects even though the shock
is observed. In the regional literature, \citet{nakamura2014fiscal} estimate
a relative multiplier that differences out the aggregate policy responses
common to all regions. Such relative responses serve as identified moments
for discriminating between classes of models
\citep{nakamura2018identification}; \citet{ottonellowinberry2020}, for
example, estimate their empirical specification on model-simulated data to
compare the model with the data.} and they provide a unified basis for
comparing the convergence rates of the panel VAR and panel LP estimators.
Level responses are a distinct object of interest in their own
right.\footnote{See, e.g., \citet{cloyne2020}, \citet{holm2021transmission},
and \citet{cloyne2023}, who estimate group-specific level responses.} We discuss how heterogeneous and level responses differ in terms of estimation and
inference in Section~\ref{subsec:parametric-scope}.

\subsection{Panel VAR}
\label{subsec:scalar-var}
 
In the scalar model, the recursive estimator is technically a panel autoregression. We refer to it as the panel VAR throughout to keep the terminology consistent with the general model discussed in Section~\ref{sec:parametric}. 
 
The panel VAR fits \eqref{eq:scalar-dgp} by two-way within least squares,
applies the split-panel jackknife (SPJ) correction defined in
Section~\ref{subsec:parametric-var}, and sets
$\widehat\beta_{1,h}^{\VAR}=\widehat\beta_1\widehat\rho^{\,h}$, where
$\widehat\rho$ and $\widehat\beta_1$ are the SPJ-corrected estimates.
 
To make the source of the rate comparison explicit, let
$\widetilde y_{i,t-1}=y_{i,t-1}-\bar y_{t-1}$, where
$\bar y_{t-1}=N^{-1}\sum_{i=1}^N y_{i,t-1}$, denote the residual from removing
time effects. The leading term of $\widehat\beta_1-\beta_1$ is proportional to
\begin{align}
\frac{1}{NT}
\sum_{t=1}^{T}
\sum_{i=1}^{N}
\widetilde y_{i,t-1}\varepsilon_tu_{it}.
\label{eq:var-score}
\end{align}
The one-step error is idiosyncratic, so this average shrinks over both units
and periods.\footnote{The displayed score is illustrative. The actual estimator applies
the two-way within transformation and residualizes the interaction
regressor with respect to the lag controls. The resulting one-step
coefficients are then corrected for their order-$T^{-1}$ dynamic
fixed-effect bias using the SPJ before the response is
constructed recursively. See
equations~\eqref{pa:eq:var-exact} and
\eqref{pa:eq:var-decomposition} of Appendix~\ref{pa:sec:var-decomposition}.}
Under regularity conditions and $N=o(T^2)$,
$\widehat\beta_{1,h}^{\VAR}-\beta_{1,h}=O_p((NT)^{-1/2})$ for every fixed
$h\ge0$. 
 
\begin{remark}[Fixed $T$, large $N$]
\label{rem:nickell}
Much of the dynamic panel literature considers the fixed-$T$, large-$N$
setting. In that setting, the within estimator of \eqref{eq:scalar-dgp}
is inconsistent \citep{nickell1981,alvarezarellano2003}, motivating
estimators that difference out the unit effects and use lagged variables
as instruments \citep{andersonhsiao1982,arellanobond1991}; see also
\citet{blundellbond1998}. Because the aggregate shock varies only over time,
the comparison requires $T\to\infty$. We therefore consider joint
asymptotics with $N,T\to\infty$ \citep{phillipsmoon1999}.
\end{remark}
 
\subsection{Panel LP}
\label{subsec:scalar-lp}
 
Fix $h\ge1$. Iterating \eqref{eq:scalar-dgp} gives
\begin{align}
y_{i,t+h}
&=
\rho^{h+1}y_{i,t-1}
+
\rho^h(\beta_0+\beta_1y_{i,t-1})\varepsilon_t
\nonumber\\
&\quad+
\sum_{j=1}^{h}
\rho^{h-j}(\beta_0+\beta_1y_{i,t+j-1})\varepsilon_{t+j}
+
\sum_{j=0}^{h}
\rho^{h-j}(\alpha_i+u_{i,t+j}).
\label{eq:scalar-iterate}
\end{align}
The corresponding horizon-$h$ projection is
\begin{align}
y_{i,t+h}
&=
\alpha_{i,h}
+
\rho^{h+1}y_{i,t-1}
+
\beta_{0,h}\varepsilon_t
+
\beta_{1,h}y_{i,t-1}\varepsilon_t
+
v_{i,t,h},
\nonumber\\
v_{i,t,h}
&=
\underbrace{
\sum_{j=1}^{h}
\rho^{h-j}(\beta_0+\beta_1y_{i,t+j-1})\varepsilon_{t+j}
}_{\text{aggregate component}}
+
\underbrace{
\sum_{j=0}^{h}
\rho^{h-j}u_{i,t+j}
}_{\text{idiosyncratic component}},
\label{eq:scalar-lp-error}
\end{align}
where $\alpha_{i,h}=\alpha_i\sum_{j=0}^{h}\rho^{h-j}$ and
$\beta_{0,h}=\beta_0\rho^h$. The projection error $v_{i,t,h}$ has mean zero
given $\varepsilon_t$ and all variables dated $t-1$ or earlier. 
%so the coefficient on $y_{i,t-1}\varepsilon_t$ is again $\beta_{1,h}$.
Thus, the
panel VAR and the panel LP target the same heterogeneous response.
The panel LP estimates the horizon-$h$ projection by two-way within least
squares. The two estimators differ in their errors. The one-step error $u_{it}$
is idiosyncratic, whereas the projection error $v_{i,t,h}$ contains
post-impact aggregate shocks, and units differ in their exposure to them.
 
The two components of $v_{i,t,h}$ split the panel LP score into an aggregate
term
\begin{align}
\frac{1}{T}
\sum_{t=1}^{T}
\sum_{j=1}^{h}
\rho^{h-j}\varepsilon_t\varepsilon_{t+j}
\left[
\frac{1}{N}
\sum_{i=1}^{N}
\widetilde y_{i,t-1}
(\beta_0+\beta_1y_{i,t+j-1})
\right]
\label{eq:scalar-lp-score-agg}
\end{align}
and an idiosyncratic term
\begin{align}
\frac{1}{NT}
\sum_{t=1}^{T}
\sum_{i=1}^{N}
\widetilde y_{i,t-1}\varepsilon_t
\sum_{j=0}^{h}\rho^{h-j}u_{i,t+j}.
\label{eq:scalar-lp-score-idio}
\end{align}
The idiosyncratic term \eqref{eq:scalar-lp-score-idio} is the analogue of
\eqref{eq:var-score} and is $O_p((NT)^{-1/2})$. The aggregate term
\eqref{eq:scalar-lp-score-agg} behaves differently. As $N$ grows, the
bracketed cross-sectional average converges to a period-specific limit, but
$\varepsilon_t\varepsilon_{t+j}$ is common to all units, so only averaging
over periods shrinks this term. When $\beta_1=0$, the aggregate term
\eqref{eq:scalar-lp-score-agg} vanishes identically because
$\sum_i\widetilde y_{i,t-1}=0$. When $\beta_1\neq0$,
$\widehat\beta_{1,h}^{\LP}-\beta_{1,h}$ is $O_p(T^{-1/2})$ but not
$o_p(T^{-1/2})$ for every fixed $h\ge1$.
A larger $N$ measures the bracketed cross-sectional average more precisely but
does not create additional realizations of the aggregate shock. The only exception is at impact: The
projection error contains no post-impact aggregate shock, so the aggregate term
is absent and the SPJ-corrected impact estimator attains the $(NT)^{-1/2}$ rate under
$N=o(T^2)$.
 
Consequently, the two estimators differ in the dimensions over which the leading term of
the estimation error averages. The panel VAR uses $N$ to average its one-step
error. The panel LP uses $N$ to measure heterogeneous exposure, but at positive
horizons its estimation error still contains post-impact aggregate shocks,
which are realized once per period. Sections~\ref{sec:parametric}
and~\ref{sec:semiparametric} show that the same mechanism carries over to multivariate models with parametric and nonparametric exposure functions, respectively.

\section{Parametric heterogeneity}
\label{sec:parametric}
 
We now extend the scalar argument of Section~\ref{sec:motivation} to a
general multivariate dynamic panel.
 
\subsection{A vector dynamic panel} \label{sec:para_dgp}
 
Let $y_{it}\in\mathbb{R}^m$ collect microeconomic states and outcomes, with
$y_{1,it}$ the outcome of interest. Denote the aggregate shock by
$\varepsilon_{1t}$ and collect the other aggregate shocks in
$\varepsilon_{-1,t}$. Consider the dynamic model
\begin{align}
A(L)y_{it}
&=
\alpha_i
+
\lambda_t
+
B y_{i,t-1}\varepsilon_{1t}
+
u_{it},
\label{eq:parametric-var}
\end{align}
where $A(L)=I_m-\sum_{\ell=1}^{p}A_\ell L^\ell$, $\alpha_i$ is a unit
effect, $\lambda_t$ is a vector of time effects, and $u_{it}$ is the
one-step error. Note that here the homogeneous component of the shock response in the scalar model \eqref{eq:scalar-dgp} (governed by $\beta_0$) is absorbed into the time effect $\lambda_t$.  The matrix $B\in\mathbb{R}^{m\times m}$ determines how
exposure to $\varepsilon_{1t}$ varies with the predetermined state
$y_{i,t-1}$. We call $y\mapsto By$ the exposure function.

Let $Y_{i,t-1}=(y_{i,t-1}',\ldots,y_{i,t-p}')'$ denote the lag vector, 
and $A_\ell\in\mathbb R^{m\times m}$ denote the VAR coefficient
matrix on the $\ell$th lag, for $\ell=1,\ldots,p$. Define the companion matrix $\mathbf A$ and the selection matrix $J$ by
\begin{equation}\label{eq:A_J}
\mathbf A=
\begin{pmatrix}
A_1 & A_2 & \cdots & A_p\\
I_m & 0   & \cdots & 0\\
0   & I_m & \ddots & \vdots\\
\vdots & \ddots & \ddots & 0\\
0 & \cdots & I_m & 0
\end{pmatrix}
\in\mathbb R^{mp\times mp},
\qquad
J=(I_m,0,\ldots,0)\in\mathbb R^{m\times mp}.
\end{equation}
 Let
$e_1=(1,0,\ldots,0)'$ select the outcome of interest. We write
$\beta_h'=e_1'J\mathbf A^hJ'B$. For two predetermined states $y$ and
$y'$, the heterogeneous response is
\begin{align}
\theta_h(y,y')
&=
\beta_h'(y-y').
\label{eq:parametric-response}
\end{align}
The heterogeneous response function at horizon $h$ is $y\mapsto\beta_h'y$. The matrix $B$ determines impact heterogeneity, and $\mathbf A^h$
propagates it to horizon $h$.
 
Appendix~\ref{pa:sec:assumptions} gives the formal conditions. The
aggregate process $(\varepsilon_{1t},\lambda_t)$ is stationary and
ergodic, and $\varepsilon_{1t}$ is independent of the aggregate
history through date $t-1$.
Conditional on the aggregate path, the unit paths are iid under the
causal representation. The appendix states the required moment and temporal dependence
conditions. Lemma~\ref{pa:lem:population} establishes stability
of $\mathbf A$ under the maintained assumptions. Assumption~\ref{pa:ass:aggregate}
requires the aggregate score defined in Appendix~\ref{app:notation} to be nondegenerate and is
used only for the positive-horizon panel LP result.

\subsection{Panel VAR}
\label{subsec:parametric-var}
 
The panel VAR estimates the dynamic model \eqref{eq:parametric-var} by
two-way within least squares and applies the SPJ
correction to remove the leading dynamic fixed-effect bias.\footnote{The $O(T^{-1})$
dynamic fixed-effect bias of the within estimator \citep{nickell1981}
can dominate the $(NT)^{-1/2}$ sampling error, so we use the split-panel
jackknife correction \citep{hahnkuersteiner2002,DhaeneJochmans2015}.}

Write $D=(B,A_1,\ldots,A_p)$ and
$R_{it}=(\varepsilon_{1t}y_{i,t-1}',Y_{i,t-1}')'$.
For a set $I$ of origin periods, the two-way within estimator is
\begin{equation}\label{eq:para_esti}
\widehat D_I
=\left(
\sum_{i=1}^{N}\sum_{t\in I}
\ddot y_{it,I}\ddot R_{it,I}'
\right)
\left(
\sum_{i=1}^{N}\sum_{t\in I}
\ddot R_{it,I}\ddot R_{it,I}'
\right)^{-1},
\end{equation}
where, for any panel array $a$, we define
$
%\begin{equation}\label{eq :double_within}$
\ddot a_{it,I}
=
a_{it}
-\frac{1}{|I|}\sum_{s\in I}a_{is}
-\frac{1}{N}\sum_{j=1}^{N}a_{jt}
+\frac{1}{N|I|}\sum_{j=1}^{N}\sum_{s\in I}a_{js}.
$
Let $I^{(0)}$ denote the full sample and $I^{(1)},I^{(2)}$
its two equal-length time blocks. The SPJ-corrected estimator is
\begin{equation}\label{eq:hat_D}
\widehat D
=
2\widehat D_{I^{(0)}}
-\frac{1}{2}
\left(\widehat D_{I^{(1)}}+\widehat D_{I^{(2)}}\right)
=
(\widehat B,\widehat A_1,\ldots,\widehat A_p).
\end{equation}

Form $\widehat{\mathbf A}$ based on \eqref{eq:A_J} and estimated $(\widehat A_1,\ldots,\widehat A_p)$, and define $\widehat\beta_h^{\VAR\,\prime}
=e_1'J\widehat{\mathbf A}^{\,h}J'\widehat B$. Then, the panel VAR estimator is given by
\begin{align}
\widehat\theta_h^{\VAR}(y,y')
&=
\widehat\beta_h^{\VAR\,\prime}(y-y').
\label{eq:parametric-var-estimator}
\end{align}
% With an argument similar to Section~\ref{subsec:scalar-var} and under the assumptions specified in Appendix~\ref{pa:sec:assumptions}, we can show that  $\widehat\theta_h^{\VAR}(y,y')$ is
% $O_p((NT)^{-1/2})$. Additionally assuming $N=o(T^2)$, the SPJ-corrected one-step coefficients converge at rate
% $(NT)^{-1/2}$. For fixed $h$, the recursion preserves this rate.

Write
\(
\psi=
\bigl(
\mathrm{vec}(A_1)',\ldots,\mathrm{vec}(A_p)',
\mathrm{vec}(B)'
\bigr)'.
\)
Let $\Sigma_\psi$ denote the covariance matrix defined in
\eqref{eq:H_Sigma_Gh} of Appendix~\ref{app:notation}, and define
\(
g_h(\psi)'=e_1'J\mathbf A^hJ'B,
\qquad
G_h=\frac{\partial g_h(\psi)}{\partial\psi'}.
\)

% Write
% $\psi=(\mathrm{vec}(A_1)',\ldots,\mathrm{vec}(A_p)',
% \mathrm{vec}(B)')'$, and  $\Sigma_\psi$ is defined in \eqref{eq:H_Sigma_Gh} in Appendix~\ref{app:notation}. Define
% $g_h(\psi)'=e_1'J\mathbf A^hJ'B$ and
% $G_h=\partial g_h(\psi)/\partial\psi'$.
 
\begin{theorem}[Asymptotic normality of the panel VAR estimator]
\label{prop:parametric-var}
Let Assumptions~\ref{pa:ass:sampling}--\ref{pa:ass:dependence}
in Appendix~\ref{pa:sec:assumptions} hold and let $N=o(T^2)$.
For any fixed $h\ge0$ and $d=y-y'\ne0$, it holds that
\begin{align}
\sqrt{NT}
\left(
\widehat\theta_h^{\VAR}(y,y')
-
\theta_h(y,y')
\right)
&\xrightarrow{d}
N\!\left(
0,\,
d'G_h\Sigma_\psi G_h'd
\right).
\label{eq:parametric-var-limit}
\end{align}
\end{theorem}
The proof is provided in Appendix~\ref{pa:sec:proposition-proof}.
 
Theorem~\ref{prop:parametric-var} relies on the correct specification
of the dynamic model \eqref{eq:parametric-var}. See 
Section~\ref{subsec:parametric-scope} for a discussion.

\subsection{Panel LP}
\label{subsec:parametric-lp}
 
Fix $h\ge1$. Iterating \eqref{eq:parametric-var} gives the horizon-$h$
panel LP
\begin{align}
y_{1,i,t+h}
&=
\alpha_{i,h}
+
\lambda_{t,h}
+
\beta_h'y_{i,t-1}\varepsilon_{1t}
+
c_h'Y_{i,t-1}
+
v_{i,t,h},
\label{eq:parametric-lp}
\\
v_{i,t,h}
&=
\underbrace{
\sum_{j=1}^{h}
\beta_{h-j}'y_{i,t+j-1}\varepsilon_{1,t+j}
}_{\text{aggregate component}}
+
\underbrace{
\sum_{j=0}^{h}
e_1'J\mathbf A^{h-j}J'u_{i,t+j}
}_{\text{idiosyncratic component}}.
\label{eq:parametric-lp-error}
\end{align}
Here $c_h'=e_1'J\mathbf A^{h+1}$ gives the lag coefficients,
$\alpha_{i,h}$ collects the propagated unit effects, and
$\lambda_{t,h}$ collects components common to all units at date $t$.
The coefficient on $y_{i,t-1}\varepsilon_{1t}$ is $\beta_h$, so the
panel LP estimates the same heterogeneous response
\eqref{eq:parametric-response} as the panel VAR.
 
Let $I_h$ denote the set of origin periods $t$ for which $y_{1,i,t+h}$ and
$Y_{i,t-1}$ are observed, and let $T_h=|I_h|$. The panel LP estimates
\eqref{eq:parametric-lp} by two-way within least squares over $t\in I_h$.
Let $\widehat\beta_h^{\LP}$ denote the estimated coefficient on
$y_{i,t-1}\varepsilon_{1t}$, and let
$\widehat\theta_h^{\LP}(y,y')=\widehat\beta_h^{\LP\,\prime}(y-y')$.
A decomposition similar to \eqref{eq:scalar-lp-score-agg} and
\eqref{eq:scalar-lp-score-idio} gives an aggregate term of order
$O_p(T^{-1/2})$ and an idiosyncratic term of order $O_p((NT)^{-1/2})$. For
$h\geq 1$, let $\Omega_h$ and $Q_h$ be defined in Appendix~\ref{app:notation}.
 
\begin{theorem}[Asymptotic normality of the panel LP estimator]
\label{prop:parametric-lp}
Let Assumptions~\ref{pa:ass:sampling}--\ref{pa:ass:aggregate}
in Appendix~\ref{pa:sec:assumptions} hold. For any fixed $h\ge1$ and $d=y-y'\ne0$, it holds that
\begin{align}
\sqrt T
\left(
\widehat\theta_h^{\LP}(y,y')
-
\theta_h(y,y')
\right)
&\xrightarrow{d}
N\!\left(
0,\,
d'Q_h^{-1}\Omega_hQ_h^{-1}d
\right).
\label{eq:parametric-lp-limit}
\end{align}
\end{theorem}
The proof is provided in Appendix~\ref{pa:sec:proposition-proof}.
 
Theorems~\ref{prop:parametric-var} and~\ref{prop:parametric-lp}
give the multivariate counterpart of the scalar comparison. At positive
horizons, the panel VAR has effective sample size $NT$ under the maintained
dynamic model, whereas the panel LP has effective sample size $T$ when the
aggregate score is nondegenerate. At $h=0$, there is no post-impact
aggregate shock, so the panel LP estimator coincides with the panel VAR impact
estimator (both after SPJ correction) and converges at rate
$(NT)^{-1/2}$ under $N=o(T^2)$.

\subsection{Scope of the dynamic model}
\label{subsec:parametric-scope}
 
The panel LP estimates the response separately at each horizon, whereas
the panel VAR constructs it from the dynamic model
\eqref{eq:parametric-var}. The rate comparison above applies under the maintained
dynamic model and assumptions. Many data-generating processes considered in the micro-panel literature meet these conditions. In this subsection, we discuss several such settings.

% {\blue [mx: last sentence not very clear.

% The rate comparison above applies under the maintained
% dynamic model and assumptions. Many data-generating processes considered in the micro-panel literature meet these conditions. In this subsection, we discuss several such settings.]}
 
\subsubsection{Homogeneous aggregate effects}
 
Let $\pi_{t-1}$ collect predetermined aggregate states, and consider
\begin{align}
A(L)y_{it}
&=
\alpha_i
+\Phi\pi_{t-1}
+\Theta\varepsilon_{1t}
+By_{i,t-1}\varepsilon_{1t}
+u_{it},
\label{eq:parametric-var-aggregate}
\end{align}
where $\Theta\in\mathbb R^m$ is the homogeneous impact response to
$\varepsilon_{1t}$, the counterpart of $\beta_0$ in the scalar model
\eqref{eq:scalar-dgp}. The terms $\Phi\pi_{t-1}$ and
$\Theta\varepsilon_{1t}$ are common to all units and are therefore
absorbed by the time effects $\lambda_t$ in \eqref{eq:parametric-var}.
Homogeneous effects of the other aggregate shocks $\varepsilon_{-1,t}$
are absorbed in the same way. Thus, neither the aggregate variables nor
their coefficients need to be estimated.
 
This holds even if the aggregate states respond to $\varepsilon_{1t}$:
with common coefficients, their contributions remain common to all
units and are absorbed by the time effects. No law of motion for
$\pi_t$ is therefore needed. Provided the implied
$(\varepsilon_{1t},\lambda_t)$ satisfies the conditions of
Theorems~\ref{prop:parametric-var}
and~\ref{prop:parametric-lp}, their conclusions continue to apply: at positive horizons, the panel VAR has effective
sample size $NT$, whereas the panel LP has effective sample size $T$.
 
\subsubsection{Heterogeneous effects of other shocks}
\label{subsubsec:parametric-other-shocks}
 
Suppose that an element $\varepsilon_{2t}$ of $\varepsilon_{-1,t}$
has heterogeneous effects and enters the dynamic model through
$B_2y_{i,t-1}\varepsilon_{2t}$, where $B_2\in\mathbb R^{m\times m}$.
For example, a study of monetary policy shocks may also face technology
shocks whose effects on units vary with their predetermined states.
 
Assume that $\varepsilon_{1t}$ remains independent of the joint
past of the expanded system, and that $\varepsilon_{2t}$ has
mean zero conditional on that past and $\varepsilon_{1t}$. If $\varepsilon_{2t}$ is observed, the interaction
$y_{i,t-1}\varepsilon_{2t}$ should be added to the regressors of the
panel VAR. The one-step error then remains idiosyncratic, and the
panel VAR converges at rate $(NT)^{-1/2}$ as in
Theorem~\ref{prop:parametric-var}.
 
If $\varepsilon_{2t}$ is unobserved, omitting its interaction leaves
the population values of $B$ and $A_\ell$ unchanged, so the panel VAR
remains consistent. Its
averaged score for $B$, however, contains
\begin{align}
\frac{1}{T}\sum_{t=1}^{T}
\varepsilon_{1t}\varepsilon_{2t}B_2
\left[
\frac{1}{N}\sum_{i=1}^{N}
\widetilde y_{i,t-1}\widetilde y_{i,t-1}'
\right],
\label{eq:parametric-omitted-score}
\end{align}
where $\widetilde y_{i,t-1}=y_{i,t-1}-\bar y_{t-1}$ as in
Section~\ref{subsec:scalar-var}. As in \eqref{eq:var-score}, the
displayed score is illustrative: it omits the removal of unit effects
and the residualization with respect to the lag vector $Y_{i,t-1}$.
As $N$ grows, the bracketed average converges to a period-specific
limit, but $\varepsilon_{1t}\varepsilon_{2t}$ is common to all units,
so this term shrinks only through averaging over periods. If the aggregate component in
\eqref{eq:parametric-omitted-score} has a nonzero long-run
variance, the panel VAR therefore converges at rate $T^{-1/2}$,
including at impact. At positive horizons, the panel
LP still identifies $\beta_h$ and converges at rate $T^{-1/2}$.
 
\subsubsection{Heterogeneous propagation}
\label{subsubsec:parametric-propagation}
 
In \eqref{eq:parametric-var}, exposure to $\varepsilon_{1t}$ varies
with the predetermined state, but the lag coefficients $A_\ell$ are
common to all units and periods.\footnote{Dynamic panels with
unit-specific lag coefficients $A_{\ell i}$, as in
\citet{pesaransmith1995}, are also outside the specification. Iterating
pooled lag coefficients need not recover the heterogeneous
response.} Let $\pi_{t-1}\in\mathbb R^k$ and suppose the dynamic
model contains $C(\pi_{t-1}\otimes y_{i,t-1})$, where
$C\in\mathbb R^{m\times km}$. Because
$C(\pi_{t-1}\otimes y_{i,t-1})=C(\pi_{t-1}\otimes I_m)y_{i,t-1}$, this
interaction makes the first-lag coefficient depend on the aggregate
state:
\begin{align}
A_{1t}=A_1+C(\pi_{t-1}\otimes I_m).
\label{eq:parametric-state-lag}
\end{align}
Post-impact aggregate shocks $\varepsilon_{1,t+j}$ and
$\varepsilon_{2,t+j}$, $j\ge1$, also make the lag coefficient vary over
time, but have mean zero conditional on the
past. Their interactions therefore do not change the expected
propagation of the response to $\varepsilon_{1t}$. The aggregate state
$\pi_{t-1}$, in contrast, can be persistent and can respond to
$\varepsilon_{1t}$.
 
If $\pi_t$ responds to $\varepsilon_{1t}$, the shock also changes
$A_{1,t+1}$, and $C(\pi_t\otimes y_{it})$ contains the product of the
responses of $\pi_t$ and $y_{it}$ to the shock. The conditional mean of
$y_{1,i,t+h}$ can therefore contain a term in $\varepsilon_{1t}^2$
whose coefficient varies with $y_{i,t-1}$, and need not be linear in
$\varepsilon_{1t}$. For the panel VAR, including the interaction keeps the one-step error idiosyncratic, but iterating
the dynamic model requires a law of motion for $\pi_t$. For the panel
LP, exogeneity of $\varepsilon_{1t}$ does not identify the
state-dependent response when the conditional mean is nonlinear in the
shock \citep{david2026causal}.
 
With common propagation, the heterogeneous component of the conditional mean is linear in
$\varepsilon_{1t}$, satisfying the identification condition of
\citet{david2026causal}. Because the response is also affine in
$y_{i,t-1}$, the linear interaction recovers the response itself rather
than a projection of it.
 
\subsubsection{Heterogeneous and level responses}
\label{subsubsec:parametric-level}
 
Section~\ref{sec:motivation} distinguishes the heterogeneous
response $\theta_h(y,y')$ from the level response
$b_h(y)$. Suppose the homogeneous component of the response is linear
in $\varepsilon_{1t}$ and does not depend on the aggregate state before
the shock. Then $b_h(y)=b_h(0)+\beta_h'y$ and
$\theta_h(y,y')=\beta_h'(y-y')$.
 
Time effects absorb $b_h(0)$, so both estimators identify
$\theta_h(y,y')$ but not $b_h(y)$. Estimating the level response
requires a coefficient on $\varepsilon_{1t}$ itself, such as $\Theta$
in \eqref{eq:parametric-var-aggregate}, which is collinear with
unrestricted time effects. This is the missing intercept problem:
cross-sectional differences do not identify effects common to all
units
\citep{nakamura2014fiscal,guren2021crossregional,wolf2023missing}.
 
Without time effects, $\Phi\pi_{t-1}$ in
\eqref{eq:parametric-var-aggregate} and the homogeneous effects of the
other aggregate shocks $\varepsilon_{-1,t}$ are no longer absorbed, and
the panel VAR must account for them. When $\pi_t$ responds to
$\varepsilon_{1t}$, constructing the level response recursively
requires a law of motion for $\pi_t$. Estimating this law from
aggregate time series limits the level response to rate $T^{-1/2}$.
Aggregate terms that are omitted enter the one-step error as a
component common to all units, which also limits the rate to
$T^{-1/2}$, and omitted aggregate states that are correlated with the
lagged outcomes make the estimated lag coefficients inconsistent.
 
The panel LP identifies $b_h(y)$ by dropping $\lambda_{t,h}$ from
\eqref{eq:parametric-lp} and adding $\varepsilon_{1t}$ as a regressor.
Because the homogeneous component is linear in $\varepsilon_{1t}$,
the coefficient on $\varepsilon_{1t}$ recovers $b_h(0)$, including the effect through the aggregate states,
without estimating a law of motion for $\pi_t$. At positive horizons,
post-impact aggregate shocks remain in the projection error, so the
panel LP converges at rate $T^{-1/2}$ whether or not the other aggregate
terms are included.

\section{Nonparametric heterogeneity}
\label{sec:semiparametric}
 
Section~\ref{sec:parametric} restricts heterogeneous exposure to be linear in the
predetermined state. We now replace $B y_{i,t-1}\varepsilon_{1t}$ by
$\mu(y_{i,t-1})\varepsilon_{1t}$, where $\mu:\mathcal Y\to\mathbb{R}^m$ is an unknown bounded
smooth function on the state space $\mathcal Y$. The dynamic model becomes
\begin{align}
A(L)y_{it}
=
\alpha_i
+
\lambda_t
+
\mu(y_{i,t-1})\varepsilon_{1t}
+
u_{it}.
\label{eq:semi-var-model}
\end{align}
This leaves the propagation law unchanged and relaxes only the shape of
heterogeneous exposure. \citet{david2026causal} give conditions under which
state-dependent projections in micro--macro panels admit a causal reading and
propose a sieve estimator; \citet{goncalves2026semiparametric} develop the
corresponding time-series methods.  Our focus is different: we compare the convergence rates of
nonparametric response estimates from panel VARs and panel LPs, allowing responses to aggregate shocks to vary
nonlinearly with the micro state.
 
Because $\lambda_t$ absorbs any component of $\mu(\cdot)\varepsilon_{1t}$ that is common
across units, $\mu(\cdot)$ is identified only up to an additive constant.
As in Section~\ref{sec:parametric}, $\mu$ is the exposure function: $\mu(y)$ is the impact loading on
$\varepsilon_{1t}$ of a unit in state $y$. With $\mathbf A$, $J$, and $e_1$ as
in Section~\ref{sec:para_dgp}, the heterogeneous response function at horizon
$h$ is $\mu_h(y)=e_1'J\mathbf A^hJ'\mu(y)$, which is also identified only up
to an additive constant. The heterogeneous response
$\theta_h(y,y')=\mu_h(y)-\mu_h(y')$ does not depend on this constant and is
no longer linear in $y-y'$.
 
% {\red The formal conditions are in Appendix~\ref{np:sec:assumptions}, using
% the common sampling condition from Appendix~\ref{app:parametric}.
% They require a compact invariant state space, raw-series approximation,
% within identification, conditional design bounds, and temporal moment
% conditions. Compact invariance is a substantive support restriction,
% not merely a finite-moment assumption. The conditional design bound
% does not follow from unconditional density comparability alone.
% A separate series-block nondegeneracy condition is used only for the
% nonvanishing LP component.}
 
\subsection{Nonparametric panel VAR and panel LP}
\label{subsec:semi-lp}
\label{subsec:semi-rates}
Let $F_y$ be the stationary marginal distribution of the predetermined
state $y_{i,t-1}$.
The nonparametric panel VAR approximates $\mu$ by a deterministic raw basis 
$p_K(y)\in\mathbb R^K$ without constant direction.
 
Replace $R_{it}$ defined in Section~\ref{subsec:parametric-var} by 
$
R_{it,K}=\bigl(\varepsilon_{1t}p_K(y_{i,t-1})',\,Y_{i,t-1}'\bigr)'. 
$  Let
$
B_K=\left[\int\mu(y)\{p_K(y)-\bar p_K\}'\,dF_y(y)\right]M_K^{-1}
\in\mathbb R^{m\times K},\quad
\bar p_K=\int p_K\,dF_y,\quad
M_K=\int(p_K-\bar p_K)(p_K-\bar p_K)'\,dF_y
$
denote the population slope of the projection with an intercept. The intercept is absorbed by time effects, while $B_Kp_K(y)$ approximates the varying part of $\mu(y)$.
We estimate $B_K$ by $\widehat B_K$, obtained as $\widehat B$ in
\eqref{eq:para_esti}--\eqref{eq:hat_D} with $R_{it}$ replaced by $R_{it,K}$.
The nonparametric panel VAR estimator is given by
\begin{equation}
\widehat\theta_h^{\VAR}(y,y')
=\widehat\mu_h^{\VAR}(y)-\widehat\mu_h^{\VAR}(y'),
\end{equation}
where $\widehat\mu_h^{\VAR}(y)
=e_1'J\widehat{\mathbf A}^{\,h}J'\widehat B_Kp_K(y)$.

For the nonparametric panel LP, stacking \eqref{eq:semi-var-model} in companion form gives the one-step recursion
$Y_{it}=\mathbf AY_{i,t-1}+J'[\alpha_i+\lambda_t+\mu(y_{i,t-1})\varepsilon_{1t}+u_{it}]$,
in which the shock interaction is a date-$t$ innovation to the state whose loading
is predetermined given $y_{i,t-1}$. Iterating $h+1$ times from $t-1$ to $t+h$,
\begin{align*}
Y_{i,t+h}
&=
\mathbf A^{h+1}Y_{i,t-1}
+
\sum_{j=0}^{h}\mathbf A^{h-j}J'
\left[\alpha_i+\lambda_{t+j}+\mu(y_{i,t+j-1})\varepsilon_{1,t+j}+u_{i,t+j}\right].
\end{align*}
With $\mu_{h-j}(\cdot)=e_1'J\mathbf A^{h-j}J'\mu(\cdot)$, the first element
of $Y_{i,t+h}$ satisfies
\begin{align}
y_{1,i,t+h}
&=
\alpha_{i,h}
+
\lambda_{t,h}
+
c_h'Y_{i,t-1}
+
\mu_h(y_{i,t-1})\varepsilon_{1t}
+
v_{i,t,h},
\label{eq:semi-lp}
\\[2pt]
v_{i,t,h}
&=
\underbrace{
\sum_{j=1}^{h}\mu_{h-j}(y_{i,t+j-1})\varepsilon_{1,t+j}
}_{\text{aggregate component}}
+
\underbrace{
\sum_{j=0}^{h}e_1'J\mathbf A^{h-j}J'u_{i,t+j}
}_{\text{idiosyncratic component}} .
\nonumber
\end{align}
% Setting $\mu(y)=By$ returns $\mu_{h-j}(y)=\beta_{h-j}'y$ and recovers
% \eqref{eq:parametric-lp} and \eqref{eq:parametric-lp-error} term by term, so the
% nonparametric loading replaces the linear one without changing which components are
% common across units and which are not. In particular the aggregate component is
% again not absorbed by any additive date effect: the cross-sectional mean of
% $\mu_{h-j}$ is common and is absorbed by the date effect in estimation,
% whereas the deviations survive. This removal is distinct from the fixed
% $\nu$-normalization of the function. Note also that the projection never separates $\mathbf A$ from
% $\mu(\cdot)$: it estimates the composite $\mu_h$ in one step, which is the usual
% robustness argument for direct estimation and the counterpart of the precision cost
% isolated here.
 
The panel LP uses the same $R_{it,K}$ and two-way within regression as in
Section~\ref{subsec:parametric-lp}.
Let $\widehat b_{h,K}^{\LP}$ denote the $K$ coefficients on
$\varepsilon_{1t}p_K(y_{i,t-1})$ in the horizon-$h$ regression
\eqref{eq:semi-lp}. Its response estimator is
\begin{equation}
\widehat\theta_h^{\LP}(y,y')
 =\{p_K(y)-p_K(y')\}'\widehat b_{h,K}^{\LP},\qquad h\geq1.
\label{eq:semi-lp-estimator}
\end{equation}
At impact, we set $\widehat\theta_0^{\LP}=\widehat\theta_0^{\VAR}$.
 
% For the population score, write $r_{it,K}$ for the additive residual
% of the stacked interaction and lag regressors, defined in the appendix.
% Its aggregate component is
% \begin{align*}
% G_{h,K}
% =\frac1{T_h}\sum_{t=1}^{T_h}\sum_{j=1}^h
% \varepsilon_{1,t+j}\mathbb{E}[r_{it,K}\mu_{h-j}(y_{i,t+j-1})\mid\mathcal A].
% \end{align*}
% The corresponding raw micro score is
% \begin{align*}
% \frac1{NT_h}\sum_{t=1}^{T_h}\sum_{i=1}^N r_{it,K}
% \sum_{j=0}^h e_1'J\mathbf A^{h-j}J'u_{i,t+j}.
% \end{align*}
% The inverse Gram matrix maps these stacked scores into the function
% estimate. The conditional design bound controls the norm of the
% aggregate loading uniformly in $K$, giving $O_p(T^{-1/2})$ without an
% extra $\sqrt K$. Conditional cross-sectional averaging gives the
% micro term order $\sqrt{K/(NT)}$, and approximation contributes
% $K^{-r/m}$. The appendix also controls sample demeaning and inverse-Gram
% remainders. At impact the sum over future aggregate innovations is
% empty, and the bias-corrected one-step fit defines the impact estimator.
 
For integrated errors, write
$\|\theta\|_{L^2(\mathcal Y\times\mathcal Y)}^2
=\iint|\theta(y,y')|^2dF_y(y)dF_y(y')$.
 
\begin{theorem}[Rates of the nonparametric panel VAR and panel LP]
\label{prop:semiparametric-rates}
Suppose Assumption~\ref{pa:ass:sampling} and Assumptions~\ref{np:ass:solution}--\ref{np:ass:technical}
in Appendix~\ref{np:sec:assumptions} hold. If additionally $N=o(T^2)$,
then for every fixed $h\geq0$,
\begin{equation}
\|\widehat\theta_h^{\VAR}-\theta_h\|_{L^2(\mathcal Y\times\mathcal Y)}
 =O_p(\sqrt{K/(NT)}+K^{-r/m}).
\label{np:eq:var-rate}
\end{equation}
For every fixed $h\geq1$, without requiring $N=o(T^2)$,
\begin{equation}
\|\widehat\theta_h^{\LP}-\theta_h\|_{L^2(\mathcal Y\times\mathcal Y)}
 =O_p(T^{-1/2}+\sqrt{K/(NT)}+K^{-r/m}).
\label{np:eq:lp-rate}
\end{equation}
At impact, under $N=o(T^2)$, the panel LP estimator coincides with the
SPJ-corrected impact estimator and satisfies \eqref{np:eq:var-rate}.
 
For a fixed $h\geq1$, if additionally Assumption~\ref{np:ass:floor} holds and
\begin{equation}
\sqrt{K/(NT)}+K^{-r/m}=o(T^{-1/2}),
\label{np:eq:floor-regime}
\end{equation}
then there exist constants $c_1,c_2>0$, independent of $N,T,K$, such that
\begin{equation}
\liminf_{N,T\to\infty}
 \mathbb P\!\left(\sqrt T\,
 \|\widehat\theta_h^{\LP}-\theta_h\|_{L^2(\mathcal Y\times\mathcal Y)}>c_1\right)\geq c_2.
\label{np:eq:floor}
\end{equation}
 
\end{theorem}
The proof is provided in Appendix~\ref{np:sec:rates-proof}. The LP bound separates three sources of error.
The term $T^{-1/2}$ reflects uncertainty from post-impact aggregate
shocks, $\sqrt{K/(NT)}$ reflects estimation of the series coefficients
from micro observations, and $K^{-r/m}$ is the approximation error.
A larger series order reduces approximation error but increases
coefficient estimation error. When both series terms are
$o(T^{-1/2})$, neither changes the leading convergence rate of
the panel LP. This is the source of the nonparametric
``free lunch'': the additional estimation cost is smaller than
the aggregate-shock uncertainty already present in the LP.
 
Appendix~\ref{np:sec:integrated-rates} gives admissible choices of the series order.
Pointwise normality requires additional conditions on approximation
and the scalar scores.
 
\begin{theorem}[Normality of the nonparametric panel VAR and panel LP]
\label{prop:semiparametric-normality}
Fix $y\ne y'$ and $h$. Suppose Assumption~\ref{pa:ass:sampling},
Assumptions~\ref{np:ass:solution}--\ref{np:ass:technical}, and the
applicable parts of Assumption~\ref{np:ass:pointwise} hold.
With the variances defined in Appendix~\ref{np:sec:pointwise},
for $h\geq0$,
\begin{equation}
\frac{\sqrt{NT}}{s_{h,N,K}}
\{\widehat\theta_h^{\VAR}(y,y')-\theta_h(y,y')\}
 \xrightarrow{d}N(0,1).
\label{np:eq:var-clt}
\end{equation}
For $h\geq1$,
\begin{equation}
\sqrt T\{\widehat\theta_h^{\LP}(y,y')-\theta_h(y,y')\}
 \xrightarrow{d}N(0,\tau_h^2).
\label{np:eq:lp-clt}
\end{equation}
At $h=0$, the panel LP estimator equals the SPJ-corrected impact estimator and obeys
\eqref{np:eq:var-clt}.
\end{theorem}
The proof is provided in Appendix~\ref{np:sec:pointwise-proof}.
Here $s_{h,N,K}/\sqrt{NT}$ is the pointwise standard deviation of the panel VAR estimator,
which can depend on the series order, while $\tau_h^2>0$ is the limiting
aggregate variance for the selected pair $(y,y')$.

\section{Hausman-type test, efficiency, and robustness}
\label{sec:misspecification}
% {\blue [mx: This title only concerns LP. Might change to "Choice between the efficiency and robustness"]}

% The preceding sections establish differences in precision between panel VARs and panel LPs. This raises a natural question: how should researchers choose between them? The choice entails a trade-off between the efficiency of panel VARs and the robustness of panel LPs. In this section, we provide a Hausman-type test to assist researchers to make such a choice. 

The preceding sections establish a precision advantage for the
panel VAR under the maintained dynamic model. We now examine
the specification restrictions underlying  that advantage and
develop a Hausman-type test to inform the choice between
panel VAR and panel LP. Section~\ref{subsec:hausman} introduces the test and establishes its
asymptotic size and power under fixed alternatives.
Sections~\ref{subsec:dynamic-misspecification} and~\ref{subsec:finite-memory}
establish its power against a panel VAR with omitted variables and a
panel VARMA, respectively. Section~\ref{subsec:lp-variance}
provides a feasible estimator of the panel LP asymptotic variance used
in the test.

\subsection{A Hausman-type test}
\label{subsec:hausman}

The comparison has the structure of \citet{hausman1978}. 
% Under the
% correctly specified one-step law and at a positive horizon $h\geq 1$, both estimators recover the same
% response coefficients, while the panel VAR converges faster
% (Section~\ref{sec:parametric}). Because $\theta_h(y,y')=\beta_h'(y-y')$ is linear in
% $y-y'$, testing the coefficients $\beta_h$ tests the response at every $(y,y')$.
Fix $h\geq1$. Let $\widehat\beta_h^{\LP}$ and
$\widehat\beta_h^{\VAR}$ denote the estimators in
Section~\ref{sec:parametric}. Under the null, write
$V_h^{\LP}=Q_h^{-1}\Omega_hQ_h^{-1}$ and let $\widehat V_h^{\LP}$
be its feasible estimator from Section~\ref{subsec:lp-variance}.
Define
\begin{align}
\mathcal H_h
&=
T(\widehat\beta_h^{\LP}-\widehat\beta_h^{\VAR})'
(\widehat V_h^{\LP})^{-1}
(\widehat\beta_h^{\LP}-\widehat\beta_h^{\VAR}).
\label{eq:hausman-statistic}
\end{align}
Note that under a correctly specified dynamic model, both estimators recover the same
target, while the panel VAR converges faster
(Section~\ref{sec:parametric}). This faster convergence makes its estimation error negligible at
the LP scale, so the test requires only the panel LP asymptotic variance.
\begin{theorem}[Size and fixed-alternative power]
\label{thm:hausman}
Fix $h\ge1$. Under the null that \eqref{eq:parametric-var} is correctly specified,
suppose Assumptions~\ref{pa:ass:sampling}--\ref{pa:ass:aggregate}
and the variance conditions of
Lemma~\ref{lem:lp-hac} hold. With $\widehat V_h^{\LP}$ defined in
\eqref{eq:lp-feasible-variance},
\[
\mathcal H_h\xrightarrow{d}\chi_m^2.
\]
Consequently, the test that rejects when $\mathcal H_h>\chi^2_{m,1-\alpha}$
has asymptotic size $\alpha$ for any $\alpha\in(0,1)$.

Under a fixed alternative where
$\widehat{\beta}_h^{\LP}-\widehat{\beta}_h^{\VAR}
\xrightarrow{p}\Delta_h\neq 0$
and $\widehat V_h^{\LP}\xrightarrow{p}V_{h,\mathrm{alt}}\succ 0$
with $V_{h,\mathrm{alt}}$ finite, 
\begin{equation}
\frac{\mathcal H_h}{T}
\xrightarrow{p}
\Delta_h'V_{h,\mathrm{alt}}^{-1}\Delta_h>0,
\label{eq:hausman-fixed-alternative}
\end{equation}
and the rejection probability tends to one.
\end{theorem}
The proof is provided in Appendix~\ref{app:hausman-proof}. \begin{remark}
\label{rem:hausman-scope}
The null model~\eqref{eq:parametric-var} assumes an idiosyncratic
one-step error.
An omitted aggregate shock with heterogeneous exposure can
violate this assumption while leaving the panel VAR consistent,
as discussed in Section~\ref{subsubsec:parametric-other-shocks}.
\end{remark}
%Rejection need not imply inconsistency of the panel VAR response estimator: such an omitted shock may preserve consistency while invalidating the faster-rate argument underlying the test. 

\subsection{Misspecification: panel VAR with omitted variables}
\label{subsec:dynamic-misspecification}

Let $y_{it}^*\in\mathbb R^{m^*}$ denote the full vector of states, with
$y_{1,it}$ the outcome of interest, and let $y_{it}\in\mathbb R^m$ denote
the observed subvector. The true DGP is
\begin{align}
A^*(L)y_{it}^*
&=
\alpha_i^*+\lambda_t^*
+B^*y_{i,t-1}\varepsilon_{1t}
+u_{it}^*,
\label{eq:dynamic-misspecification-dgp}
\end{align}
where $A^*(L)=I_{m^*}-\sum_{\ell=1}^{p^*}A_\ell^*L^\ell$, the lag order
$p^*$ is unknown, and $B^*\in\mathbb R^{m^*\times m}$. We refer to this
alternative as a panel VAR with omitted variables: a panel VAR fitted to
$y_{it}$ with $p$ lags omits the variables in $y_{it}^*$ outside $y_{it}$
and, if $p<p^*$, the lags beyond $p$. Let
$\mathbf A^*$ and $J^*$ denote the companion matrix and selection matrix formed
from $A_1^*,\ldots,A_{p^*}^*$ as in \eqref{eq:A_J}, and let $e_1^*$
select $y_{1,it}$ from $y_{it}^*$. Averaging over post-impact aggregate
shocks, the horizon-$h$ coefficient on $y_{i,t-1}\varepsilon_{1t}$ is
$\beta_h^{*\,\prime}=e_1^{*\,\prime}J^*(\mathbf A^*)^hJ^{*\,\prime}B^*$,
and the true heterogeneous response is
\begin{align}
\theta_h^*(y,y')
&=
\beta_h^{*\,\prime}(y-y')
=
e_1^{*\,\prime}J^*(\mathbf A^*)^hJ^{*\,\prime}B^*(y-y').
\label{eq:dynamic-misspecification-response}
\end{align}

The practitioner estimates the panel LP of
Section~\ref{subsec:parametric-lp} on the observed vector with $p$ lags,
\begin{align}
y_{1,i,t+h}
&=
\alpha_{i,h}+\lambda_{t,h}
+\beta_h^{\LP\,\prime}y_{i,t-1}\varepsilon_{1t}
+c_h'Y_{i,t-1}
+v_{i,t,h},
\label{eq:dynamic-misspecification-lp}
\end{align}
where $Y_{i,t-1}=(y_{i,t-1}',\ldots,y_{i,t-p}')'$. Here $\beta_h^{\LP}$ and $c_h$ denote the population projection coefficients on
$y_{i,t-1}\varepsilon_{1t}$ and $Y_{i,t-1}$ in the regression
\eqref{eq:dynamic-misspecification-lp}, after removing the two fixed
effects and partialling out the stated controls.

At the target coefficient $\beta_h^*$, the DGP
\eqref{eq:dynamic-misspecification-dgp} gives the candidate LP error
\begin{align}
v_{i,t,h}
={}&
\underbrace{e_1^{*\,\prime}J^*(\mathbf A^*)^{h+1}Y^*_{i,t-1}
-c_h'Y_{i,t-1}}_{P_{i,t,h}}\nonumber\\
&+\underbrace{\sum_{j=1}^{h}\beta_{h-j}^{*\,\prime}
 y_{i,t+j-1}\varepsilon_{1,t+j}}_{G_{i,t,h}}
+\underbrace{\sum_{j=0}^{h}e_1^{*\,\prime}J^*
 (\mathbf A^*)^{h-j}J^{*\,\prime}u^*_{i,t+j}}_{U_{i,t,h}},
\label{eq:dynamic-lp-error}
\end{align}
up to additive unit and time effects, where
$Y^*_{i,t-1}=(y_{i,t-1}^{*\,\prime},\ldots,y_{i,t-p^*}^{*\,\prime})'$.
For the population residual $z_{it}$ defined in
\eqref{eq:Q_CLC block}, evaluated under the present DGP, 
the term $P_{i,t,h}$ is predetermined; the shocks in $G_{i,t,h}$ are
fresh at their respective periods, and the micro innovations are
independent. Hence
\[
\mathbb E[z_{it}v_{i,t,h}]
 =\mathbb E[z_{it}P_{i,t,h}]
  +\mathbb E[z_{it}G_{i,t,h}]
  +\mathbb E[z_{it}U_{i,t,h}]=0.
\]
Thus $\beta_h^{\LP}=\beta_h^*$, and the stated moment convergence
conditions give $\widehat\beta_h^{\LP}\xrightarrow{p}\beta_h^*$.
Lemma~\ref{lem:dynamic-lp-identification} in
Appendix~\ref{app:dynamic-misspecification-proof} establishes these claims.

The panel VAR instead fits \eqref{eq:parametric-var} to $y_{it}$ with
$p$ lags. Let $A_1^\dagger,\ldots,A_p^\dagger$ and $B^\dagger$ denote the resulting
population projection coefficients, and let $\mathbf A^\dagger$ be their
companion matrix, so $\beta_h^{\VAR\,\prime}=e_1'J(\mathbf A^\dagger)^hJ'B^\dagger$.
%Let $A_1,\ldots,A_p$ and $B$ now denote its population
% projection coefficients and $\mathbf A$ the associated companion matrix.
% Since the omitted terms are predetermined, $B$ still recovers the impact
% exposure of the observed state, but the panel VAR propagates it through
% the fitted observed-state law,
Omitted lags or variables are generally correlated with the
included lag controls, leading to omitted-variable bias in the
estimated lag coefficients.
%$\beta_h^{\VAR\,\prime}=e_1'J\mathbf A^hJ'B$. 
In general,
% \begin{align}
% \theta_h^{\VAR}(y,y')
% &=
% \beta_h^{\VAR\,\prime}(y-y')
% =
% e_1'J\mathbf A^hJ'B(y-y')
% \neq
% \theta_h^*(y,y').
% \label{eq:dynamic-misspecification-var-response}
% \end{align}
\begin{align}
\theta_h^{\VAR}(y,y')
=e_1'J(\mathbf A^\dagger)^hJ'B^\dagger(y-y')
\neq\theta_h^*(y,y').
\label{eq:dynamic-misspecification-var-response}
\end{align}

\begin{corollary}[Power against omitted variables]
\label{cor:dynamic-misspecification}
Suppose \eqref{eq:dynamic-misspecification-dgp} and the
fixed-alternative regularity conditions in
Appendix~\ref{app:hausman-proof} hold. 
If, for $h\geq1$,
$
\Delta_h=\beta_h^*-\beta_h^{\VAR}\ne0
$, then \eqref{eq:hausman-fixed-alternative} holds and the rejection
probability tends to one.
\end{corollary}
The proof is provided in Appendix~\ref{app:dynamic-misspecification-proof}.

% \begin{remark}[Relation to LP--VAR equivalence and LP-IV]
% \label{rem:lp-var-equivalence}
% The comparison does not contradict \citet{plagborgmollerwolf2021}, whose
% LP--VAR equivalence requires the VAR to include lags of the shock. The
% panel VAR \eqref{eq:parametric-var} excludes lags of
% $y_{i,t-1}\varepsilon_{1t}$; adding them would restore agreement up to
% horizon $p$. Because $\varepsilon_{1t}$ is observed, the panel LP is the
% LP-IV of \citet{stockwatson2018} with the shock as its own instrument,
% and likewise does not require the observed vector to span the full
% state. {\blue [mx:seems not very much relevant]}
% \end{remark}

\subsection{Misspecification: panel VARMA}
\label{subsec:finite-memory}

The second departure keeps $A(L)$ correctly specified but adds a
moving-average component of order $q$ to the shock interaction:
\begin{align}
A(L)y_{it}
&=
\alpha_i+\lambda_t
+\sum_{j=0}^{q}B_jy_{i,t-j-1}\varepsilon_{1,t-j}
+u_{it}.
\label{eq:finite-memory-dgp}
\end{align}
We refer to \eqref{eq:finite-memory-dgp} as a panel VARMA. Through the
$j$th moving-average term, $\varepsilon_{1t}$ enters the equation for date
$t+j$ as $B_jy_{i,t-1}\varepsilon_{1t}$, for $j=0,\ldots,q$; each of these
terms
depends on the same pre-shock state $y_{i,t-1}$. Iterating
\eqref{eq:finite-memory-dgp} shows that
$\beta_h^{*\,\prime}=e_1'\sum_{j=0}^{\min\{h,q\}}J\mathbf A^{h-j}J'B_j$
is the horizon-$h$ coefficient on $y_{i,t-1}\varepsilon_{1t}$, so that
\begin{align}
\theta_h^*(y,y')
&=
\beta_h^{*\,\prime}(y-y')
=
e_1'\sum_{j=0}^{\min\{h,q\}}J\mathbf A^{h-j}J'B_j\,(y-y').
\label{eq:finite-memory-response}
\end{align}
When $q=0$, this reduces to the baseline response
$e_1'J\mathbf A^hJ'B_0(y-y')$.

Let $\beta_h^{\LP}$ denote the population projection coefficient on
$y_{i,t-1}\varepsilon_{1t}$ in the panel LP
\eqref{eq:parametric-lp}, under \eqref{eq:finite-memory-dgp}.
With some abuse of notation, we reuse $P_{i,t,h}$, $G_{i,t,h}$,
and $U_{i,t,h}$ for the corresponding components under
\eqref{eq:finite-memory-dgp}.
At the target coefficient $\beta_h^*$,
\eqref{eq:finite-memory-dgp} gives the candidate LP error, up to
additive unit and time effects,
\begin{align}
v_{i,t,h}
={}&\underbrace{(e_1'J\mathbf A^{h+1}-c_h')Y_{i,t-1}
+\sum_{k=0}^{h}\sum_{j=h-k+1}^{q}
 e_1'J\mathbf A^kJ'B_j
 y_{i,t+h-k-j-1}\varepsilon_{1,t+h-k-j}}_{P_{i,t,h}}\nonumber\\
&+\underbrace{\sum_{k=0}^{h-1}\sum_{j=0}^{\min\{h-k-1,q\}}
 e_1'J\mathbf A^kJ'B_j
 y_{i,t+h-k-j-1}\varepsilon_{1,t+h-k-j}}_{G_{i,t,h}}
+\underbrace{\sum_{k=0}^{h}e_1'J\mathbf A^kJ'u_{i,t+h-k}}_{U_{i,t,h}},
\label{eq:finite-memory-lp-error}
\end{align}
where empty sums are zero. All shock terms in $P_{i,t,h}$ are dated
before $t$, whereas those in $G_{i,t,h}$ are dated after $t$ and have
predetermined loadings. With $z_{it}$ defined under this DGP,
the same timing and independence arguments give
\[
\mathbb E[z_{it}v_{i,t,h}]
 =\mathbb E[z_{it}P_{i,t,h}]
  +\mathbb E[z_{it}G_{i,t,h}]
  +\mathbb E[z_{it}U_{i,t,h}]=0.
\]
Therefore $\beta_h^{\LP}=\beta_h^*$ and
$\widehat\beta_h^{\LP}\xrightarrow{p}\beta_h^*$ under the stated
moment convergence conditions, as established in
Lemma~\ref{lem:finite-memory-lp-identification} in
Appendix~\ref{app:finite-memory-proof}.

The panel VAR \eqref{eq:parametric-var} omits the moving-average terms
$B_jy_{i,t-j-1}\varepsilon_{1,t-j}$, $j\geq1$. Because these terms are generally
correlated with $Y_{i,t-1}$, they can distort its lag
coefficients; because they are predetermined, its impact coefficient
remains $B_0$. Let $\mathbf A^\dagger$ denote the companion matrix
of the fitted population lag coefficients, so that
$\beta_h^{\VAR\,\prime}=e_1'J(\mathbf A^\dagger)^hJ'B_0$.
In general,
\begin{align}
\theta_h^{\VAR}(y,y')
&\neq
\theta_h^*(y,y').
\label{eq:finite-memory-var-response}
\end{align}
There are two potential sources of response bias.
The omitted moving-average terms can distort the fitted lag
coefficients, and they also generate delayed effects of the
impact shock that the baseline recursion omits.
Even if the fitted lag coefficients coincide with those in
$A(L)$, the panel VAR constructs
$e_1'J\mathbf A^hJ'B_0(y-y')$, leaving the response gap
\begin{align}
\theta_h^*(y,y')-e_1'J\mathbf A^hJ'B_0(y-y')
&=
e_1'\sum_{j=1}^{\min\{h,q\}}J\mathbf A^{h-j}J'B_j\,(y-y'),
\label{eq:finite-memory-gap}
\end{align}
which is exactly the contribution of the omitted moving-average terms.

\begin{corollary}[Power against a panel VARMA]
\label{cor:finite-memory}
Suppose \eqref{eq:finite-memory-dgp} and the fixed-alternative
regularity conditions in Appendix~\ref{app:hausman-proof} hold.
If, for $h\geq1$,
$
\Delta_h=\beta_h^*-\beta_h^{\VAR}\ne0
$, then \eqref{eq:hausman-fixed-alternative} holds and the rejection
probability tends to one.
\end{corollary}
The proof is provided in Appendix~\ref{app:finite-memory-proof}.

\subsection{Estimating the LP asymptotic variance}
\label{subsec:lp-variance}

Fix $h\geq1$ and let $L=T_h$, with $I_h$ and $T_h$ as in
Section~\ref{subsec:parametric-lp}. Using the origin periods $I_h$ of the
panel LP, apply the two-way within transformation to
$x_{it}=\varepsilon_{1t}y_{i,t-1}$, $Y_{i,t-1}$, and
$w_{it,h}=y_{1,i,t+h}$. Recall
\[
\begin{pmatrix}
\widehat\beta_h^{\LP}\\
\widehat c_h
\end{pmatrix}
=
\left(
\sum_{i,t\in I_h}\ddot R_{it}\ddot R_{it}'
\right)^{-1}
\sum_{i,t\in I_h}\ddot R_{it}\ddot w_{it,h},
\qquad
R_{it}=(x_{it}',Y_{i,t-1}')'.
\]

Let
\begin{align*}
\widehat C_h
 &=\left(\sum_{i,t\in I_h}\ddot x_{it}\ddot Y_{i,t-1}'\right)
   \left(\sum_{i,t\in I_h}\ddot Y_{i,t-1}\ddot Y_{i,t-1}'\right)^{-1},
&\quad
\widehat z_{it}
 &=\ddot x_{it}-\widehat C_h\ddot Y_{i,t-1},
\\
\widehat v_{it,h}
 &=\ddot w_{it,h}
   -\widehat\beta_h^{\LP\,\prime}\ddot x_{it}
   -\widehat c_h'\ddot Y_{i,t-1},
&\quad
\widehat Q_h
 &=\frac1{NL}\sum_{i,t\in I_h}
   \widehat z_{it}\widehat z_{it}'.
\end{align*}
Form the date score $\widehat g_{t,h}
=N^{-1}\sum_i\widehat z_{it}\widehat v_{it,h}$.
For an integer bandwidth $b<L$, define the Bartlett HAC estimator
\begin{align}
\widehat\Gamma_{k,h}
 &=\frac1L\sum_{t=k+1}^{L}\widehat g_{t,h}\widehat g_{t-k,h}',
 \qquad 0\leq k\leq b,\nonumber\\
\widehat\Lambda_h
 &=\widehat\Gamma_{0,h}
 +\sum_{k=1}^{b}\left(1-\frac{k}{b+1}\right)
   (\widehat\Gamma_{k,h}+\widehat\Gamma_{k,h}'),\quad
\widehat V_h^{\LP}
 =\frac{T}{L}\widehat Q_h^{-1}
              \widehat\Lambda_h\widehat Q_h^{-1}.
\label{eq:lp-feasible-variance}
\end{align}
The origin periods are indexed $1,\ldots,L$.
Note that this is a date-score HAC construction of the
\citet{driscollkraay1998} type.
% taking cross-sectional averages
% retains common-shock uncertainty, while HAC accounts for serial
% dependence. 
The covariance estimate for $\widehat\beta_h^{\LP}$ itself
is $\widehat V_h^{\LP}/T$.

\begin{lemma}[Consistency of the feasible LP variance]
\label{lem:lp-hac}
Suppose the conditions in Assumption~\ref{ass:lp-hac} hold.
Let $Q_h=\mathbb E[z_{it}z_{it}']$ and let $\Lambda_h$ be the
long-run covariance of the population date score defined in
Appendix~\ref{app:lp-variance-proof}, under the DGP in question.
If $b\to\infty$ and
\begin{equation}\label{eq:band_HAC}
b\left(T^{-1/2}+N^{-1/2}\right)\to0,
\end{equation}
then
$\widehat V_h^{\LP}\xrightarrow{p}Q_h^{-1}\Lambda_hQ_h^{-1}$.
% Under the null of Theorem~\ref{thm:hausman},
% $\Lambda_h=\Omega_h$, giving consistency for $V_h^{\LP}$.
% Under either fixed alternative, $\Lambda_h\succ0$ gives the finite
% positive-definite limit required for the power result.
\end{lemma}
For example, $b=\lfloor\min\{N,L\}^{1/4}\rfloor$ satisfies the
condition \eqref{eq:band_HAC}. The proof is provided in Appendix~\ref{app:lp-variance-proof}. 
% The additional moment and dependence conditions
% in Appendix~\ref{app:lp-variance-proof} control the use of estimated
% residuals and sample fixed effects as the bandwidth grows.
 
\section{Empirical application (in progress).}
In ongoing work, we apply the proposed framework to study the heterogeneous impulse responses of monetary policy to firms in T\"urkiye. We link four administrative datasets available at the Central Bank of the Republic of T\"urkiye---annual balance sheets and income statements, monthly firm-to-firm VAT and export--import records, monthly social security employment records, and the monthly firm--bank credit register---through unique firm identifiers into a firm-month panel spanning January 2009 to April 2025. Monetary policy shocks are identified from high-frequency changes in the two-year government bond yield around scheduled Monetary Policy Committee announcements \cite{gertler2015monetary, jarocinski2020deconstructing}. We then estimate the heterogeneous impulse responses of firm sales, employment, and borrowing as functions of firm characteristics such as leverage, age, liquidity, and collateral type.

\section{Conclusion}
\label{sec:conclusion}

Our analysis yields three main conclusions about the estimation of
heterogeneous responses to aggregate shocks from micro panels.

\begin{enumerate}
\item The panel VAR and the panel LP converge at different rates. In time
series, LPs and VARs estimate the same impulse responses at a common rate, so
the choice between them is a trade-off between bias and variance. In micro panels, the two estimators
still target the same heterogeneous response, but under a correctly specified
dynamic model the panel VAR has effective sample size $NT$, whereas at
positive horizons the panel LP has effective sample size $T$. The projection
error of the panel LP contains post-impact aggregate shocks. Because units
differ in their exposure to these shocks, time effects remove only their
common component, and a larger cross section cannot average them away.

\item Nonparametric exposure need not slow down the panel LP.
Under suitable smoothness and panel-growth conditions, the series
estimation and approximation errors can be made smaller than
$T^{-1/2}$, leaving the LP's convergence rate unchanged.
The panel VAR, by contrast, converges more slowly than its
parametric counterpart because it is estimated nonparametrically. The LP's ``free lunch'' therefore comes from
absorbing the nonparametric estimation cost within its existing
aggregate-shock uncertainty.

\item The faster rate of the panel VAR relies on correct specification of the
dynamic model. Omitted lags or variables, or a moving-average component in the shock interaction, bias the panel VAR response but leave the panel LP consistent,
mirroring the robustness of local projections to misspecified dynamics in
time series.
\end{enumerate}

To inform the choice between the two estimators, we have proposed a
Hausman-type test of the dynamic model. Because the panel VAR converges
faster under the null, the test requires only the asymptotic variance of the
panel LP. It has asymptotically correct size and is consistent against both
departures above.

Two extensions remain open. The first is to shocks identified through
external instruments rather than observed directly. The second is to
propagation that depends on aggregate states responding to the shock, which
would require a law of motion for these states.

\section*{Acknowledgements}
The views expressed here are those of the authors and do not necessarily represent those of the Central Bank of the Republic of T\"urkiye.
% [Personal acknowledgements first, then institutions and agencies. The views
% expressed here are those of the authors and do not necessarily represent those of
% the Central Bank of the Republic of T\"urkiye.]

% \section*{Funding}
% [Grant numbers and funding bodies, or a statement that no funding was received.]

% \section*{Conflict of Interest}
% [Declaration, or a statement that the authors declare no conflict of interest.]

\bibliography{biblio,proof_references}

@incollection{su2016semiparametric,
  author    = {Su, Liangjun and Zhang, Yonghui},
  title     = {Semiparametric Estimation of Partially Linear Dynamic Panel Data Models with Fixed Effects},
  booktitle = {Essays in Honor of {Aman} {Ullah}},
  editor    = {Gonz{\'a}lez-Rivera, Gloria and Hill, R. Carter and Lee, Tae-Hwy},
  series    = {Advances in Econometrics},
  volume    = {36},
  pages     = {137--204},
  publisher = {Emerald Group Publishing},
  year      = {2016},
  doi       = {10.1108/S0731-905320160000036014}
}

@article{su2013nonparametric,
  author  = {Su, Liangjun and Lu, Xun},
  title   = {Nonparametric Dynamic Panel Data Models: Kernel Estimation and Specification Testing},
  journal = {Journal of Econometrics},
  year    = {2013},
  volume  = {176},
  number  = {2},
  pages   = {112--133},
  doi     = {10.1016/j.jeconom.2013.04.020}
}

@article{gertler2015monetary,
  author  = {Gertler, Mark and Karadi, Peter},
  title   = {Monetary Policy Surprises, Credit Costs, and Economic Activity},
  journal = {American Economic Journal: Macroeconomics},
  year    = {2015},
  volume  = {7},
  number  = {1},
  pages   = {44--76}
}

@article{jarocinski2020deconstructing,
  author  = {Jaroci{\'n}ski, Marek and Karadi, Peter},
  title   = {Deconstructing Monetary Policy Surprises---The Role of Information Shocks},
  journal = {American Economic Journal: Macroeconomics},
  year    = {2020},
  volume  = {12},
  number  = {2},
  pages   = {1--43}
}

@article{kashyap2000banks,
  author  = {Kashyap, Anil K. and Stein, Jeremy C.},
  title   = {What Do a Million Observations on Banks Say about the Transmission of Monetary Policy?},
  journal = {American Economic Review},
  year    = {2000},
  volume  = {90},
  number  = {3},
  pages   = {407--428},
  doi     = {10.1257/aer.90.3.407}
}

@article{nakamura2014fiscal,
  author  = {Nakamura, Emi and Steinsson, J{\'o}n},
  title   = {Fiscal Stimulus in a Monetary Union: Evidence from {US} Regions},
  journal = {American Economic Review},
  year    = {2014},
  volume  = {104},
  number  = {3},
  pages   = {753--792},
  doi     = {10.1257/aer.104.3.753}
}

@article{nakamura2018identification,
  author  = {Nakamura, Emi and Steinsson, J{\'o}n},
  title   = {Identification in Macroeconomics},
  journal = {Journal of Economic Perspectives},
  year    = {2018},
  volume  = {32},
  number  = {3},
  pages   = {59--86},
  doi     = {10.1257/jep.32.3.59}
}

@article{wolf2023missing,
  author  = {Wolf, Christian K.},
  title   = {The Missing Intercept: A Demand Equivalence Approach},
  journal = {American Economic Review},
  year    = {2023},
  volume  = {113},
  number  = {8},
  pages   = {2232--2269},
  doi     = {10.1257/aer.20211751}
}

@article{guren2021crossregional,
  author  = {Guren, Adam and McKay, Alisdair and Nakamura, Emi and Steinsson, J{\'o}n},
  title   = {What Do We Learn from Cross-Regional Empirical Estimates in Macroeconomics?},
  journal = {NBER Macroeconomics Annual},
  year    = {2021},
  volume  = {35},
  pages   = {175--223}
}

@article{cloyne2023,
  author  = {Cloyne, James and Ferreira, Clodomiro and Froemel, Maren and Surico, Paolo},
  title   = {Monetary Policy, Corporate Finance, and Investment},
  journal = {Journal of the European Economic Association},
  year    = {2023},
  volume  = {21},
  number  = {6},
  pages   = {2586--2634},
  doi     = {10.1093/jeea/jvad009}
}

@article{kaplanmollviolante2018,
  author  = {Kaplan, Greg and Moll, Benjamin and Violante, Giovanni L.},
  title   = {Monetary Policy According to {HANK}},
  journal = {American Economic Review},
  year    = {2018},
  volume  = {108},
  number  = {3},
  pages   = {697--743},
  doi     = {10.1257/aer.20160042}
}

@article{jeenas2023firm,
  author  = {Jeenas, Priit},
  title   = {Firm Balance Sheet Liquidity, Monetary Policy Shocks, and Investment Dynamics},
  journal = {Journal of Political Economy},
  year    = {2026},
  note    = {Forthcoming}
}

@article{mianraosufi2013,
  author  = {Mian, Atif and Rao, Kamalesh and Sufi, Amir},
  title   = {Household Balance Sheets, Consumption, and the Economic Slump},
  journal = {The Quarterly Journal of Economics},
  year    = {2013},
  volume  = {128},
  number  = {4},
  pages   = {1687--1726},
  doi     = {10.1093/qje/qjt020}
}

@article{chodorowreich2019,
  author  = {Chodorow-Reich, Gabriel},
  title   = {Geographic Cross-Sectional Fiscal Spending Multipliers: What Have We Learned?},
  journal = {American Economic Journal: Economic Policy},
  year    = {2019},
  volume  = {11},
  number  = {2},
  pages   = {1--34},
  doi     = {10.1257/pol.20160465}
}

@article{jorda2005,
  author  = {Jord{\`a}, {\`O}scar},
  title   = {Estimation and Inference of Impulse Responses by Local Projections},
  journal = {American Economic Review},
  year    = {2005},
  volume  = {95},
  number  = {1},
  pages   = {161--182},
  doi     = {10.1257/0002828053828518}
}

@article{plagborgmollerwolf2021,
  author  = {Plagborg-M{\o}ller, Mikkel and Wolf, Christian K.},
  title   = {Local Projections and {VAR}s Estimate the Same Impulse Responses},
  journal = {Econometrica},
  year    = {2021},
  volume  = {89},
  number  = {2},
  pages   = {955--980},
  doi     = {10.3982/ECTA17813}
}

@article{li2024,
  author  = {Li, Dake and Plagborg-M{\o}ller, Mikkel and Wolf, Christian K.},
  title   = {Local Projections vs. {VAR}s: Lessons From Thousands of {DGP}s},
  journal = {Journal of Econometrics},
  year    = {2024},
  volume  = {244},
  number  = {2},
  pages   = {105722},
  doi     = {10.1016/j.jeconom.2024.105722}
}

@article{montielolea2026primer,
  author  = {Montiel Olea, Jos{\'e} Luis and Plagborg-M{\o}ller, Mikkel and Qian, Eric and Wolf, Christian K.},
  title   = {Local Projections or Vector Autoregressions? {A} Primer for Macroeconomists},
  journal = {NBER Macroeconomics Annual},
  year    = {2026},
  volume  = {40},
  pages   = {111--152},
  doi     = {10.1086/738945}
}

@article{montielolea2024double,
  author  = {Montiel Olea, Jos{\'e} Luis and Plagborg-M{\o}ller, Mikkel and Qian, Eric and Wolf, Christian K.},
  title   = {Double Robustness of Local Projections and Some Unpleasant {VAR}ithmetic},
  journal = {Econometrica},
  year    = {2026},
  volume  = {94},
  number  = {4},
  pages   = {1313--1343}
}

@techreport{almuzara2024,
  author      = {Almuzara, Mart{\'i}n and Sancibri{\'a}n, V{\'i}ctor},
  title       = {Micro Responses to Macro Shocks},
  institution = {Federal Reserve Bank of New York},
  type        = {Staff Report},
  number      = {1090},
  year        = {2024},
  doi         = {10.59576/sr.1090}
}

@article{hahnkuersteiner2002,
  author  = {Hahn, Jinyong and Kuersteiner, Guido},
  title   = {Asymptotically Unbiased Inference for a Dynamic Panel Model with Fixed Effects when Both $n$ and {$T$} Are Large},
  journal = {Econometrica},
  year    = {2002},
  volume  = {70},
  number  = {4},
  pages   = {1639--1657},
  doi     = {10.1111/1468-0262.00344}
}

@article{alvarezarellano2003,
  author  = {Alvarez, Javier and Arellano, Manuel},
  title   = {The Time Series and Cross-Section Asymptotics of Dynamic Panel Data Estimators},
  journal = {Econometrica},
  year    = {2003},
  volume  = {71},
  number  = {4},
  pages   = {1121--1159},
  doi     = {10.1111/1468-0262.00441}
}

@article{arellanobond1991,
  author  = {Arellano, Manuel and Bond, Stephen},
  title   = {Some Tests of Specification for Panel Data: Monte Carlo Evidence and an Application to Employment Equations},
  journal = {Review of Economic Studies},
  year    = {1991},
  volume  = {58},
  number  = {2},
  pages   = {277--297},
  doi     = {10.2307/2297968}
}

@article{lee2014,
  author  = {Lee, Yoonseok},
  title   = {Nonparametric Estimation of Dynamic Panel Models with Fixed Effects},
  journal = {Econometric Theory},
  year    = {2014},
  volume  = {30},
  number  = {6},
  pages   = {1315--1347},
  doi     = {10.1017/S0266466614000188}
}

@article{ottonellowinberry2020,
  author  = {Ottonello, Pablo and Winberry, Thomas},
  title   = {Financial Heterogeneity and the Investment Channel of Monetary Policy},
  journal = {Econometrica},
  year    = {2020},
  volume  = {88},
  number  = {6},
  pages   = {2473--2502},
  doi     = {10.3982/ECTA15949}
}

@article{cloyne2020,
  author  = {Cloyne, James and Ferreira, Clodomiro and Surico, Paolo},
  title   = {Monetary Policy when Households have Debt: New Evidence on the Transmission Mechanism},
  journal = {Review of Economic Studies},
  year    = {2020},
  volume  = {87},
  number  = {1},
  pages   = {102--129},
  doi     = {10.1093/restud/rdy074}
}

@article{holm2021transmission,
  author  = {Holm, Martin Blomhoff and Paul, Pascal and Tischbirek, Andreas},
  title   = {The Transmission of Monetary Policy under the Microscope},
  journal = {Journal of Political Economy},
  year    = {2021},
  volume  = {129},
  number  = {10},
  pages   = {2861--2904},
  doi     = {10.1086/715416}
}

@article{gulyas2024,
  author  = {Gulyas, Andreas and Meier, Matthias and Ryzhenkov, Mykola},
  title   = {Labor Market Effects of Monetary Policy across Workers and Firms},
  journal = {European Economic Review},
  year    = {2024},
  volume  = {166},
  pages   = {104756},
  doi     = {10.1016/j.euroecorev.2024.104756}
}

@article{paranhos2025,
  author  = {Paranhos, Livia},
  title   = {How Do Firms' Financial Conditions Influence the Transmission of Monetary Policy? {A} Non-Parametric Local Projection Approach},
  journal = {Journal of Econometrics},
  year    = {2025},
  volume  = {249},
  pages   = {105886},
  doi     = {10.1016/j.jeconom.2024.105886}
}

@techreport{goncalves2024nonparametric,
  author      = {Gon{\c{c}}alves, S{\'i}lvia and Herrera, Ana Mar{\'i}a and Kilian, Lutz and Pesavento, Elena},
  title       = {Nonparametric Local Projections},
  institution = {Federal Reserve Bank of Dallas},
  type        = {Working Paper},
  number      = {2414},
  year        = {2024},
  doi         = {10.24149/wp2414}
}

@article{newey1997,
  author  = {Newey, Whitney K.},
  title   = {Convergence Rates and Asymptotic Normality for Series Estimators},
  journal = {Journal of Econometrics},
  year    = {1997},
  volume  = {79},
  number  = {1},
  pages   = {147--168}
}

@techreport{david2026causal,
  author      = {David, Joel M. and Giacomini, Raffaella and Jiao, Xiyu and Wang, Weining},
  title       = {Causal State-Dependent Local Projections},
  institution = {Federal Reserve Bank of Chicago},
  type        = {Working Paper},
  number      = {2026-18},
  year        = {2026}
}

@techreport{goncalves2026semiparametric,
  author      = {Gon{\c{c}}alves, S{\'i}lvia and Herrera, Ana Mar{\'i}a and Kilian, Lutz and Pesavento, Elena and Kelanemer Holban, Iones},
  title       = {Semiparametric Local Projections},
  institution = {Federal Reserve Bank of Dallas},
  type        = {Working Paper},
  number      = {2616},
  year        = {2026},
  doi         = {10.24149/wp2616}
}

@article{auclert2019,
  author  = {Auclert, Adrien},
  title   = {Monetary Policy and the Redistribution Channel},
  journal = {American Economic Review},
  year    = {2019},
  volume  = {109},
  number  = {6},
  pages   = {2333--2367},
  doi     = {10.1257/aer.20160137}
}

@article{moulton1990,
  author  = {Moulton, Brent R.},
  title   = {An Illustration of a Pitfall in Estimating the Effects of Aggregate Variables on Micro Units},
  journal = {The Review of Economics and Statistics},
  year    = {1990},
  volume  = {72},
  number  = {2},
  pages   = {334--338},
  doi     = {10.2307/2109724}
}

@article{abadie2023,
  author  = {Abadie, Alberto and Athey, Susan and Imbens, Guido W. and Wooldridge, Jeffrey M.},
  title   = {When Should You Adjust Standard Errors for Clustering?},
  journal = {The Quarterly Journal of Economics},
  year    = {2023},
  volume  = {138},
  number  = {1},
  pages   = {1--35},
  doi     = {10.1093/qje/qjac038}
}

@article{andrews2005,
  author  = {Andrews, Donald W. K.},
  title   = {Cross-Section Regression with Common Shocks},
  journal = {Econometrica},
  year    = {2005},
  volume  = {73},
  number  = {5},
  pages   = {1551--1585},
  doi     = {10.1111/j.1468-0262.2005.00629.x}
}

@article{phillipsmoon1999,
  author  = {Phillips, Peter C. B. and Moon, Hyungsik R.},
  title   = {Linear Regression Limit Theory for Nonstationary Panel Data},
  journal = {Econometrica},
  year    = {1999},
  volume  = {67},
  number  = {5},
  pages   = {1057--1111},
  doi     = {10.1111/1468-0262.00070}
}

@article{driscollkraay1998,
  author  = {Driscoll, John C. and Kraay, Aart C.},
  title   = {Consistent Covariance Matrix Estimation with Spatially Dependent Panel Data},
  journal = {Review of Economics and Statistics},
  year    = {1998},
  volume  = {80},
  number  = {4},
  pages   = {549--560},
  doi     = {10.1162/003465398557825}
}

@article{pesaransmith1995,
  author  = {Pesaran, M. Hashem and Smith, Ron},
  title   = {Estimating Long-Run Relationships from Dynamic Heterogeneous Panels},
  journal = {Journal of Econometrics},
  year    = {1995},
  volume  = {68},
  number  = {1},
  pages   = {79--113},
  doi     = {10.1016/0304-4076(94)01644-F}
}

@article{nickell1981,
  author  = {Nickell, Stephen},
  title   = {Biases in Dynamic Models with Fixed Effects},
  journal = {Econometrica},
  year    = {1981},
  volume  = {49},
  number  = {6},
  pages   = {1417--1426},
  doi     = {10.2307/1911408}
}

@article{andersonhsiao1982,
  author  = {Anderson, T. W. and Hsiao, Cheng},
  title   = {Formulation and Estimation of Dynamic Models Using Panel Data},
  journal = {Journal of Econometrics},
  year    = {1982},
  volume  = {18},
  number  = {1},
  pages   = {47--82},
  doi     = {10.1016/0304-4076(82)90095-1}
}

@article{blundellbond1998,
  author  = {Blundell, Richard and Bond, Stephen},
  title   = {Initial Conditions and Moment Restrictions in Dynamic Panel Data Models},
  journal = {Journal of Econometrics},
  year    = {1998},
  volume  = {87},
  number  = {1},
  pages   = {115--143},
  doi     = {10.1016/S0304-4076(98)00009-8}
}

@article{hausman1978,
  author  = {Hausman, Jerry A.},
  title   = {Specification Tests in Econometrics},
  journal = {Econometrica},
  year    = {1978},
  volume  = {46},
  number  = {6},
  pages   = {1251--1271},
  doi     = {10.2307/1913827}
}

@article{stockwatson2018,
  author  = {Stock, James H. and Watson, Mark W.},
  title   = {Identification and Estimation of Dynamic Causal Effects in Macroeconomics Using External Instruments},
  journal = {The Economic Journal},
  year    = {2018},
  volume  = {128},
  number  = {610},
  pages   = {917--948},
  doi     = {10.1111/ecoj.12593}
}

@article{DhaeneJochmans2015,
  author = {Dhaene, Geert and Jochmans, Koen},
  title = {Split-panel jackknife estimation of fixed-effect models},
  journal = {Review of Economic Studies},
  year = {2015},
  volume = {82},
  pages = {991--1030}
}

@book{HallHeyde1980,
  author = {Hall, Peter and Heyde, C. C.},
  title = {Martingale Limit Theory and Its Application},
  publisher = {Academic Press},
  year = {1980}
}

\newpage
\appendix
\numberwithin{equation}{section}

\section{Parametric results}\label{app:parametric}
\subsection{Estimators}
We observe $y_{it}$ for $t=1-p,\ldots,T$ and use
$t=1,\ldots,T$ for estimation. For a contiguous set $I$ of $L$ origin periods, define, for any panel array $a$,
\begin{align*}
\bar a_{i,I}&=L^{-1}\sum_{t\in I}a_{it},&
\bar a_{N,t}&=N^{-1}\sum_i a_{it},&
\bar a_{N,I}&=(NL)^{-1}\sum_i\sum_{t\in I}a_{it},\\
\ddot a_{it,I}&=a_{it}-\bar a_{i,I}-\bar a_{N,t}+\bar a_{N,I}.
\end{align*}
Unit-constant and date-constant columns transform to zero and are omitted.
Let $R_{it}=(\varepsilon_{1t}y_{i,t-1}',Y_{i,t-1}')'\in\mathbb R^k$, $k=m+mp$, and define
\begin{equation}
\widehat Q_I=(NL)^{-1}\sum_i\sum_{t\in I}\ddot R_{it,I}\ddot R_{it,I}'.
\label{pa:eq:sample-gram}
\end{equation}
For $h\geq1$, using \eqref{pa:eq:sample-gram}, let
$I_h=\{1,\ldots,T_h\}$, $T_h=T-h$, and define the LP estimator by
\[
\begin{pmatrix}\widehat\beta_h^{LP}\\\widehat c_h\end{pmatrix}
=\widehat Q_{I_h}^{-1}(NT_h)^{-1}
 \sum_i\sum_{t\in I_h}\ddot R_{it,I_h}\ddot y_{1,i,t+h,I_h},
\qquad \widehat\theta_h^{LP}(y,y')=d'\widehat\beta_h^{LP}.
\]
The within transformation of the outcome uses the same origin-date set $I_h$.

For the one-step system write $D=(B,A_1,\ldots,A_p)\in\mathbb R^{m\times k}$.
For any $I$, set
\[
\widehat D_I=\left\{(NL)^{-1}\sum_i\sum_{t\in I}
                  \ddot y_{it,I}\ddot R_{it,I}'\right\}\widehat Q_I^{-1}.
\]
For a matrix, $\mathrm{vec}$ stacks its columns. Let $P$ be the fixed
permutation matrix such that
\[
\psi=(\mathrm{vec}(A_1)',\ldots,\mathrm{vec}(A_p)',\mathrm{vec}(B)')'
      =P\mathrm{vec}(D),\qquad \widehat\psi_I=P\mathrm{vec}(\widehat D_I).
\]
Assume $T$ is even and partition $I^{(0)}=\{1,\ldots,T\}$ into two
contiguous halves $I^{(1)},I^{(2)}$ of length $T/2$. Define the
bias-corrected one-step coefficient matrix by
\begin{equation}
\widehat D=2\widehat D_{I^{(0)}}-\tfrac12
 (\widehat D_{I^{(1)}}+\widehat D_{I^{(2)}})
 =(\widehat B,\widehat A_1,\ldots,\widehat A_p),\qquad
\widehat\psi=P\mathrm{vec}(\widehat D).
\label{pa:eq:spj}
\end{equation}
For $p\geq2$, form the estimated companion matrix
\[
\widehat{\mathbf A}=
\begin{pmatrix}
(\widehat A_1,\ldots,\widehat A_{p-1})&\widehat A_p\\
I_{m(p-1)}&0_{m(p-1)\times m}
\end{pmatrix};
\]
for $p=1$, set $\widehat{\mathbf A}=\widehat A_1$.
The recursive response estimator is
\begin{equation}
\widehat\beta_h^{VAR\,\prime}
 =e_1'J\widehat{\mathbf A}^hJ'\widehat B,\qquad
\widehat\theta_h^{VAR}(y,y')=\widehat\beta_h^{VAR\,\prime}(y-y').
\label{pa:eq:recursive-estimator}
\end{equation}
Equivalently, the $mp\times m$ response matrices start at
$\widehat{\mathcal R}_0=J'\widehat B$ and satisfy
$\widehat{\mathcal R}_{h+1}=\widehat{\mathbf A}\widehat{\mathcal R}_h$;
then $\widehat\beta_h^{VAR\,\prime}=e_1'J\widehat{\mathcal R}_h$.
At impact this reduces to $\widehat\beta_0^{VAR\,\prime}=e_1'\widehat B$.
The notation $g_h(\widehat\psi)=\widehat\beta_h^{VAR}$ abbreviates
this explicitly defined recursion.

The split correction is applied to the one-step parameters before
recursive propagation. Each half-panel fit re-estimates both sets of
fixed effects and retains the observed lags preceding that half.
The same $m$ and $p$ are used in all fits. Dropping one end date
permits the same construction when $T$ is odd.
At impact, define $\widehat\beta_0^{LP}=\widehat\beta_0^{VAR}$.
Thus the reported impact limit uses the corrected one-step fit,
not the uncorrected within estimator.

\subsection{Population notation} \label{app:notation}
\begin{gather*}
\mathcal A_t=\sigma(\varepsilon_{1s},\lambda_s:s\leq t),\qquad
\mathcal A=\sigma(\varepsilon_{1s},\lambda_s:s\in\mathbb Z),\\
\mathcal F_t=
 \sigma(\alpha_i,\varepsilon_{1s},\lambda_s,u_{is}:i\geq1,\ s\leq t).
\end{gather*}
The sigma field $\mathcal A$ contains the full path of the
aggregate shock and date effect, but no individual outcomes, permanent
effects, or micro innovations.

Under the sampling conditions below, define
\begin{equation}
a_t=\mathbb{E}[R_{it}\mid\mathcal A],\qquad
b_i=\mathbb{E}[R_{it}\mid\alpha_i]-\mathbb{E} R_{it},\qquad
r_{it}=R_{it}-a_t-b_i.
\label{pa:eq:population-within}
\end{equation}
These are population projections; sample two-way within continues to
use the observed $R_{it}$. Partition $r_{it}=(r_{x,it}',r_{Y,it}')'$ and put
\begin{equation} \label{eq:Q_CLC block}
    Q=\mathbb{E}[r_{it}r_{it}']
=\begin{pmatrix}Q_{xx}&Q_{xY}\\Q_{Yx}&Q_{YY}\end{pmatrix},\qquad
C_0=Q_{xY}Q_{YY}^{-1},\qquad L_0=(I_m,-C_0),\qquad z_{it}=L_0r_{it}.
\end{equation} 
Inverses are defined on the positive-definite parameter class specified below.
Define $Q_h=\mathbb{E}[z_{it}z_{it}']=Q_{xx}-Q_{xY}Q_{YY}^{-1}Q_{Yx}$.
The subscript records the LP horizon; its population value is common across
fixed horizons under stationarity. Write $\sigma_\varepsilon^2=\mathbb E\varepsilon_{1t}^2$.
For $h\geq1$, let
\begin{align}
\Psi_{j,h,t}&=\mathbb{E}[z_{it}y_{i,t+j-1}'\mid\mathcal A],\nonumber\\
s_{t,h}&=\sum_{j=1}^h\varepsilon_{1,t+j}\Psi_{j,h,t}\beta_{h-j},&
k_{r,h}&=\sum_{j=1}^h\Psi_{j,h,r-j}\beta_{h-j},\nonumber\\
\Omega_h&=\sigma_\varepsilon^2\mathbb{E}[k_{0,h}k_{0,h}'].
\label{pa:eq:aggregate-notation}
\end{align}
Let $\mathcal K$ satisfy $\mathrm{vec}(M)=\mathcal K\mathrm{vec}(M')$
for every $m\times k$ matrix $M$. Let $\Sigma_u=\mathbb{E}[u_{it}u_{it}']$
and define
\begin{equation}\label{eq:H_Sigma_Gh}
H=P\mathcal K(I_m\otimes Q^{-1}),\qquad
\Sigma_\psi=H(\Sigma_u\otimes Q)H',\qquad
G_h=\frac{\partial g_h(\psi)}{\partial\psi'}.
\end{equation}

\subsection{Assumptions}\label{pa:sec:assumptions}
\begin{assumption}[Sampling and aggregate process]\label{pa:ass:sampling}
The process $\{(\varepsilon_{1t},\lambda_t):t\in\mathbb Z\}$ is strictly
stationary and ergodic, with
\[
\varepsilon_{1t}\ \perp\!\!\!\perp\ \mathcal A_{t-1},\qquad
\mathbb{E}\varepsilon_{1t}=0,\qquad
0<\mathbb{E}\varepsilon_{1t}^2=\sigma_\varepsilon^2<\infty.
\]
The $\alpha_i$ are iid across units and the $u_{it}$ are iid across
$(i,t)$. The full aggregate process, the collection of permanent
effects, and the collection of micro innovations are mutually
independent. Moreover
$\mathbb{E} u_{it}=0$ and $\mathbb{E}[u_{it}u_{it}']=\Sigma_u\succ0$.
\end{assumption}
The date effect $\lambda_t\in\mathbb R^m$ may be serially dependent
and contemporaneously dependent on $\varepsilon_{1t}$. The independence
condition on $\varepsilon_{1t}$ includes all past time effects.
It implies that the shocks $\varepsilon_{1t}$ are iid over time, but
does not require the pairs $(\varepsilon_{1t},\lambda_t)$ to be iid.

\begin{assumption}[Causal solution and moments]\label{pa:ass:solution}
Equation \eqref{eq:parametric-var} has a time-homogeneous causal stationary solution
\[
y_{it}=F\!\left(\alpha_i,
 \{(\varepsilon_{1s},\lambda_s,u_{is}):s\leq t\}\right),
\]
where $F$ is common across units. For some $q=4+\eta$, $\eta>0$,
\[
\mathbb{E}|\varepsilon_{1t}|^q+\mathbb{E}\|u_{it}\|^q
 +\mathbb{E}\|y_{it}\|^q+\mathbb{E}\|R_{it}\|^q<\infty.
\]
\end{assumption}
This representation specifies the permanent heterogeneity through
$\alpha_i$. Conditional on $\mathcal A$, complete unit paths are iid.
Lemma~\ref{pa:lem:population} derives stability and the lag-covariance
summability from these conditions.

\begin{assumption}[Identification and panel dimensions]\label{pa:ass:identification}
$Q\succ0$, $N,T\to\infty$, and $h,m,p$ are fixed.
\end{assumption}

For use in the dependence condition, write
\[
r_{it}=(r_{it,1},\ldots,r_{it,k})',\qquad
u_{it}=(u_{it,1},\ldots,u_{it,m})'.
\]
Thus $r_{it,a}$ is coordinate $a\in\{1,\ldots,k\}$ and $u_{it,b}$
is coordinate $b\in\{1,\ldots,m\}$. For $a,b=1,\ldots,k$, define
\[
F_{it}^{ab}=r_{it,a}r_{it,b},\qquad
f_t^{ab}=\mathbb{E}[F_{it}^{ab}\mid\mathcal A].
\]
Let $\mathcal V_h$ be the collection of scalar coordinate processes
\[
\mathcal V_h=
 \bigl\{(r_{it,a})_{t\in\mathbb Z}:1\leq a\leq k\bigr\}
 \cup\bigl\{(u_{it,b})_{t\in\mathbb Z}:1\leq b\leq m\bigr\}
 \cup\bigl\{(v_{i,t,h})_{t\in\mathbb Z}\bigr\}.
\]
Selecting $V^{(j)}\in\mathcal V_h$ selects a coordinate process;
$V^{(j)}_{it}$ denotes its observation at date $t$.
For centered scalar variables $V_1,V_2,V_3,V_4$, define
\begin{align*}
\mathrm{cum}(V_1,V_2,V_3,V_4)
={}&\mathbb{E}[V_1V_2V_3V_4]
 -\mathbb{E}[V_1V_2]\mathbb{E}[V_3V_4]\\
&-\mathbb{E}[V_1V_3]\mathbb{E}[V_2V_4]
 -\mathbb{E}[V_1V_4]\mathbb{E}[V_2V_3].
\end{align*}
For variables centered conditional on $\alpha_i$, the conditional
cumulant uses $\mathbb{E}[\cdot\mid\alpha_i]$ in each of the seven
expectations in this formula. The coordinates used below satisfy
this conditional centering under the sampling and causal-solution
assumptions.

\begin{assumption}[Temporal dependence]\label{pa:ass:dependence}
For any $V^{(1)},\ldots,V^{(4)}\in\mathcal V_h$,
\[
\begin{split}
\sum_{s\in\mathbb Z}
 |\mathrm{Cov}(V^{(1)}_{i0},V^{(2)}_{is}\mid\alpha_i)|&\leq C_i,\\
\sum_{s,r,v\in\mathbb Z}
 |\mathrm{cum}(V^{(1)}_{i0},V^{(2)}_{is},
              V^{(3)}_{ir},V^{(4)}_{iv}\mid\alpha_i)|&\leq C_i,
\qquad \mathbb{E} C_i^2\leq C.
\end{split}
\]
For all regressor coordinates $a,b$,
$\sum_{s\in\mathbb Z}|\mathrm{Cov}(f_0^{ab},f_s^{ab})|\leq C$.
For every contiguous $I$ of length $L$, $a=1,\ldots,k$, $b=1,\ldots,m$,
and units $i,\ell$,
\begin{equation}
\sum_{t,s,r,v\in I}
\left|\mathrm{Cov}(r_{it,a}u_{is,b},r_{\ell r,a}u_{\ell v,b})\right|
\leq
\begin{cases}CL^2,&i=\ell,\\CL,&i\ne\ell.\end{cases}
\label{pa:eq:four-date-condition}
\end{equation}
\end{assumption}
All bounds are uniform over the indicated coordinates and periods.
The four-date bound controls the dynamic mean correction.

\begin{assumption}[Aggregate-score nondegeneracy]\label{pa:ass:aggregate}
For the fixed horizon $h\geq1$, $\Omega_h\succ0$.
\end{assumption}
This additional condition is used for positive-horizon LP inference.
The corrected VAR conclusion and impact result require only
Assumptions~\ref{pa:ass:sampling}--\ref{pa:ass:dependence}, together
with the stated VAR restriction $N=o(T^2)$.

\subsection{Notation}\label{pa:sec:appendix-notation}
All sums over $i$ run from $1$ to $N$. The norm of a vector is Euclidean;
matrix bounds may use the Frobenius norm, since dimensions are fixed.
For an interval $I$ of length $L$, define
\begin{align*}
U_I&=(NL)^{-1}\sum_i\sum_{t\in I}u_{it}r_{it}',&
D_I^u&=N^{-1}\sum_i\bar u_{i,I}\bar r_{i,I}',\\
E_I^u&=L^{-1}\sum_{t\in I}\bar u_{N,t}\bar r_{N,t}',&
F_I^u&=\bar u_{N,I}\bar r_{N,I}'.
\end{align*}
Define the lag matrices and the dynamic bias coefficient by
\begin{equation}
\begin{aligned}
\Gamma_\ell&=\mathbb{E}[u_{i0}r_{i\ell}'],&
\Phi&=\sum_{\ell=1}^{\infty}\Gamma_\ell,\\
\Phi_L&=\sum_{\ell=1}^{L-1}(1-\ell/L)\Gamma_\ell,&
b_\psi&=-P\mathrm{vec}(\Phi Q^{-1}).
\end{aligned}
\label{pa:eq:bias-notation}
\end{equation}
The lag formula and weighted summability in Lemma~\ref{pa:lem:population}
ensure that the objects in \eqref{pa:eq:bias-notation} exist and that
$\|\Phi_L-\Phi\|=O(L^{-1})$.

For LP, partition $\widehat Q_{I_h}$ as $Q$ was partitioned and define
\[
\widehat C_h=\widehat Q_{xY,I_h}\widehat Q_{YY,I_h}^{-1},\quad
\widehat L_h=(I_m,-\widehat C_h),\quad
\widehat z_{it,h}=\widehat L_h\ddot r_{it,I_h},\quad
\widehat Q_h=\widehat L_h\widehat Q_{I_h}\widehat L_h'.
\]
For $j=0,\ldots,h$, define $\pi_{j,h}'=e_1'J\mathbf A^{h-j}J'$.
Set
\[
\Psi_{N,j,h,t}=N^{-1}\sum_i z_{it}y_{i,t+j-1}',\qquad
v_{i,t,h}^u=\sum_{j=0}^h \pi_{j,h}'u_{i,t+j}.
\]

For any interval-indexed array $B_I$, define the time split combination
\[
\mathcal J_T[B]=2B_{I^{(0)}}-\tfrac12(B_{I^{(1)}}+B_{I^{(2)}}),\qquad
\widehat D=(\widehat B,\widehat A_1,\ldots,\widehat A_p).
\]
Thus $\widehat\psi=P\mathrm{vec}(\widehat D)$.

\subsection{Decomposition of the one-step VAR coefficients}\label{pa:sec:var-decomposition}
Two-way within is an orthogonal projection and removes $a_t+b_i$ exactly.
The one-step normal equations therefore give the exact identity
\begin{equation}
\widehat D_I-D=(U_I-D_I^u-E_I^u+F_I^u)\widehat Q_I^{-1}.
\label{pa:eq:var-exact}
\end{equation}
Apply the split weights to \eqref{pa:eq:var-exact} before taking bounds,
and then add and subtract the common inverse $Q^{-1}$:
\begin{align}
\widehat D-D
={}&U_{I^{(0)}}Q^{-1}
 -\mathcal J_T[D^u]Q^{-1}
 -\mathcal J_T[E^u]Q^{-1}
 +\mathcal J_T[F^u]Q^{-1}\nonumber\\
&+\mathcal J_T\!\left[
 (U_I-D_I^u-E_I^u+F_I^u)(\widehat Q_I^{-1}-Q^{-1})\right]\nonumber\\
={}&I^{VAR}+II^{VAR}+III^{VAR}+IV^{VAR}+V^{VAR}.
\label{pa:eq:var-decomposition}
\end{align}

In particular,
\begin{equation}
\mathcal J_T[D^u]
=\frac{2}{NT^2}\sum_i
 \left\{\sum_{\substack{s\in I^{(1)}\\t\in I^{(2)}}}
       +\sum_{\substack{s\in I^{(2)}\\t\in I^{(1)}}}\right\}u_{is}r_{it}'.
\label{pa:eq:cross-block}
\end{equation}
All within-half pairs cancel, including their realized aggregate-path
fluctuations. Lemma \ref{pa:lem:split} records the exact weighting identities
and the cross-block bounds used below.

\subsection{Recursive propagation of the VAR coefficients}
Let $\Delta \mathbf A=\widehat{\mathbf A}-\mathbf A$ and $\Delta B=\widehat B-B$. The estimator
\eqref{pa:eq:recursive-estimator} satisfies the exact identity
\begin{equation}
\widehat\beta_h^{VAR\,\prime}-\beta_h'
 =e_1'J(\widehat{\mathbf A}^h-\mathbf A^h)J'B
  +e_1'J\widehat{\mathbf A}^hJ'\Delta B,
\label{pa:eq:recursive-error}
\end{equation}
where, for $h\geq1$,
\begin{equation}
\widehat{\mathbf A}^h-\mathbf A^h
 =\sum_{\ell=0}^{h-1}\widehat{\mathbf A}^\ell\Delta \mathbf A \mathbf A^{h-1-\ell}.
\label{pa:eq:power-difference}
\end{equation}

\subsection{Decomposition of the LP estimator}\label{pa:sec:lp-decomposition}
For $L=T_h$, define
\begin{align*}
\mathcal S_h^N&=L^{-1}\sum_{t=1}^L\sum_{j=1}^h
          \varepsilon_{1,t+j}\Psi_{N,j,h,t}\beta_{h-j},&
\mathcal S_h&=L^{-1}\sum_{t=1}^L s_{t,h},\\
M_h&=(NL)^{-1}\sum_i\sum_{t=1}^L z_{it}v_{i,t,h}^u,&
D_h^v&=N^{-1}\sum_i\bar z_{i,I_h}\bar v_{i,I_h},\\
E_h^v&=L^{-1}\sum_{t=1}^L\bar z_{N,t}\bar v_{N,t},&
F_h^v&=\bar z_{N,I_h}\bar v_{N,I_h},\\
S_h^r&=(NL)^{-1}\sum_i\sum_{t=1}^L\ddot r_{it,I_h}\ddot v_{i,t,h,I_h},&
J_h&=(\widehat L_h-L_0)S_h^r.
\end{align*}
Since the FWL residual is orthogonal to both fixed effects and the sample
lag controls, its normal equation and the same mean expansion imply
\begin{align}
\widehat\beta_h^{LP}-\beta_h
={}&Q_h^{-1}\mathcal S_h+Q_h^{-1}(\mathcal S_h^N-\mathcal S_h)+Q_h^{-1}M_h
\nonumber\\
&+Q_h^{-1}(-D_h^v-E_h^v+F_h^v)+Q_h^{-1}J_h
\nonumber\\
&+(\widehat Q_h^{-1}-Q_h^{-1})
  \{\mathcal S_h^N+M_h-D_h^v-E_h^v+F_h^v+J_h\}
\nonumber\\
={}&I_h^{LP}+II_h^{LP}+III_h^{LP}+IV_h^{LP}+V_h^{LP}+VI_h^{LP}.
\label{pa:eq:lp-decomposition}
\end{align}
In particular, the martingale argument below is applied to the population
residual $z_{it}$, not to the full-sample residual $\widehat z_{it,h}$.

\subsection{Lemmas}\label{pa:sec:lemmas}
Proofs of all lemmas are in Appendix \ref{pa:sec:lemma-proofs}.

\begin{lemma}[Population projections and aggregate moments]\label{pa:lem:population}
Under Assumptions~\ref{pa:ass:sampling}--\ref{pa:ass:solution}, the population projections satisfy
\[
\mathbb{E}[r_{it}\mid\mathcal A]
=\mathbb{E}[r_{it}\mid\alpha_i]=0,\qquad
\ddot r_{it,I}=\ddot R_{it,I}.
\]
The variable $r_{it}$ is measurable before $u_{it}$ is drawn, conditional
on $\mathcal F_{t-1}\vee\sigma(\varepsilon_{1t},\lambda_t)$; $\Psi_{j,h,t}$ is
$\mathcal A_{t+j-1}$-measurable.
Conditional on $\alpha_i$, the coordinates in $\mathcal V_h$ are centered,
stationary and ergodic.
Whenever $Q\succ0$, the aggregate sequences $f_t^{ab}$, $k_{r,h}$ and
$s_{t,h}$ are stationary and ergodic, and, with $\delta=\eta/2$,
\[
\mathbb{E}\|z_{it}y_{i,t+j-1}'\|^{2+\delta}<\infty,\qquad
\mathbb{E}\|k_{0,h}\|^{2+\delta}<\infty .
\]
In particular $Q$ and $\Omega_h$ are finite. The companion matrix $\mathbf A$ is stable.
For the lag matrices in \eqref{pa:eq:bias-notation},
\[
\Gamma_\ell=(0,\Sigma_u J(\mathbf A^{\ell-1})'),\qquad \ell\geq1,
\qquad
\sum_{\ell=1}^\infty(1+\ell)\|\Gamma_\ell\|<\infty.
\]
The zero block corresponds to the $m$ shock-interaction regressors.
\end{lemma}
The proof is provided in Appendix~\ref{pa:sec:lemma-proofs}.

\begin{lemma}[Time means and Gram matrices]\label{pa:lem:gram}
For $V=r,u,v_h$ and each interval of length $L\asymp T$,
$\mathbb{E}\|\bar V_{i,I}\|^2\leq C/L$ and
$\mathbb{E}\|\bar V_{i,I}\|^4\leq C/L^2$.
Moreover, for $I^{(0)},I^{(1)},I^{(2)},I_h$,
\begin{align*}
\widehat Q_I-Q&=O_p(T^{-1/2}+N^{-1/2}),\\
\widehat Q_I^{-1}-Q^{-1}&=O_p(T^{-1/2}+N^{-1/2}),\qquad
\|\widehat Q_I^{-1}\|=O_p(1).
\end{align*}
Consequently $\widehat L_h-L_0=o_p(1)$ and
$\widehat Q_h^{-1}-Q_h^{-1}=o_p(1)$.
\end{lemma}
The proof is provided in Appendix~\ref{pa:sec:lemma-proofs}.

\begin{lemma}[Split weights and cross-block correction]\label{pa:lem:split}
For an array $\xi_t$ common to the full and half samples and any real $\gamma$,
\[
\mathcal J_T\left[L^{-\gamma}\sum_{t\in I}\xi_t\right]
=(2-2^{\gamma-1})T^{-\gamma}\sum_{t=1}^T\xi_t.
\]
This identity is pathwise and does not require $\mathbb{E}\xi_t=0$.
The mean correction obeys \eqref{pa:eq:cross-block}, and
\begin{align}
\mathbb{E}\mathcal J_T[D^u]
&=\frac{2}{T^2}\sum_{\ell=1}^{T-1}n_{\ell,T}\Gamma_\ell
=O(T^{-2}),\nonumber\\
n_{\ell,T}
&=\begin{cases}
\ell,&1\leq\ell\leq T/2,\\
T-\ell,&T/2<\ell<T,
\end{cases}\nonumber\\
\mathcal J_T[D^u]-\mathbb{E}\mathcal J_T[D^u]
&=O_p(N^{-1/2}T^{-1}+T^{-3/2}).
\label{pa:eq:split-bounds}
\end{align}
\end{lemma}
The proof is provided in Appendix~\ref{pa:sec:lemma-proofs}.

\begin{lemma}[One-step terms and bias correction]\label{pa:lem:var}
For each full or half interval with $L\asymp T$,
\begin{align*}
U_I&=O_p((NL)^{-1/2}),&
\mathbb{E} D_I^u&=\Phi_L/L,\\
D_I^u-\mathbb{E} D_I^u&=O_p(N^{-1/2}L^{-1}+L^{-3/2}),&
E_I^u&=O_p(N^{-1}L^{-1/2}),\\
F_I^u&=O_p(N^{-1}L^{-1}),&
(\Phi_L-\Phi)/L&=O(L^{-2}).
\end{align*}
In \eqref{pa:eq:var-decomposition}, $I^{VAR}=O_p((NT)^{-1/2})$ and
\begin{align*}
II^{VAR}&=O_p(T^{-2}+N^{-1/2}T^{-1}+T^{-3/2}),\\
III^{VAR}&=O_p(N^{-1}T^{-1/2}),\qquad
IV^{VAR}=O_p(N^{-1}T^{-1}),\\
V^{VAR}&=O_p\!\left[
 \{(NT)^{-1/2}+T^{-1}\}(T^{-1/2}+N^{-1/2})\right].
\end{align*}
These bounds imply, without a relative growth restriction on $N$ and $T$,
\[
\|\widehat D-D\|=O_p((NT)^{-1/2}+T^{-3/2}).
\]
Under $N=o(T^2)$, the last four terms are $o_p((NT)^{-1/2})$.
Consequently
\begin{equation}
\sqrt{NT}(\widehat\psi-\psi)
=H\frac1{\sqrt{NT}}\sum_i\sum_{t=1}^T(u_{it}\otimes r_{it})+o_p(1)
\ \xrightarrow{d}\ N(0,\Sigma_\psi).
\label{pa:eq:var-linear}
\end{equation}
The uncorrected interval estimator also satisfies
\begin{equation}
\widehat\psi_I-\psi
=H\frac1{NL}\sum_i\sum_{t\in I}(u_{it}\otimes r_{it})
 +\frac{b_\psi}{L}+o_p((NT)^{-1/2}).
\label{pa:eq:interval-expansion}
\end{equation}
\end{lemma}
The proof is provided in Appendix~\ref{pa:sec:lemma-proofs}.

\begin{lemma}[LP aggregate term]\label{pa:lem:aggregate}
For fixed $h\geq1$,
\[
\sqrt{T_h}\mathcal S_h\xrightarrow{d}N(0,\Omega_h),\qquad
\mathcal S_h^N-\mathcal S_h=O_p((NT_h)^{-1/2}).
\]
The autocovariance of $s_{t,h}$ vanishes at $|\ell|\geq h$, and its
long-run variance equals \eqref{pa:eq:aggregate-notation}.
Consequently $\sqrt T I_h^{LP}\xrightarrow{d}
N(0,Q_h^{-1}\Omega_hQ_h^{-1})$ and
$II_h^{LP}=O_p((NT)^{-1/2})$.
\end{lemma}
The proof is provided in Appendix~\ref{pa:sec:lemma-proofs}.

\begin{lemma}[Remaining LP terms]\label{pa:lem:lp-rest}
At every fixed $h\geq1$,
\[
M_h=O_p((NT)^{-1/2}),\quad
D_h^v,F_h^v=O_p(T^{-1}),\quad E_h^v=O_p((NT)^{-1/2}),\quad
S_h^r=O_p(T^{-1/2}).
\]
In \eqref{pa:eq:lp-decomposition},
$III_h^{LP}=O_p((NT)^{-1/2})$,
$IV_h^{LP}=O_p(T^{-1}+(NT)^{-1/2})$, and
$V_h^{LP},VI_h^{LP}=o_p(T^{-1/2})$.
\end{lemma}
The proof is provided in Appendix~\ref{pa:sec:lemma-proofs}.

\subsection{Proofs of Theorems~\ref{prop:parametric-var} and~\ref{prop:parametric-lp}}\label{pa:sec:proposition-proof}
For VAR, Lemma~\ref{pa:lem:var} gives
$\widehat\psi-\psi=O_p((NT)^{-1/2})$ under $N=o(T^2)$.
For fixed $h$, \eqref{pa:eq:recursive-error}--\eqref{pa:eq:power-difference}
give
\begin{align}
\widehat\beta_h^{VAR\,\prime}-\beta_h'
={}&e_1'J\left\{\sum_{\ell=0}^{h-1}
 \mathbf A^\ell\Delta\mathbf A\mathbf A^{h-1-\ell}\right\}J'B
 +e_1'J\mathbf A^hJ'\Delta B+\mathcal E_{h,NT}' ,
\label{pa:eq:recursive-linear}
\end{align}
where
$\|\mathcal E_{h,NT}\|\leq
C_h(\|\Delta\mathbf A\|^2+\|\Delta\mathbf A\|\,\|\Delta B\|)
=O_p((NT)^{-1})$; at $h=0$ it is zero.
The linear part is $G_h(\widehat\psi-\psi)$, hence
\[
\sqrt{NT}\{\widehat\theta_h^{VAR}(y,y')-\theta_h(y,y')\}
=d'G_h\sqrt{NT}(\widehat\psi-\psi)+o_p(1).
\]
Lemma~\ref{pa:lem:var} proves \eqref{eq:parametric-var-limit}.
For the uncorrected estimator with $N/T\to\kappa\in(0,\infty)$,
\eqref{pa:eq:interval-expansion} instead gives limiting mean
$\sqrt\kappa\,d'G_hb_\psi$.

For the LP part, use the exact six-term decomposition
\eqref{pa:eq:lp-decomposition}. Lemma \ref{pa:lem:aggregate} gives the normal limit
of its first term, and Lemmas \ref{pa:lem:aggregate} and \ref{pa:lem:lp-rest} make
the other five terms negligible after multiplication by $\sqrt T$.
Contraction by $d'$ proves \eqref{eq:parametric-lp-limit}. At $h=0$, the LP equation
is the first equation of the one-step system, with the same regressors and
the same two fixed effects. The impact estimator identity defined above
is exact, so Lemma \ref{pa:lem:var} also proves the corrected impact claim.
\hfill$\square$

\subsection{Proofs of the lemmas}\label{pa:sec:lemma-proofs}
\begin{proof}[Proof of Lemma \ref{pa:lem:population}]
Given $\mathcal A$, each unit path is a common measurable function of its
own $\alpha_i$ and micro innovations. The paths are therefore conditionally
iid. By independence of $\alpha_i$ and $\mathcal A$,
$\mathbb{E}[b_i\mid\mathcal A]=0$ and
$\mathbb{E}[a_t\mid\alpha_i]=\mathbb{E} R_{it}$.
Substitution into \eqref{pa:eq:population-within} proves both centering
identities. Since $a_t$ is date-specific and $b_i$ is unit-specific, sample
two-way within removes them exactly.

The regressor uses only lagged micro outcomes and the current aggregate
shock. Independence of future aggregate innovations implies that its
conditional expectation given $\mathcal A$ depends only on $\mathcal A_t$.
Similarly, $z_{it}y_{i,t+j-1}'$ depends on aggregate innovations only through
date $t+j-1$, establishing the stated measurability.
Every term of $v_{i,t,h}$ has zero expectation conditional on $\alpha_i$:
for its aggregate terms condition first on $\mathcal F_{t+j-1}$, and for
its micro terms use the zero-mean innovations. The corresponding centering
for $r$ was just proved, and for $u$ is immediate.

Conditional on $\alpha_i$, the causal coordinates in $\mathcal V_h$
are factors of the stationary ergodic aggregate process and the
independent iid micro process. Their product is ergodic.
Integrating out the unit inputs shows that $f_t^{ab}$, $\Psi_{j,h,t}$,
$k_{r,h}$ and $s_{t,h}$ are stationary ergodic functions of the aggregate
path.

Conditional Jensen gives $\mathbb E\|r_{it}\|^q<\infty$ and,
when $Q\succ0$, $\mathbb E\|z_{it}\|^q<\infty$.
H\"older's inequality yields
\[
\mathbb E\|z_{it}y_{i,t+j-1}'\|^{q/2}
\leq(\mathbb E\|z_{it}\|^q)^{1/2}
     (\mathbb E\|y_{i,t+j-1}\|^q)^{1/2}<\infty.
\]
Jensen and fixed $h$ transfer this bound to $\Psi_{j,h,t}$ and $k_{r,h}$.
Independence of the current shock gives the same $2+\eta/2$ moment
for $\varepsilon_{1r}k_{r,h}$.

To verify stability, let $Y_{it}=(y_{it}',\ldots,y_{i,t-p+1}')'$.
Innovation independence and $\mathbb{E}\varepsilon_{1t}=0$ give
$\mathbb{E}[(Y_{it}-\mathbb{E} Y_{it})u_{i,t-\ell}']=\mathbf A^\ell J'\Sigma_u$
for $\ell\geq0$, by recursion. The standardized innovations
$\Sigma_u^{-1/2}u_{i,t-\ell}$ are orthonormal in $L^2$ across coordinates
and periods. Bessel's inequality therefore implies
\[
\sum_{\ell=0}^{\infty}\|\mathbf A^\ell J'\Sigma_u^{1/2}\|_F^2
\leq\mathbb{E}\|Y_{it}-\mathbb{E} Y_{it}\|^2<\infty.
\]
The companion pair $(\mathbf A,J')$ is controllable. Since $\Sigma_u\succ0$,
an eigenvalue of $\mathbf A$ with modulus at least one would make this sum diverge.
Hence $\rho(\mathbf A)<1$.

For $\ell\geq1$, the interaction block of
$\mathbb{E}[u_{i0}r_{i\ell}']$ is zero because the current
$\varepsilon_{1\ell}$ is fresh and mean zero. The control block is
the transpose of the covariance recursion just established:
\[
\mathbb{E}[u_{i0}Y_{i,\ell-1}']
 =\Sigma_u J(\mathbf A^{\ell-1})'.
\]
Subtracting the population date component and unit component from
$R_{i\ell}$ does not change this covariance: the former depends only
on common innovations, and the latter on the independent permanent
effect. This proves the formula for $\Gamma_\ell$. Stability of $\mathbf A$
makes $\|\mathbf A^\ell\|$ decay geometrically up to a fixed constant and
therefore proves the claimed weighted summability. 
\end{proof}

\begin{proof}[Proof of Lemma \ref{pa:lem:gram}]
Conditional on $\alpha_i$, the relevant scalar coordinates are centered.
Expanding the second moment of a sum and using the covariance sum in
Assumption~\ref{pa:ass:dependence} gives $CL$ after taking expectations.
For the fourth moment, the cumulant expansion has one fourth-cumulant sum
of order $L$ and three products of covariance sums of order $L^2$.
The bound on $\mathbb{E} C_i^2$ therefore gives
$\mathbb{E}|\sum_{t\in I}V_{it}|^4\leq CL^2$.
Division by $L^2$ and $L^4$ proves the mean bounds.

For a raw Gram entry, suppress the coordinate labels and write
$F_{it}=F_{it}^{ab}$, $f_t=f_t^{ab}$ and $\zeta_{it}=F_{it}-f_t$.
The fourth regressor moments and conditional Jensen give
$\mathbb{E}\zeta_{it}^2\leq C$. Use the exact centering identity
\[
\frac1{NL}\sum_{i,t}F_{it}-\mathbb{E} F_{it}
=\frac1L\sum_t(f_t-\mathbb{E} f_t)
 +\frac1{NL}\sum_{i,t}\zeta_{it}.
\]
The first variance is $O(L^{-1})$ by Assumption
\ref{pa:ass:dependence}. Given $\mathcal A$, different unit paths and hence
their centered averages are independent. Jensen's inequality bounds the
second variance by $C/N$, even if the unit-specific time averages retain
permanent heterogeneity in their second moments. Thus the raw Gram differs
from $Q$ by $O_p(L^{-1/2}+N^{-1/2})$.

Its two-way within counterpart subtracts
$N^{-1}\sum_i\bar r_{i,I}\bar r_{i,I}'$ and
$L^{-1}\sum_t\bar r_{N,t}\bar r_{N,t}'$ and adds
$\bar r_{N,I}\bar r_{N,I}'$. The first is $O_p(L^{-1})$ by the mean bound.
Conditional cross-sectional centering gives
$\mathbb{E}\|\bar r_{N,t}\|^2\leq C/N$, so the second is $O_p(N^{-1})$.
The grand-mean term is bounded by either corresponding positive matrix
average and is negligible. This proves the Gram result.
Positive definiteness of $Q$ implies, with probability tending to one,
$\lambda_{\min}(\widehat Q_I)\geq\lambda_{\min}(Q)/2$.
Use $\widehat Q_I^{-1}-Q^{-1}
=-Q^{-1}(\widehat Q_I-Q)\widehat Q_I^{-1}$ and continuity of the Schur
complement to obtain all remaining statements.
\end{proof}

\begin{proof}[Proof of Lemma \ref{pa:lem:split}]
Since the halves partition $I^{(0)}$ and each has length $T/2$,
\[
\tfrac12\sum_{j=1}^2 (T/2)^{-\gamma}\sum_{t\in I^{(j)}}\xi_t
=2^{\gamma-1}T^{-\gamma}\sum_{t=1}^T\xi_t.
\]
Subtracting this expression from twice the full-sample expression proves
the weighting rule. In particular, averages with $\gamma=1$ are retained,
while terms with $\gamma=2$ cancel, regardless of their means.

For $D_I^u=(NL^2)^{-1}\sum_i\sum_{s,t\in I}u_{is}r_{it}'$,
the full-sample coefficient after multiplication by two is $2/(NT^2)$.
Each half-sample coefficient after multiplication by one half is also
$2/(NT^2)$. Thus the within-half pairs cancel and only the two cross-block
rectangles in \eqref{pa:eq:cross-block} remain.
The rectangle $s\in I^{(2)},t\in I^{(1)}$ has zero expectation by innovation timing.
For the other rectangle, stationarity gives
$\mathbb{E}[u_{is}r_{it}']=\Gamma_{t-s}$.
Counting pairs with $t-s=\ell$ gives $n_{\ell,T}$ in
\eqref{pa:eq:split-bounds}. Since $n_{\ell,T}\leq\ell$, the weighted
summability proved in Lemma~\ref{pa:lem:population} bounds the mean by $CT^{-2}$.

For the centered correction, expand each coordinate's variance over the
cross-block rectangles. Their sums of absolute covariances are no larger
than the corresponding sums over all four periods in $I^{(0)}$.
Condition \eqref{pa:eq:four-date-condition} therefore gives
\[
\mathbb{E}\|\mathcal J_T[D^u]-\mathbb{E}\mathcal J_T[D^u]\|^2
\leq \frac{C}{N^2T^4}\{NT^2+N(N-1)T\}
\leq C\{(NT^2)^{-1}+T^{-3}\}.
\]
The $T^{-3/2}$ remainder is negligible at the root-$NT$ scale
under the sufficient restriction $N=o(T^2)$.
\end{proof}

\begin{proof}[Proof of Lemma \ref{pa:lem:var}]
Conditional on $\mathcal F_{t-1}\vee\sigma(\varepsilon_{1t},\lambda_t)$, $r_{it}$ is known and $u_{it}$
is fresh, mean zero and independent across units. Hence
$\sum_i u_{it}r_{it}'$ is a martingale difference over periods, with second
moment bounded by $CN$. Summing over $L$ periods proves the rate of $U_I$.

To compute the bias, expand the actual unit-mean product:
\[
\mathbb{E} D_I^u=L^{-2}\sum_{s,t\in I}\mathbb{E}[u_{is}r_{it}'].
\]
If $s\geq t$, the expectation is zero by innovation independence and timing.
If $t=s+\ell$, it equals $\Gamma_\ell$ by stationarity. There are exactly
$L-\ell$ such pairs. Therefore
\[
\mathbb{E} D_I^u=L^{-2}\sum_{\ell=1}^{L-1}(L-\ell)\Gamma_\ell
=\Phi_L/L.
\]
The weighted summability in Lemma~\ref{pa:lem:population} gives
\[
\|\Phi_L-\Phi\|
\leq\sum_{\ell\geq L}\|\Gamma_\ell\|
  +L^{-1}\sum_{\ell<L}\ell\|\Gamma_\ell\|\leq C/L.
\]
For each centered coordinate of $D_I^u$, expand its variance over
$(i,\ell,t,s,r,v)$. Condition \eqref{pa:eq:four-date-condition} gives
\[
\mathbb{E}\|D_I^u-\mathbb{E} D_I^u\|^2
\leq\frac{C}{N^2L^4}\{NL^2+N(N-1)L\}
\leq C\{(NL^2)^{-1}+L^{-3}\}.
\]

For the date correction, $\bar r_{N,t}$ is known before $u_{it}$ is drawn,
$\mathbb{E}\|\bar r_{N,t}\|^2\leq C/N$, and the conditional covariance of
$\bar u_{N,t}$ is $\Sigma_u/N$. Consequently the summands of $E_I^u$ are
martingale differences and
$\mathbb{E}\|E_I^u\|^2\leq C/(N^2L)$.
For the grand mean, conditional independence and centering across units,
together with the fourth-moment bound of Lemma \ref{pa:lem:gram}, give
\[
\mathbb{E}\|\bar r_{N,I}\|^4\leq C/(N^2L^2),\qquad
\mathbb{E}\|\bar u_{N,I}\|^4\leq C/(N^2L^2).
\]
Cauchy--Schwarz now proves $F_I^u=O_p((NL)^{-1})$.

For the joint decomposition \eqref{pa:eq:var-decomposition},
Lemma \ref{pa:lem:split} controls $II^{VAR}$, and the preceding date-mean and
grand-mean bounds control $III^{VAR}$ and $IV^{VAR}$.
Furthermore, uniformly over the three intervals,
\[
U_I-D_I^u-E_I^u+F_I^u
=O_p((NT)^{-1/2}+T^{-1}).
\]
Lemma \ref{pa:lem:gram} gives
$\widehat Q_I^{-1}-Q^{-1}=O_p(T^{-1/2}+N^{-1/2})$,
so the stated bound for $V^{VAR}$ follows without equating the sample
inverses. Summing the five bounds gives the stated unrestricted rate
for $\widehat D-D$.

The normalized bounds for the four remainders are respectively
\[
\begin{split}
\sqrt{NT}\|II^{VAR}\|
 &=O_p\!\left(\frac{\sqrt N}{T^{3/2}}
             +\frac1{\sqrt T}+\frac{\sqrt N}{T}\right),\\
\sqrt{NT}\|III^{VAR}\|&=O_p(N^{-1/2}),\qquad
\sqrt{NT}\|IV^{VAR}\|=O_p((NT)^{-1/2}),\\
\sqrt{NT}\|V^{VAR}\|
 &=O_p\!\left(T^{-1/2}+N^{-1/2}+\frac{\sqrt N}{T}\right).
\end{split}
\]
All tend to zero when $N,T\to\infty$ and $N=o(T^2)$.
Thus $\widehat D-D=U_{I^{(0)}}Q^{-1}+o_p((NT)^{-1/2})$.
Since
$\mathrm{vec}(U_{I^{(0)}}')=(NT)^{-1}\sum_{i,t}(u_{it}\otimes r_{it})$,
vectorization gives the representation in \eqref{pa:eq:var-linear}.

For the uncorrected estimator, insert
$D_I^u=\Phi/L+(\Phi_L-\Phi)/L+(D_I^u-\mathbb{E} D_I^u)$ into
\eqref{pa:eq:var-exact}. The bounds already proved give
\eqref{pa:eq:interval-expansion} under the same growth condition.

Finally, define $W_{N,t}=N^{-1/2}\sum_i(u_{it}\otimes r_{it})$ and
$Q_{N,t}=N^{-1}\sum_i r_{it}r_{it}'$. The sequence $W_{N,t}$ is a
martingale difference for $\mathcal F_t$. Conditional independence of the
micro innovations gives the exact predictable covariance
\[
\mathbb{E}[W_{N,t}W_{N,t}'\mid\mathcal F_{t-1}]
=\Sigma_u\otimes\mathbb{E}[Q_{N,t}\mid\mathcal F_{t-1}].
\]
The raw Gram argument in Lemma \ref{pa:lem:gram} gives
$T^{-1}\sum_t Q_{N,t}\to_p Q$.
Each coordinate of $Q_{N,t}-\mathbb{E}[Q_{N,t}\mid\mathcal F_{t-1}]$
is a martingale difference with uniformly bounded second moment, by
Jensen's inequality and the fourth regressor moments. Its time average
is $O_p(T^{-1/2})$. The averaged predictable covariance therefore
converges to $\Sigma_u\otimes Q$.

Conditional on $\mathcal F_{t-1}\vee\sigma(\varepsilon_{1t},\lambda_t)$, the independent centered micro
innovations and their finite fourth moments give
\[
\mathbb{E}[\|W_{N,t}\|^4\mid
 \mathcal F_{t-1}\vee\sigma(\varepsilon_{1t},\lambda_t)]
\leq C\left\{N^{-1}\sum_i\|r_{it}\|^2\right\}^2.
\]
Taking expectations and applying Jensen shows
$\sup_{N,t}\mathbb{E}\|W_{N,t}\|^4<\infty$.
For any $\epsilon>0$, the expectation of the conditional Lindeberg sum is at most
\[
\frac1T\sum_{t=1}^T
\mathbb{E}[\|W_{N,t}\|^2 1\{\|W_{N,t}\|>\epsilon\sqrt T\}]
\leq C\epsilon^{-2}T^{-1}\to0.
\]
Markov's inequality gives conditional Lindeberg in probability. The
martingale central limit theorem and Cram\'er--Wold give
$T^{-1/2}\sum_t W_{N,t}\xrightarrow{d}N(0,\Sigma_u\otimes Q)$; see
\citet[Chapter 3]{HallHeyde1980}. Multiplication by $H$ finishes the proof.
\end{proof}

\begin{proof}[Proof of Lemma \ref{pa:lem:aggregate}]
First use the population residual, whose timing is explicit. Put $L=T_h$.
Changing the order of the finite sum, with $r=t+j$, gives
\[
\sum_{t=1}^L s_{t,h}
=\sum_{r=2}^{L+h}\varepsilon_{1r}
  \sum_{j=1}^h1\{1\leq r-j\leq L\}
               \Psi_{j,h,r-j}\beta_{h-j}.
\]
For the interior periods the inner sum is $k_{r,h}$. Completing it to
$\sum_{r=1}^L\varepsilon_{1r}k_{r,h}$ changes only a fixed number of
boundary terms, of bounded second moment. After division by $\sqrt L$,
the boundary remainder is $O_p(L^{-1/2})$.
The sequence $\varepsilon_{1r}k_{r,h}$ is a martingale difference because
$k_{r,h}$ is $\mathcal A_{r-1}$-measurable and the current aggregate shock
is independent of that field. Its conditional covariance is
$\sigma_\varepsilon^2 k_{r,h}k_{r,h}'$.
Ergodicity and the moment in Lemma \ref{pa:lem:population} imply variance
convergence and Lindeberg, giving the stated normal limit by the same
martingale theorem as in Lemma \ref{pa:lem:var}.

Each summand of $s_{t,h}$ has zero conditional mean given $\mathcal F_t$.
Moreover $s_{t-\ell,h}$ is $\mathcal F_t$-measurable when $\ell\geq h$.
Iterated expectations give zero autocovariance at those lags. The variance
of a normalized sum of $s_{t,h}$ thus converges to its finite covariance
sum. The calendar-time martingale representation has limiting variance
$\sigma_\varepsilon^2\mathbb{E}[k_{0,h}k_{0,h}']$; the boundary error tends
to zero in $L^2$, so the two formulas are equal. 

For the cross-sectional replacement, conditional on $\mathcal A$ the unit
products in $\Psi_{N,j,h,t}$ are identically distributed and independent.
Lemma \ref{pa:lem:population} gives
\[
\mathbb{E}\|\Psi_{N,j,h,t}-\Psi_{j,h,t}\|^2\leq C/N.
\]
Crucially, this difference is measurable at $t+j-1$. Reindex its sum by
$r=t+j$ as above. The resulting summands are martingale differences,
and the squared norm of their loadings has expectation at most $C/N$
because $h$ is fixed. Thus
\[
\mathbb{E}\|\mathcal S_h^N-\mathcal S_h\|^2\leq C/(NL).
\]
The assertions about $I_h^{LP}$ and
$II_h^{LP}$ follow by multiplication by $Q_h^{-1}$ and $T/T_h\to1$.
\end{proof}

\begin{proof}[Proof of Lemma \ref{pa:lem:lp-rest}]
Reindex the micro score by $r=t+j$:
\[
\sum_i\sum_{t=1}^L z_{it}v_{i,t,h}^u
=\sum_{r=1}^{L+h}\sum_i\sum_{j=0}^h
 1\{1\leq r-j\leq L\}z_{i,r-j}\pi_{j,h}'u_{ir}.
\]
For $j\geq1$ the loading is $\mathcal F_{r-1}$-measurable. For $j=0$
it is measurable conditional on $\mathcal F_{r-1}\vee\sigma(\varepsilon_{1r},\lambda_r)$, before the
micro innovations at $r$ are drawn. Independent micro innovations make
the date sums martingale differences with second moments bounded by $CN$.
This proves $\mathbb{E}\|M_h\|^2\leq C/(NL)$.

By Lemma \ref{pa:lem:gram} and Cauchy--Schwarz,
\[
\mathbb{E}\|D_h^v\|
\leq N^{-1}\sum_i
 (\mathbb{E}\|\bar z_{i,I_h}\|^2\,
  \mathbb{E}|\bar v_{i,I_h}|^2)^{1/2}\leq C/L.
\]
Jensen's inequality applied to the unit averages gives the same $C/L$
bound for $\mathbb{E}\|F_h^v\|$; a faster rate is unnecessary here.

For $E_h^v$, split $\bar v_{N,t}$ using \eqref{eq:parametric-lp-error}.
The aggregate part has summands
$\varepsilon_{1,t+j}\bar z_{N,t}\bar y_{N,t+j-1}'\beta_{h-j}$.
Their loadings are measurable at $t+j-1$. Conditional cross-sectional
centering and the finite fourth moments give
$\mathbb{E}\|\bar z_{N,t}\|^4\leq C/N^2$ and
$\mathbb{E}\|\bar y_{N,t+j-1}\|^4\leq C$.
Hence their loading second moments are bounded by $C/N$.
Calendar-time reindexing gives an $O((NL)^{-1})$ second-moment bound for
the aggregate date correction. For the micro part, the same reindexing
and conditional covariance $\Sigma_u/N$ for $\bar u_{N,r}$ give a bound
$C/(N^2L)$. Thus $E_h^v=O_p((NL)^{-1/2})$.

The preceding score arguments also hold with $r_{it}$ in place of
$z_{it}$: the aggregate raw loadings have bounded second moments by
the $q>4$ regressor and state moments, and each future shock remains
orthogonal to its loading. Together with the mean bounds this gives
$S_h^r=O_p(L^{-1/2})$. Consequently
$J_h=(\widehat L_h-L_0)S_h^r=o_p(L^{-1/2})$ by Lemma \ref{pa:lem:gram}.
Finally the full score in braces in \eqref{pa:eq:lp-decomposition} is
$O_p(L^{-1/2})$, and the inverse-Gram difference is $o_p(1)$.
This proves each of the remaining term bounds.
\end{proof}

\section{Nonparametric results}\label{app:nonparametric}
\subsection{Series construction and estimators}

Let $F_y$ denote the stationary marginal law of $y_{i,t-1}$. 
Let $p_K(y)\in\mathbb R^K$ be a deterministic, computable raw basis
chosen without knowledge of $F_y$, with no constant direction. Write
$\bar p_K=\int p_K\,dF_y$, $p_K^c=p_K-\bar p_K$, and
$M_K=\int p_K^c(p_K^c)'\,dF_y$. These population-centered quantities
are used only in the analysis; the estimator evaluates the raw $p_K$.
Define the population projection with an intercept before any within transformation:
\begin{equation}
B_K=\left\{\int\mu(y)p_K^c(y)'\,dF_y(y)\right\}M_K^{-1},\quad
c_K=\int\mu\,dF_y-B_K\bar p_K,\quad
\varrho_K(y)=\mu(y)-c_K-B_Kp_K(y).
\label{np:eq:projection}
\end{equation}
Write $b_{h,K}'=e_1'J\mathbf A^hJ'B_K$,
$c_{h,K}=e_1'J\mathbf A^hJ'c_K$, and
$\varrho_{h,K}=\mu_h-c_{h,K}-b_{h,K}'p_K$.
The intercepts $c_K$ and $c_{h,K}$ are absorbed by time effects and cancel
from response contrasts. The notation $K$ always denotes series order;
$m$ denotes the state dimension. Set
\begin{gather*}
R_{it,K}=\begin{pmatrix}\varepsilon_{1t}p_K(y_{i,t-1})\\Y_{i,t-1}\end{pmatrix},
\qquad \xi_{h,K}=\begin{pmatrix}b_{h,K}\\c_h\end{pmatrix},\\
D_K=(B_K,A_1,\ldots,A_p),\qquad S_K=(I_K,0),\qquad P_Y=(0,I_{mp}).
\end{gather*}
The one-step approximation error is $\varrho_{it,K}=\varrho_K(y_{i,t-1})\varepsilon_{1t}$;
the LP approximation error is $\varrho_{it,h,K}=\varrho_{h,K}(y_{i,t-1})\varepsilon_{1t}$.

For a vector-valued function $f$, let
$\|f\|_{F_y}^2=\int\|f(y)\|^2dF_y(y)$, and for a response contrast use
$\|\theta\|_{L^2(\mathcal Y\times\mathcal Y)}^2=\iint|\theta(y_1,y_2)|^2dF_y(y_1)dF_y(y_2)$.
Matrix norms without a subscript
are spectral norms; $\|\cdot\|_F$ is the Frobenius norm. The integrated
norms use $F_y$ and $F_y\times F_y$, respectively.

\subsection{Two-way within estimators}
For a contiguous set $I$ of $L$ origin periods and any panel array $z$, define
\begin{gather*}
\bar z_{i,I}=L^{-1}\sum_{t\in I}z_{it},\quad
\bar z_{N,t}=N^{-1}\sum_i z_{it},\quad
\bar z_{N,I}=(NL)^{-1}\sum_i\sum_{t\in I}z_{it},\\
\ddot z_{it,I}=z_{it}-\bar z_{i,I}-\bar z_{N,t}+\bar z_{N,I}.
\end{gather*}
Columns constant across units or across periods transform to zero and are
omitted. In particular, a constant series interaction is a date effect.
Define
\[
\widehat Q_{I,K}=(NL)^{-1}\sum_i\sum_{t\in I}\ddot R_{it,K,I}\ddot R_{it,K,I}',
\qquad
\widehat D_I=\left\{(NL)^{-1}\sum_i\sum_{t\in I}
\ddot y_{it,I}\ddot R_{it,K,I}'\right\}\widehat Q_{I,K}^{-1}.
\]
Let $I^{(0)}=\{1,\ldots,T\}$ and, for even $T$, partition it into
contiguous halves $I^{(1)},I^{(2)}$ of length $T/2$. Set
\[
\widehat D=2\widehat D_{I^{(0)}}-\tfrac12
 (\widehat D_{I^{(1)}}+\widehat D_{I^{(2)}})
 =(\widehat B_K,\widehat A_1,\ldots,\widehat A_p).
\]
For $p\geq2$, form the estimated companion matrix
\[
\widehat{\mathbf A}=
\begin{pmatrix}
(\widehat A_1,\ldots,\widehat A_{p-1})&\widehat A_p\\
I_{m(p-1)}&0_{m(p-1)\times m}
\end{pmatrix};
\]
for $p=1$, set $\widehat{\mathbf A}=\widehat A_1$.
Define the recursive series estimator by
\begin{equation}
\widehat\mu_h^{VAR}(y)
 =e_1'J\widehat{\mathbf A}^hJ'\widehat B_Kp_K(y).
\label{np:eq:recursive-estimator}
\end{equation}
Equivalently, start with
$\widehat{\mathcal R}_0(y)=J'\widehat B_K p_K(y)$ and iterate
$\widehat{\mathcal R}_{h+1}(y)=\widehat{\mathbf A}\widehat{\mathcal R}_h(y)$;
then $\widehat\mu_h^{VAR}(y)=e_1'J\widehat{\mathcal R}_h(y)$.
At impact, $\widehat\mu_0^{VAR}(y)=e_1'\widehat B_K p_K(y)$.
The one-step parameters are corrected before the recursion is applied.

Each half uses the same $K$ and raw basis, re-estimates both sets of
fixed effects, and retains observed lags preceding that half. One end
date may be dropped when $T$ is odd. This is the split-panel jackknife
(SPJ) construction of \citet{DhaeneJochmans2015}; all three fits use
the same realized panel.

For $h\geq1$, let $T_h=T-h$ and $I_h=\{1,\ldots,T_h\}$, and define
\[
\widehat\xi_{h,K}^{LP}=\widehat Q_{I_h,K}^{-1}(NT_h)^{-1}
\sum_i\sum_{t\in I_h}\ddot R_{it,K,I_h}\ddot y_{1,i,t+h,I_h},\qquad
\widehat b_{h,K}^{LP}=S_K\widehat\xi_{h,K}^{LP},\qquad
\widehat\mu_h^{LP}(y)=p_K(y)'\widehat b_{h,K}^{LP}.
\]
The outcome is transformed over the same origin periods, so its individual
mean averages $y_{1,i,t+h}$ over $t\in I_h$. At $h=0$ use
$\widehat\mu_0^{LP}=\widehat\mu_0^{VAR}$, namely the bias-corrected impact fit.
For either method $E$, put
$\widehat\theta_h^E(y_1,y_2)=\widehat\mu_h^E(y_1)-\widehat\mu_h^E(y_2)$.

\subsection{Population notation}
Use the aggregate history $\mathcal A_t$, full path $\mathcal A$, and joint filtration $\mathcal F_t$ defined in Appendix~\ref{app:parametric}.
Let $\mathcal C_N=\sigma(\alpha_1,\ldots,\alpha_N)$.
Under the sampling and causal-solution conditions stated below,
the population additive residual is
\[
a_{t,K}^R=\mathbb{E}[R_{it,K}\mid\mathcal A],\quad
b_{i,K}^R=\mathbb{E}[R_{it,K}\mid\alpha_i]-\mathbb{E} R_{it,K},\quad
r_{it,K}=R_{it,K}-a_{t,K}^R-b_{i,K}^R.
\]
Define the cross-sectionally centered residual, its quadratic array, and
a reference Gram conditional on the sampled permanent effects:
\[
\rho_{it,K}=r_{it,K}-\bar r_{N,t,K},\qquad
F_{N,t,K}=N^{-1}\sum_i\rho_{it,K}\rho_{it,K}',\qquad
Q_{N,K}=\mathbb{E}[F_{N,t,K}\mid\mathcal C_N].
\]
The reference Gram is a proof device, not an input to estimation. It is
common to the full sample and the two halves, but may remain random through
$\mathcal C_N$. This avoids imposing a dimension-dependent law of large
numbers over the permanent effects just to replace it by a deterministic
matrix. No restriction such as $K^2/N\to0$ is imposed here.

\subsection{Assumptions}\label{np:sec:assumptions}
Maintain Assumption~\ref{pa:ass:sampling}. The following conditions apply to the nonparametric model; the parametric solution and temporal conditions are not imposed in addition.

\begin{assumption}[Causal solution and moments]\label{np:ass:solution}
Equation \eqref{eq:semi-var-model} has a time-homogeneous causal stationary solution
\[
y_{it}=F\!\left(\alpha_i,
 \{(\varepsilon_{1s},\lambda_s,u_{is}):s\leq t\}\right),
\]
where $F$ is common across units and $y_{it}$ takes values in a
possibly unbounded state space $\mathcal Y$. The exposure function is
bounded on $\mathcal Y$. For some $q=4+\eta$, $\eta>0$,
\[
\mathbb{E}|\varepsilon_{1t}|^q+\mathbb{E}\|u_{it}\|^q
 +\mathbb{E}\|\alpha_i\|^q+\mathbb{E}\|y_{it}\|^q<\infty.
\]
\end{assumption}
The causal representation gives conditional iid unit paths and
specifies the permanent heterogeneity through $\alpha_i$.
Lemma~\ref{np:lem:gram} derives stability of $\mathbf A$.

\begin{assumption}[Series approximation, identification, and dimensions]
\label{np:ass:identification}
For some $r>0$, put $a=r/m$ and require, uniformly in $K$,
\[
\|\varrho_K\|_{F_y}\leq CK^{-a},\qquad
cI_K\preceq M_K\preceq CI_K.
\]
Uniformly in $K$ and $N\geq N_0$, almost surely,
\[
cI\preceq Q_{N,K}\preceq CI,\qquad
\mathbb{E}[r_{it,K}r_{it,K}'\mid\mathcal A]
 \preceq C(1+\varepsilon_{1t}^2)I.
\]
Further, $N,T,K\to\infty$, $K^2/T\to0$, and $h,m,p$ are fixed.
\end{assumption}
The approximation condition concerns the population projection of the
raw, computable basis
\eqref{np:eq:projection}, as in series estimation \citep{newey1997,lee2014}.
The $L^2(F_y)$ approximation bound replaces a global uniform bound on
an unbounded state space. The conditional design bound is separate from
the unconditional Riesz bound on $M_K$; within identification still
requires the lower bound on $Q_{N,K}$. The growth restrictions are
sufficient conditions: $K^2/T\to0$
comes from the full-Gram Frobenius bound and is a binding restriction
on the oracle choice, rather than a finite-sample cutoff.

For the following dependence condition, explicitly write
\[
r_{it,K}=(r_{it,K,1},\ldots,r_{it,K,K+mp})',\qquad
u_{it}=(u_{it,1},\ldots,u_{it,m})'.
\]
For the current $K$, collect the scalar coordinate processes in
\[
\begin{aligned}
\mathcal V_{h,K}={}&
 \{(r_{it,K,a})_{t\in\mathbb Z}:1\leq a\leq K+mp\}\\
&\cup\{(u_{it,b})_{t\in\mathbb Z}:1\leq b\leq m\}
 \cup\{(v_{i,t,h})_{t\in\mathbb Z}\}.
\end{aligned}
\]
For centered scalar variables $V_1,V_2,V_3,V_4$, define
\begin{align*}
\mathrm{cum}(V_1,V_2,V_3,V_4)
={}&\mathbb{E}[V_1V_2V_3V_4]
 -\mathbb{E}[V_1V_2]\mathbb{E}[V_3V_4]\\
&-\mathbb{E}[V_1V_3]\mathbb{E}[V_2V_4]
 -\mathbb{E}[V_1V_4]\mathbb{E}[V_2V_3].
\end{align*}
For variables centered conditional on $\alpha_i$, the conditional
cumulant uses $\mathbb{E}[\cdot\mid\alpha_i]$ in each of the seven
expectations in this formula. The coordinates used below satisfy
this conditional centering under the sampling and causal-solution
assumptions.

For the blocks $V_{it}\in\{r_{it,K},P_Yr_{it,K},u_{it},v_{i,t,h}\}$,
whose scalar coordinate processes belong to $\mathcal V_{h,K}$, put
$d_V=K$ for $r_{it,K}$ and $d_V=1$ otherwise.
Let $\Delta F_{N,t,K}=F_{N,t,K}-Q_{N,K}$,
$\langle H,G\rangle_F=\sum_{a,b}H_{ab}G_{ab}$, and let
$\|\cdot\|_*$ denote the nuclear norm, the sum of singular values.
\begin{assumption}[Temporal moments of raw arrays]
\label{np:ass:technical}
\begin{enumerate}
\item[(a)] For each $V$ just defined, conditional on $\alpha_i$, the
stationary covariance and fourth cumulant kernels satisfy
\[
\sum_{s\in\mathbb Z}\|\mathbb{E}[V_{i0}V_{is}'\mid\alpha_i]\|_*
 \leq C_i d_V,
\]
\[
\sum_{s,r,v\in\mathbb Z}\sum_{a,b}
\left|\mathrm{cum}(V_{i0,a},V_{is,a},V_{ir,b},V_{iv,b}\mid\alpha_i)\right|
 \leq C_i d_V^2,\qquad \mathbb{E} C_i^2\leq C.
\]
\item[(b)] For $P=I_{K+mp}$ and $P=P_Y$,
\[
\sum_{\ell\in\mathbb Z}
\left|\mathbb{E}\langle P\Delta F_{N,0,K},P\Delta F_{N,\ell,K}\rangle_F\right|
\leq C d_P,\qquad d_I=K^2,\quad d_{P_Y}=K.
\]
\item[(c)] For every contiguous $I$ of length $L$ and indices $i,j$,
\[
\sum_{s,t,r,v\in I}\sum_{a=1}^{K+mp}\sum_{b=1}^m
\left|\mathrm{Cov}(u_{is,b}r_{it,K,a},u_{jv,b}r_{jr,K,a})\right|
\leq
\begin{cases}
CKL^2,&i=j,\\
CKL,&i\ne j.
\end{cases}
\]
\end{enumerate}
\end{assumption}
The bounds are uniform in $K,N$ and in the indicated periods and
coordinates. They control time means, the full Gram matrix and its
control rows, and the four-date products in the dynamic mean correction.
They are imposed directly on the raw arrays.

For fixed $h\geq1$, define the population common-shock loadings
\begin{equation}
\begin{aligned}
g_{j,t,h,K}
 &=\mathbb{E}[r_{it,K}\mu_{h-j}(y_{i,t+j-1})\mid\mathcal A],
 \qquad j=1,\ldots,h,\\
\kappa_{t,h,K}
 &=\sum_{j=1}^h g_{j,t-j,h,K},\qquad t\in\mathbb Z.
\end{aligned}
\label{np:eq:stationary-loadings}
\end{equation}
These definitions contain no sample-boundary indicators. Conditional iid
sampling makes $g$ independent of $i$; causality makes
$\kappa_{t,h,K}$ measurable with respect to $\mathcal A_{t-1}$.
The process is stationary. The selector $S_K$ retains the part of this
loading relevant for function estimation rather than the control block.

\begin{assumption}[Series-block nondegeneracy]\label{np:ass:floor}
For the fixed $h\geq1$ under consideration,
\[
\liminf_{K\to\infty}
 \mathbb{E}\|S_K\kappa_{0,h,K}\|^2>0.
\]
\end{assumption}
This condition is used only for the lower bound. Positive innovation
variances or variation confined to the control block do not imply it.

For a fixed nested raw basis $p_K=(p_1,\ldots,p_K)'$ with
the same coordinates at every order, the coordinatewise population
centering preserves every existing series coordinate as $K$ increases.
Consequently there are random variables $\zeta_{k,h}$, independent
of the truncation order, such that
\[
S_K\kappa_{0,h,K}=(\zeta_{1,h},\ldots,\zeta_{K,h})',\qquad
\mathbb{E}\|S_K\kappa_{0,h,K}\|^2
 =\sum_{k=1}^K\mathbb{E}\zeta_{k,h}^2.
\]
This quantity is nondecreasing in $K$. In this nested-coordinate
case, Assumption~\ref{np:ass:floor} is equivalent to positivity for
some finite $K_0$. The argument requires consistent coordinates and
normalization across $K$; it does not apply directly to a basis
redefined or rotated at each truncation order.

\subsection{Pointwise scores and inference conditions}
\label{np:sec:pointwise}
Fix distinct states $y_1,y_2$ and a horizon $h$. Throughout this
subsection, suppress these states in variance notation and put
\[
d_K=p_K(y_1)-p_K(y_2),\qquad
\ell_K=\max\{1,\|d_K\|\},\qquad
t_K=\sqrt{K/(NT)},\qquad b_K=K^{-r/m}.
\]
The population sieve contrast is
$\theta_{h,K}=e_1'J\mathbf A^hJ'B_Kd_K$.
For a perturbation $H=(H_B,H_1,\ldots,H_p)$ of $D_K$, let
$\mathbf H_A$ have first block row $(H_1,\ldots,H_p)$ and all other
rows zero. Define the row vector $\varphi_{h,K}$ by
\begin{equation}
\begin{split}
\varphi_{h,K}\mathrm{vec}(H)
={}&e_1'J\left\{\sum_{j=0}^{h-1}
 \mathbf A^j\mathbf H_A\mathbf A^{h-1-j}\right\}J'B_Kd_K\\
&+e_1'J\mathbf A^hJ'H_Bd_K .
\end{split}
\label{np:eq:contrast-derivative}
\end{equation}
The sum is zero at $h=0$. This derivative includes estimation of both
the transition and exposure coefficients. The bound on $\|B_K\|_F$
in Lemma~\ref{np:lem:gram} gives $\|\varphi_{h,K}\|\leq C_h\ell_K$.

Write $Q=Q_{N,K}$ and use the stationary loadings
$\kappa_{t,h,K}$ in \eqref{np:eq:stationary-loadings}. Define the scalar
date scores and their unconditional variances by
\begin{align}
z^{VAR}_{t,h,N,K}
 &=\varphi_{h,K}(Q^{-1}\otimes I_m)\frac1{\sqrt N}
       \sum_{i=1}^N(r_{it,K}\otimes u_{it}),
 &s_{h,N,K}^2&=\mathbb E[(z^{VAR}_{0,h,N,K})^2],
 \label{np:eq:var-pointwise-score}\\
z^{LP}_{t,h,N,K}
 &=\varepsilon_{1t}d_K'S_KQ^{-1}\kappa_{t,h,K},
 &\tau_{h,N,K}^2&=\mathbb E[(z^{LP}_{0,h,N,K})^2].
 \label{np:eq:lp-pointwise-score}
\end{align}
The LP definition applies to $h\geq1$.
Let $\mathcal H_{N,t}=\mathcal C_N\vee\mathcal A_t
\vee\sigma(u_{is}:i\leq N,\ s\leq t)$. Both scores are martingale
differences for this filtration. In particular, the random reference
Gram is measurable at the start of the sample.

For explicit sufficient remainder conditions, put
\begin{align*}
r^{VAR}_{N,T,K}
 &=T^{-2}+\frac{\sqrt K}{\sqrt N T}
   +\frac{\sqrt K}{T^{3/2}}+\frac{\sqrt K}{N\sqrt T}
   +\frac{\sqrt K}{NT}+\frac{K}{\sqrt T}t_K,\\
r^{LP}_{N,T,K}
 &=b_K+t_K+\frac{\sqrt K}{T}
   +\frac K{\sqrt T}\left(T^{-1/2}+t_K+\frac{\sqrt K}{T}\right).
\end{align*}
These orders follow from the coefficient and score bounds below; they
do not impose a normal limit as an assumption.

\begin{assumption}[Pointwise inference]\label{np:ass:pointwise}
\begin{enumerate}
\item[(a)] At the two selected states,
$\|\varrho_K(y_1)\|+\|\varrho_K(y_2)\|\leq Cb_K$.
For each method $E$ under consideration, let
$n_{VAR}=T$, $n_{LP}=T_h$, $\sigma_{VAR}=s_{h,N,K}$, and
$\sigma_{LP}=\tau_{h,N,K}$. Its score has positive finite variance and
satisfies
\[
\frac1{n_E\sigma_E^2}\sum_{t=1}^{n_E}
 \mathbb E[(z^E_{t,h,N,K})^2\mid\mathcal H_{N,t-1}]
 \xrightarrow{p}1,
\qquad
\frac{\mathbb E|z^E_{0,h,N,K}/\sigma_E|^{2+\delta}}
 {n_E^{\delta/2}}\longrightarrow0
\]
for some $\delta>0$.
\item[(b)] For VAR, $N=o(T^2)$, $s_{h,N,K}/\sqrt{NT}\to0$, and
\[
\frac{\sqrt{NT}\ell_K}{s_{h,N,K}}
 \{b_K+r^{VAR}_{N,T,K}+(t_K+b_K)^2\}\longrightarrow0.
\]
\item[(c)] For positive-horizon LP,
$\tau_{h,N,K}^2\to\tau_h^2\in(0,\infty)$ and
$\sqrt T\,\ell_Kr^{LP}_{N,T,K}\to0$.
No restriction $N=o(T^2)$ is imposed for this part.
\end{enumerate}
\end{assumption}
Part (a) controls approximation at the reported states, conditional
variance stabilization, and a Lyapunov condition for the growing score
array. Parts (b) and (c) make approximation, within corrections, and
inverse-Gram remainders negligible at the relevant pointwise scale.
They are stronger than the conditions for integrated rates.
For example, if $N\asymp T$, $\ell_K\asymp s_{h,N,K}\asymp\sqrt K$,
and $r/m>3$, choosing $K\asymp T^\kappa$ with
$m/r<\kappa<1/3$ satisfies the growth requirements in (b) and (c);
the score conditions in (a) and the LP variance limit remain separate
requirements.

\subsection{Integrated rates and series choices}
\label{np:sec:integrated-rates}
Write $a=r/m$. The following choices concern the upper bounds in
Theorem~\ref{prop:semiparametric-rates}, not pointwise inference.
The lower bound \eqref{np:eq:floor} requires
Assumption~\ref{np:ass:floor} and \eqref{np:eq:floor-regime};
it is specific to this LP estimator, not a minimax bound.

\subsubsection{Balanced series order and admissibility}
Balancing $K/(NT)$ and $K^{-2a}$ gives
\begin{equation}
K_*\asymp(NT)^{1/(2a+1)},\qquad
\|\widehat\theta_h^{VAR}-\theta_h\|_{L^2(\mathcal Y\times\mathcal Y)}
=O_p\bigl((NT)^{-a/(2a+1)}\bigr).
\label{np:eq:oracle}
\end{equation}
This oracle choice is admissible only if $N=o(T^2)$ and
\begin{equation}
\frac{(NT)^{2/(2a+1)}}{T}\longrightarrow0
\quad\Longleftrightarrow\quad N=o(T^{a-1/2}).
\label{np:eq:oracle-feasible}
\end{equation}
For $N\asymp T$, this requires $a>3/2$.
Otherwise one may take
$K=\min\{K_*,\sqrt T/\ell_T\}$, rounded to an integer, with
$\ell_T\to\infty$ slowly enough that $K\to\infty$, and evaluate
\eqref{np:eq:var-rate} at that choice.

\subsubsection{LP choices giving the parametric time-series order}
The nonparametric terms are $o(T^{-1/2})$ when
\begin{equation}
T^{1/(2a)}\ll K\ll N,\qquad K\ll\sqrt T.
\label{np:eq:free-lunch}
\end{equation}
Such choices exist if and only if $a>1$ and
$N/T^{1/(2a)}\to\infty$. For example, take
$K=T^{1/(2a)}\ell_T$ with
$1\ll\ell_T\ll\min\{N/T^{1/(2a)},T^{1/2-1/(2a)}\}$.
The upper bound is then $O_p(T^{-1/2})$, and
Assumption~\ref{np:ass:floor} rules out $o_p(T^{-1/2})$.
This conclusion concerns convergence order, not equal variances or
pointwise inference.

\subsubsection{LP oracle balance and the three error components}
When \eqref{np:eq:oracle-feasible} holds, the same $K_*$ gives
\[
\|\widehat\theta_h^{LP}-\theta_h\|_{L^2(\mathcal Y\times\mathcal Y)}
=O_p\bigl(T^{-1/2}+(NT)^{-a/(2a+1)}\bigr),
\]
without the VAR restriction $N=o(T^2)$.
For $a>1$ and $N\gg T^{1/(2a)}$, the common-shock term dominates, and
\eqref{np:eq:free-lunch} suffices even if $K_*$ is inadmissible.
For $N\asymp T^{1/(2a)}$, $K_*\asymp N$ and all three components
have order $T^{-1/2}$; this choice is admissible when $a>1$, but
does not satisfy the strict floor regime.
For $N\ll T^{1/(2a)}$, the balanced series terms dominate.
An inadmissible oracle choice must instead be replaced by an admissible
$K$ and evaluated in \eqref{np:eq:lp-rate}.

\subsection{Score notation}
All constants below are uniform over the finitely many intervals
of length proportional to $T$.
For any scalar or vector error $e_{it}$, define the row-oriented scores
\begin{align*}
U_I(e)&=(NL)^{-1}\sum_i\sum_{t\in I}e_{it}r_{it,K}',&
D_I(e)&=N^{-1}\sum_i\bar e_{i,I}\bar r_{i,I,K}',\\
E_I(e)&=L^{-1}\sum_{t\in I}\bar e_{N,t}\bar r_{N,t,K}',&
F_I(e)&=\bar e_{N,I}\bar r_{N,I,K}',\\
W_I(e)&=U_I(e)-D_I(e)-E_I(e)+F_I(e).
\end{align*}
Since $R-r$ consists of additive unit and date components,
\begin{equation}
(NL)^{-1}\sum_i\sum_{t\in I}\ddot e_{it,I}\ddot R_{it,K,I}'=W_I(e).
\label{np:eq:score-identity}
\end{equation}
For an interval-indexed quantity $Z$, write
$\mathcal J_T Z=2Z_{I^{(0)}}-(Z_{I^{(1)}}+Z_{I^{(2)}})/2$.
In the remainder of the appendix abbreviate $Q=Q_{N,K}$ and
$\widehat Q_I=\widehat Q_{I,K}$.

\subsection{Decomposition of the one-step VAR coefficients}
Applying \eqref{np:eq:score-identity} to the one-step system gives the exact
finite-sample identity
\begin{equation}
\widehat D_I-D_K=\{W_I(u)+W_I(\varrho_K)\}\widehat Q_I^{-1},
\label{np:eq:var-exact}
\end{equation}
where $W_I(\varrho_K)$ uses $e_{it}=\varrho_{it,K}$. Applying the SPJ weights to
\eqref{np:eq:var-exact} gives
\begin{align}
\widehat D-D_K={}&U_{I^{(0)}}(u)Q^{-1}
 -(\mathcal J_T D(u))Q^{-1}\nonumber\\
&-(\mathcal J_T E(u))Q^{-1}
 +(\mathcal J_T F(u))Q^{-1}\nonumber\\
&+\mathcal J_T\{W_I(u)(\widehat Q_I^{-1}-Q^{-1})\}
 +\mathcal J_T\{W_I(\varrho_K)\widehat Q_I^{-1}\}\nonumber\\
 =:{}& I^{VAR}+II^{VAR}+III^{VAR}+IV^{VAR}+V^{VAR}+VI^{VAR}.
\label{np:eq:var-decomposition}
\end{align}
The equality uses $\mathcal J_T U(u)=U_{I^{(0)}}(u)$.

\subsection{Recursive propagation of the VAR coefficients}
Let $\Delta \mathbf A=\widehat{\mathbf A}-\mathbf A$ and $\Delta B_K=\widehat B_K-B_K$.
The estimator \eqref{np:eq:recursive-estimator} satisfies, with
$\Delta p_K=p_K(y_1)-p_K(y_2)$ and
$\Delta\varrho_{h,K}=\varrho_{h,K}(y_1)-\varrho_{h,K}(y_2)$,
\begin{align}
\widehat\theta_h^{VAR}(y_1,y_2)-\theta_h(y_1,y_2)
={}&e_1'J(\widehat{\mathbf A}^h-\mathbf A^h)J'B_K\Delta p_K\nonumber\\
&+e_1'J\widehat{\mathbf A}^hJ'\Delta B_K\Delta p_K-\Delta\varrho_{h,K},
\label{np:eq:recursive-error}
\end{align}
and, for fixed $h\geq1$,
\begin{equation}
\widehat{\mathbf A}^h-\mathbf A^h
 =\sum_{\ell=0}^{h-1}\widehat{\mathbf A}^\ell\Delta \mathbf A \mathbf A^{h-1-\ell}.
\label{np:eq:power-difference}
\end{equation}

\subsection{LP decomposition}
Using the stationary loadings in \eqref{np:eq:stationary-loadings}, define
\[
G_{h,K}=T_h^{-1}\sum_{t\in I_h}\sum_{j=1}^h
                 g_{j,t,h,K}\varepsilon_{1,t+j}.
\]
The loading $g_{j,t,h,K}$ depends on common innovations only through
$t+j-1$. Define the centered raw score remainder
$M_{h,K}=U_{I_h}(v_h)'-G_{h,K}$ and
$C_{h,K}=-D_{I_h}(v_h)'-E_{I_h}(v_h)'+F_{I_h}(v_h)'$.
Using \eqref{eq:semi-lp} and \eqref{np:eq:score-identity}, and writing
$\Delta p_K=p_K(y_1)-p_K(y_2)$ and
$\Delta\varrho_{h,K}=\varrho_{h,K}(y_1)-\varrho_{h,K}(y_2)$,
\begin{align}
\widehat\theta_h^{LP}(y_1,y_2)-\theta_h(y_1,y_2)={}&-\Delta\varrho_{h,K}
 +\Delta p_K'S_K\widehat Q_{I_h}^{-1}W_{I_h}(\varrho_{h,K})'\nonumber\\
&+\Delta p_K'S_KQ^{-1}G_{h,K}
 +\Delta p_K'S_KQ^{-1}M_{h,K}\nonumber\\
&+\Delta p_K'S_KQ^{-1}C_{h,K}\nonumber\\
&+\Delta p_K'S_K(\widehat Q_{I_h}^{-1}-Q^{-1})
             (G_{h,K}+M_{h,K}+C_{h,K})\nonumber\\
 =:{}& I_h^{LP}(y_1,y_2)+II_h^{LP}(y_1,y_2)
       +III_h^{LP}(y_1,y_2)+IV_h^{LP}(y_1,y_2)\nonumber\\
&+V_h^{LP}(y_1,y_2)+VI_h^{LP}(y_1,y_2).
\label{np:eq:lp-decomposition}
\end{align}

\subsection{Lemmas}
\begin{lemma}[Means, Gram matrices, and approximation]
\label{np:lem:gram}
The companion matrix $\mathbf A$ is stable under Assumptions~\ref{pa:ass:sampling} and~\ref{np:ass:solution}.
Under Assumption~\ref{pa:ass:sampling} and Assumptions~\ref{np:ass:solution}--\ref{np:ass:technical}, for $L\asymp T$,
\[
\mathbb{E}\|\bar r_{i,I,K}\|^2\leq CK/L,\quad
\mathbb{E}\|\bar r_{i,I,K}\|^4\leq CK^2/L^2.
\]
The corresponding bounds for $P_Yr$, $u$, and $v_h$ have $K$ replaced by
one. Moreover $Q$ is block diagonal between the series and control blocks,
and
\[
\|\widehat Q_I-Q\|_F=O_p(K/\sqrt T),\quad
\|P_Y(\widehat Q_I-Q)\|_F=O_p(\sqrt{K/T}),\quad
\|\widehat Q_I^{-1}\|=O_p(1).
\]
For either approximation error, its fitted coefficient contribution is
$O_p(K^{-a})$ in Euclidean or Frobenius norm, respectively. Also
$\|B_K\|_F\leq C\|\mu-\int\mu\,dF_y\|_{F_y}$.
\end{lemma}
The proof is provided in Appendix~\ref{np:sec:lemma-proofs}.

The remaining lemmas maintain Assumption~\ref{pa:ass:sampling} and
Assumptions~\ref{np:ass:solution}--\ref{np:ass:technical}.

\begin{lemma}[SPJ algebra and one-step scores]
\label{np:lem:spj}
For any array $\xi_t$ used unchanged on all three intervals,
\begin{equation}
\mathcal J_T\left\{L^{-\gamma}\sum_{t\in I}\xi_t\right\}
 =(2-2^{\gamma-1})T^{-\gamma}\sum_{t=1}^T\xi_t.
\label{np:eq:homogeneous}
\end{equation}
In particular, linear averages are retained and $L^{-2}$ sums cancel,
regardless of their means. The mean correction satisfies
\begin{equation}
\mathcal J_T D(u)=\frac{2}{NT^2}\sum_i
 \left(\sum_{s\in I^{(1)},t\in I^{(2)}}
       +\sum_{s\in I^{(2)},t\in I^{(1)}}\right)u_{is}r_{it,K}'.
\label{np:eq:cross-block}
\end{equation}
There is a bounded matrix $\Phi_K=(0,\Phi_Y)$ such that, uniformly over
these intervals,
\begin{align*}
\|U_I(u)\|_F&=O_p\bigl(\sqrt{K/(NT)}\bigr),\\
\mathbb{E} D_I(u)&=\Phi_K/L+O_F(L^{-2}),\\
\|D_I(u)-\mathbb{E} D_I(u)\|_F
 &=O_p\bigl(\sqrt K/(\sqrt N T)+\sqrt K/T^{3/2}\bigr),\\
\|\mathcal J_T D(u)\|_F
 &=O_p\bigl(T^{-2}+\sqrt K/(\sqrt N T)+\sqrt K/T^{3/2}\bigr),\\
\|E_I(u)\|_F&=O_p\bigl(\sqrt K/(N\sqrt T)\bigr),\qquad
\|F_I(u)\|_F=O_p\bigl(\sqrt K/(NT)\bigr).
\end{align*}
Here $O_F$ denotes a deterministic bound in Frobenius norm.
\end{lemma}
The proof is provided in Appendix~\ref{np:sec:lemma-proofs}.

\begin{lemma}[Corrected one-step VAR coefficients]
\label{np:lem:var}

If additionally $N=o(T^2)$, the Frobenius-norm orders in \eqref{np:eq:var-decomposition} are
\begin{gather*}
I^{VAR}=O_{p,F}(t_K),\qquad VI^{VAR}=O_{p,F}(b_K),\\
II^{VAR},III^{VAR},IV^{VAR},V^{VAR}=o_{p,F}(t_K).
\end{gather*}
Consequently
\[
\|\widehat D-D_K\|_F=O_p(t_K+b_K),\qquad
\|\Delta \mathbf A\|+\|\Delta B_K\|_F=O_p(t_K+b_K).
\]
More precisely, with $r^{VAR}_{N,T,K}$ defined in
Appendix~\ref{np:sec:pointwise},
\begin{equation}
\left\|\widehat D-D_K-U_{I^{(0)}}(u)Q^{-1}\right\|_F
 =O_p(b_K+r^{VAR}_{N,T,K}).
\label{np:eq:pointwise-coefficient-remainder}
\end{equation}
\end{lemma}
The proof is provided in Appendix~\ref{np:sec:lemma-proofs}.

\begin{lemma}[LP score and function components]
\label{np:lem:lp}
For fixed $h\geq1$,
\[
\|G_{h,K}\|=O_p(T^{-1/2}),\quad
\|M_{h,K}\|=O_p(t_K),\quad
\|C_{h,K}\|=O_p(\sqrt K/T+t_K).
\]
The six terms in \eqref{np:eq:lp-decomposition} have respective
$L^2(\mathcal Y\times\mathcal Y)$
bounds
\[
O_p(b_K),\quad O_p(b_K),\quad O_p(T^{-1/2}),\quad
O_p(t_K),\quad O_p(\sqrt K/T+t_K),\quad
o_p(T^{-1/2}+t_K).
\]
At a fixed pair satisfying the pointwise approximation bound in
Assumption~\ref{np:ass:pointwise}(a), the same decomposition gives
\begin{equation}
\widehat\theta_h^{LP}(y_1,y_2)-\theta_h(y_1,y_2)
 =d_K'S_KQ^{-1}G_{h,K}+O_p(\ell_Kr^{LP}_{N,T,K}).
\label{np:eq:lp-pointwise-remainder}
\end{equation}
\end{lemma}
The proof is provided in Appendix~\ref{np:sec:lemma-proofs}.

\begin{lemma}[A nonvanishing aggregate function component]
\label{np:lem:floor}
Under Assumption~\ref{pa:ass:sampling} and Assumptions~\ref{np:ass:solution}--\ref{np:ass:technical}, put $L=T_h$ and
\[
Z_{L,K}=L^{-1/2}\sum_{t=1}^L
 \varepsilon_{1t}S_K\kappa_{t,h,K}.
\]
Then $\sup_{L,K}\mathbb{E}\|Z_{L,K}\|^4<\infty$ and
\[
\mathbb{E}\|Z_{L,K}\|^2
 =\sigma_\varepsilon^2\mathbb{E}\|S_K\kappa_{0,h,K}\|^2,\qquad
\mathbb{E}\|\sqrt L\,S_KG_{h,K}-Z_{L,K}\|^4\leq C L^{-2}.
\]
If additionally Assumption~\ref{np:ass:floor} holds, the aggregate function
component satisfies, for some $c_3,c_4>0$,
\[
\liminf_{N,T\to\infty}
 \mathbb P\!\left(\sqrt T\,\|III_h^{LP}\|_{L^2(\mathcal Y\times\mathcal Y)}>c_3\right)\geq c_4.
\]
No independence between the random reference Gram and the score is
required for this conclusion.
\end{lemma}
The proof is provided in Appendix~\ref{np:sec:lemma-proofs}.

\subsection{Proof of Theorem~\ref{prop:semiparametric-rates}}
\label{np:sec:rates-proof}
\begin{proof}
For VAR, impose $N=o(T^2)$. Lemma~\ref{np:lem:var} gives
$\|\Delta \mathbf A\|+\|\Delta B_K\|_F=O_p(t_K+b_K)=o_p(1)$ for the corrected
one-step matrices. For fixed $h$, \eqref{np:eq:power-difference} implies
$\|\widehat{\mathbf A}^h-\mathbf A^h\|\leq C_h\|\Delta \mathbf A\|$ on a neighborhood of $\mathbf A$
whose probability tends to one. Since $\|B_K\|_F$ and the eigenvalues of
$M_K$ are bounded, the exact recursive identity
\eqref{np:eq:recursive-error} gives
\begin{align*}
\|\widehat\theta_h^{VAR}-\theta_h\|_{L^2(\mathcal Y\times\mathcal Y)}
&\leq C_h\|\Delta \mathbf A\|\|B_K\|_F
  +C_h\|\Delta B_K\|_F+\sqrt2\|\varrho_{h,K}\|_{F_y}\\
&=O_p(t_K+b_K).
\end{align*}
For positive-horizon LP, Lemma~\ref{np:lem:lp} applied to
\eqref{np:eq:lp-decomposition} proves \eqref{np:eq:lp-rate}, since
$\sqrt K/T=o(T^{-1/2})$. At impact the estimator is the corrected
one-step fit and its future aggregate error is absent, so the VAR bound
applies.

Under \eqref{np:eq:floor-regime}, Lemma~\ref{np:lem:lp} makes the sum of
all terms other than $III_h^{LP}$ in \eqref{np:eq:lp-decomposition}
$o_p(T^{-1/2})$. Lemma~\ref{np:lem:floor} and the reverse triangle
inequality therefore give \eqref{np:eq:floor}, after decreasing its
positive threshold.
\end{proof}

\subsection{Proof of Theorem~\ref{prop:semiparametric-normality}}
\label{np:sec:pointwise-proof}
\begin{proof}
All statements below concern the fixed pair $(y_1,y_2)$.
Causality makes $r_{it,K}$ measurable given $\mathcal H_{N,t-1}$ and
the current aggregate innovation. Conditional on these variables, the
current $u_{it}$ have mean zero. Also $\kappa_{t,h,K}$ is
$\mathcal A_{t-1}$-measurable and $\varepsilon_{1t}$ is independent of
the past with mean zero. Since $Q$ is $\mathcal C_N$-measurable, both
scalar scores in \eqref{np:eq:var-pointwise-score}--\eqref{np:eq:lp-pointwise-score}
are martingale differences relative to $\mathcal H_{N,t}$.

For either score, part (a) of Assumption~\ref{np:ass:pointwise}
supplies conditional variance convergence. For every $\epsilon>0$,
the expectation of its conditional Lindeberg sum is bounded by
\[
\frac{1}{n_E\sigma_E^2}\sum_{t=1}^{n_E}
 \mathbb E\!\left[(z_t^E)^2
 1\{|z_t^E|>\epsilon\sqrt{n_E}\sigma_E\}\right]
\leq
\frac{\mathbb E|z_0^E/\sigma_E|^{2+\delta}}
 {\epsilon^\delta n_E^{\delta/2}}\longrightarrow0 .
\]
Markov's inequality gives the conditional Lindeberg condition in
probability. The martingale central limit theorem
\citep[Chapter 3]{HallHeyde1980} therefore gives
\[
\frac1{\sqrt{n_E}\sigma_E}\sum_{t=1}^{n_E}z^E_{t,h,N,K}
 \xrightarrow{d}N(0,1).
\]

For VAR, use the common coefficient expansion
\eqref{np:eq:pointwise-coefficient-remainder} in Lemma~\ref{np:lem:var}.
The fixed-horizon map $D\mapsto e_1'J\mathbf A(D)^hJ'Bd_K$
has derivative \eqref{np:eq:contrast-derivative}. On a bounded
neighborhood of $D_K$, its Taylor remainder is at most
$C_h\ell_K\|\widehat D-D_K\|_F^2$; this follows by expanding the
finite matrix power, including products of transition and exposure
errors. Since $\|\widehat D-D_K\|_F=O_p(t_K+b_K)$ and
$|\theta_h(y_1,y_2)-\theta_{h,K}|\leq C_hb_K$ by part (a),
\[
\widehat\theta_h^{VAR}(y_1,y_2)-\theta_h(y_1,y_2)
 =\frac1{\sqrt N T}\sum_{t=1}^T z^{VAR}_{t,h,N,K}
 +O_p\!\left(\ell_K\{b_K+r^{VAR}_{N,T,K}+(t_K+b_K)^2\}\right).
\]
Part (b) makes the last term $o_p(s_{h,N,K}/\sqrt{NT})$.
The score CLT proves \eqref{np:eq:var-clt}, including $h=0$.

For positive-horizon LP, use \eqref{np:eq:lp-pointwise-remainder}.
Put $L=T_h$. Reindexing the aggregate score by innovation date
changes only a fixed number of endpoint terms:
\[
d_K'S_KQ^{-1}G_{h,K}
 =L^{-1}\sum_{t=1}^L z^{LP}_{t,h,N,K}+O_p(\ell_K/L).
\]
Indeed, \eqref{np:eq:bessel}, fixed $h$, and $\|Q^{-1}\|\leq C$
bound each endpoint loading in $L^2$ by $C\ell_K$.
The endpoint order is included in $\ell_Kr^{LP}_{N,T,K}$.
Part (c), the score CLT, and $T/T_h\to1$ prove
\eqref{np:eq:lp-clt}. At impact the two estimators coincide by
definition.
\end{proof}

\subsection{Proofs of the lemmas}
\label{np:sec:lemma-proofs}
\begin{proof}[Proof of Lemma~\ref{np:lem:gram}]
The stability argument in Lemma~\ref{pa:lem:population} applies:
the nonlinear loading is predetermined and multiplies a fresh mean-zero
shock, so
\[
\mathbb E[(Y_{it}-\mathbb E Y_{it})u_{i,t-\ell}']
=\mathbf A^\ell J'\Sigma_u,\qquad \ell\geq0.
\]
Bessel's inequality, the finite second state moment, and controllability
of $(\mathbf A,J')$ therefore imply $\rho(\mathbf A)<1$.

Conditional on $\alpha_i$, expand the squared norm of the partial sum
of $V_{it}$. Each lag has at most $L$ occurrences, so part (a) of
Assumption~\ref{np:ass:technical} bounds its second moment by $C_iLd_V$.
For the fourth moment expand
$\sum_{a,b}\sum_{t,s,r,v}\mathbb{E}[V_{it,a}V_{is,a}V_{ir,b}V_{iv,b}\mid\alpha_i]$.
The cumulant part is at most $C_iLd_V^2$. The three covariance-pair
parts are at most $C_i^2L^2d_V^2$, using the covariance nuclear-norm
sum. Integrating over $\alpha_i$ and dividing by $L^2$ or $L^4$ gives
the stated mean bounds. 

Current $\varepsilon_{1t}$ is independent of past states and mean zero.
Therefore the series block of $b_{i,K}^R$ is zero, its block of $r_{it,K}$
is
$\varepsilon_{1t}\{p_K(y_{i,t-1})-\mathbb{E}[p_K(y_{i,t-1})\mid\mathcal A]\}$,
and the control block is determined by past innovations and permanent
effects. The same timing holds after cross-sectional centering. Thus
the off-diagonal blocks of $Q$ vanish, also conditional on $\mathcal C_N$.

Two-way within is unchanged when $R$ is replaced by $r$. Hence
\begin{equation}
\widehat Q_I-Q=L^{-1}\sum_{t\in I}\Delta F_{N,t,K}
 -N^{-1}\sum_i\bar\rho_{i,I,K}\bar\rho_{i,I,K}'.
\label{np:eq:gram-exact}
\end{equation}
In \eqref{np:eq:gram-exact}, part (b) of Assumption~\ref{np:ass:technical}
bounds the squared Frobenius
norm in expectation of the first term by $CK^2/L$, or $CK/L$ after
premultiplication by $P_Y$. The second term has norm at most
$N^{-1}\sum_i\|\bar r_{i,I,K}\|^2=O_p(K/L)$. Its control rows have norm
at most
\[
\left(N^{-1}\sum_i\|P_Y\bar\rho_{i,I,K}\|^2\right)^{1/2}
\left(N^{-1}\sum_i\|\bar\rho_{i,I,K}\|^2\right)^{1/2}
=O_p(\sqrt K/L).
\]
These prove both Gram bounds. Since $K/\sqrt T\to0$ and the eigenvalues
of $Q$ are uniformly bounded away from zero,
$\|\widehat Q_I^{-1}\|\leq2/c$ with probability tending to one.

For a scalar approximation array $e$, the least-squares projection
inequality gives
\[
\left\|\widehat Q_I^{-1}(NL)^{-1}\sum_{i,t\in I}\ddot R_{it,K,I}\ddot e_{it,I}\right\|^2
\leq\lambda_{\min}(\widehat Q_I)^{-1}(NL)^{-1}\sum_{i,t\in I}\ddot e_{it,I}^2
\leq O_p(1)(NL)^{-1}\sum_{i,t\in I}e_{it}^2.
\]
Orthogonal two-way within contracts the sample norm. Since current
$\varepsilon_{1t}$ is independent of the predetermined state, stationarity
and \eqref{np:eq:projection} give
$\mathbb E[(NL)^{-1}\sum_{i,t\in I}\varepsilon_{1t}^2
\|\varrho_K(y_{i,t-1})\|^2]
=\sigma_\varepsilon^2\|\varrho_K\|_{F_y}^2=O(K^{-2a})$.
Markov's inequality therefore bounds the last average by $O_p(K^{-2a})$.
Summing over the fixed $m$ equations gives the matrix version. The same calculation applies
to $\varrho_{h,K}$ for fixed $h$. The projection identity and
$M_K\succeq cI_K$ give
$\|B_K\|_F^2\leq c^{-1}
\|\mu-\int\mu\,dF_y\|_{F_y}^2$.
\end{proof}

\begin{proof}[Proof of Lemma~\ref{np:lem:spj}]
The pathwise weighting and cross-block cancellation in
Lemma~\ref{pa:lem:split} apply to $r_{it,K}$ and prove
\eqref{np:eq:homogeneous}--\eqref{np:eq:cross-block}.
Fresh-shock timing and the covariance recursion used in
Lemma~\ref{np:lem:gram} give
\[
\Gamma_{\ell,K}:=\mathbb E[u_{i0}r_{i\ell,K}']
=(0,\Sigma_uJ(\mathbf A^{\ell-1})'),\qquad \ell\geq1,
\]
and zero for $\ell\leq0$. Stability implies
$\sum_{\ell\geq1}(1+\ell)\|\Gamma_{\ell,K}\|_F\leq C$.
With $\Phi_K=\sum_{\ell\geq1}\Gamma_{\ell,K}=(0,\Phi_Y)$,
the same pair counts as in Lemma~\ref{pa:lem:split} yield
\[
\mathbb E D_I(u)=\Phi_K/L+O_F(L^{-2}),\qquad
\mathbb E[\mathcal J_TD(u)]
=\frac2{T^2}\sum_{\ell=1}^{T-1}n_{\ell,T}\Gamma_{\ell,K}
=O_F(T^{-2}).
\]

Expanding the squared Frobenius norm of the centered $D_I(u)$ and using
part (c) of Assumption~\ref{np:ass:technical} gives
\[
\mathbb{E}\|D_I(u)-\mathbb{E} D_I(u)\|_F^2
\leq \frac{C}{N^2L^4}\{NKL^2+N(N-1)KL\}
\leq C\left(\frac{K}{NL^2}+\frac{K}{L^3}\right).
\]
The same bound applies to the centered SPJ combination by the triangle
inequality. 

For $U_I(u)$, condition on past innovations, the types, and current
$(\varepsilon_{1t},\lambda_t)$. The design residual is known and the current $u_{it}$ are
independent with mean zero. Orthogonality of the calendar-time
martingale differences and $\mathbb{E}\|r_{it,K}\|^2\leq CK$ give
$\mathbb{E}\|U_I(u)\|_F^2\leq CK/(NL)$.
For $E_I(u)$, conditional iid sampling and
$\mathbb{E}[r_{it,K}\mid\mathcal A]=0$ give
$\mathbb{E}\|\bar r_{N,t,K}\|^2\leq CK/N$. Current $\bar u_{N,t}$
has covariance $\Sigma_u/N$ and is independent of this design, giving
$\mathbb{E}\|E_I(u)\|_F^2\leq CK/(N^2L)$.
Finally $\bar u_{N,I}=O_p((NL)^{-1/2})$. Since the entire residual paths
are conditionally iid and centered given $\mathcal A$,
$\bar r_{N,I,K}=O_p(\sqrt{K/(NL)})$ by Lemma~\ref{np:lem:gram}.
Their product gives the bound for $F_I(u)$ without assuming that the
two factors are independent.
\end{proof}

\begin{proof}[Proof of Lemma~\ref{np:lem:var}]
Use Lemmas~\ref{np:lem:gram} and \ref{np:lem:spj}. Relative to $t_K$, the
three components in the bound for $\mathcal J_T D(u)$ are respectively
\[
\frac{\sqrt N}{\sqrt K T^{3/2}},\qquad T^{-1/2},\qquad
\frac{\sqrt N}{T},
\]
all tending to zero under Assumption~\ref{np:ass:identification} and the
additional VAR restriction $N=o(T^2)$. The bounds for
$E$ and $F$ have ratios $N^{-1/2}$ and $(NT)^{-1/2}$. This proves the claims
for the first four terms of \eqref{np:eq:var-decomposition}.

For inverse rotation it is important not to multiply the entire
$O_p(T^{-1})$ bias by a crude full-Gram bound. Write
\[
W_I(u)=-\Phi_K/L+Z_I,\qquad
\|Z_I\|_F=O_p(t_K+\sqrt K/T^{3/2}+T^{-2})=O_p(t_K).
\]
The inverse identity
$\widehat Q_I^{-1}-Q^{-1}=-Q^{-1}(\widehat Q_I-Q)\widehat Q_I^{-1}$
bounds $Z_I$ times this inverse difference by
$O_{p,F}(Kt_K/\sqrt T)=o_{p,F}(t_K)$. Because $Q$ is block diagonal and
$\Phi_K=(0,\Phi_Y)$, the remaining term satisfies
\begin{align*}
L^{-1}\|\Phi_K(\widehat Q_I^{-1}-Q^{-1})\|_F
&\leq \frac{C}{L}\|P_Y(\widehat Q_I-Q)\|_F\|\widehat Q_I^{-1}\|\\
&=O_p(\sqrt K/T^{3/2})=o_p(t_K).
\end{align*}
It proves the
claim for $V^{VAR}$. Projection contraction in Lemma~\ref{np:lem:gram}
proves the claim for $VI^{VAR}$, including the three fits jointly.
Retaining these explicit orders and those in Lemma~\ref{np:lem:spj}
gives \eqref{np:eq:pointwise-coefficient-remainder}; the homogeneous-sum
identity makes its leading term $U_{I^{(0)}}(u)Q^{-1}$.

\end{proof}

\begin{proof}[Proof of Lemma~\ref{np:lem:lp}]
Boundedness of $\mu$ and fixed $h$ make all $\mu_{h-j}$ bounded.
Conditional Cauchy--Schwarz in every unit direction
and Assumption~\ref{np:ass:identification} give
\begin{equation}
\|g_{j,t,h,K}\|^2
\leq\|\mathbb{E}[r_{it,K}r_{it,K}'\mid\mathcal A]\|
       \mathbb{E}[\mu_{h-j}(y_{i,t+j-1})^2\mid\mathcal A]
\leq C(1+\varepsilon_{1t}^2).
\label{np:eq:bessel}
\end{equation}

Reindex by the calendar date $r=t+j$ and set
\[
\kappa_{r,h,K,I_h}=\sum_{j=1}^h
 1\{r-j\in I_h\}g_{j,r-j,h,K}.
\]
Then $T_hG_{h,K}=\sum_r\varepsilon_{1r}\kappa_{r,h,K,I_h}$.
Its loading is measurable with respect to the common history through
$r-1$. Orthogonality over $r$, the fixed number of summands in each
loading, and \eqref{np:eq:bessel} imply
$\mathbb{E}\|G_{h,K}\|^2\leq C/T_h$. 

For the centered aggregate part of $M_{h,K}$, its loading at date $r$
is a sum of at most $h$ terms of the form
\[
N^{-1}\sum_i r_{i,r-j,K}\mu_{h-j}(y_{i,r-1})
 -\mathbb{E}[r_{i,r-j,K}\mu_{h-j}(y_{i,r-1})\mid\mathcal A].
\]
Conditional iid sampling bounds its squared norm in expectation by
$CK/N$. It is known before $\varepsilon_{1r}$ is drawn. Calendar-time
orthogonality therefore bounds its averaged score variance by
$CK/(NT_h)$. The micro part can likewise be written as
\[
(NT_h)^{-1}\sum_r\sum_i
\left\{\sum_{j=0}^h1\{r-j\in I_h\}r_{i,r-j,K}\pi_{j,h}'\right\}u_{ir}.
\]
The bracket is known before current $u_{ir}$, allowing also the current
common shock when $j=0$. Conditional independence across $i$ and
calendar-time martingale orthogonality give the same variance bound.

For the individual-mean correction, Cauchy--Schwarz gives
\[
\|D_{I_h}(v_h)\|_F\leq
\left(N^{-1}\sum_i\bar v_{i,I_h,h}^2\right)^{1/2}
\left(N^{-1}\sum_i\|\bar r_{i,I_h,K}\|^2\right)^{1/2}
=O_p(\sqrt K/T).
\]
The same bound applies to $F_{I_h}(v_h)$ by Jensen. For the date
correction $E_{I_h}(v_h)$, reindex its aggregate part by $r=t+j$.
Each loading is
$\bar r_{N,r-j,K}N^{-1}\sum_i\mu_{h-j}(y_{i,r-1})$.
The second factor is bounded, and the squared norm of the first has
expectation at most $CK/N$ because $\mathbb{E}[r_{it,K}\mid\mathcal A]=0$.
The resulting score is $O_p(t_K)$ by the same martingale calculation.
The micro part has order at most $t_K$, using current independence
of the averaged micro innovation. Thus
$\|C_{h,K}\|=O_p(\sqrt K/T+t_K)$.

For any coefficient vector $z$, the contrast identity and the upper
Riesz bound give
$\|\{p_K(y_1)-p_K(y_2)\}'S_Kz\|_{L^2(\mathcal Y\times\mathcal Y)}^2
=2(S_Kz)'M_K(S_Kz)\leq2C\|z\|^2$.
The approximation contrast has norm at most
$\sqrt2\|\varrho_{h,K}\|_{F_y}$. Combine the score bounds with
$\|Q^{-1}\|\leq C$ and the approximation result of Lemma~\ref{np:lem:gram}.
The inverse-Gram remainder is bounded by
\[
O_p(K/\sqrt T)\,O_p(T^{-1/2}+t_K+\sqrt K/T)
=o_p(T^{-1/2}+t_K).
\]
Here $K/\sqrt T\to0$ and $\sqrt K/T=o(T^{-1/2})$. 
For a fixed pair, replace the integrated Riesz bound by
$|d_K'S_Kz|\leq\ell_K\|z\|$ and use the pointwise approximation
bound. The same component bounds, retaining the displayed
inverse-Gram order, give \eqref{np:eq:lp-pointwise-remainder}.

\end{proof}

\begin{proof}[Proof of Lemma~\ref{np:lem:floor}]
Write $d_{t,K}=\varepsilon_{1t}S_K\kappa_{t,h,K}$. The conditional
Cauchy--Schwarz bound \eqref{np:eq:bessel} and fixed $h$ imply
\[
\|\kappa_{t,h,K}\|^4
 \leq C_h\sum_{j=1}^h(1+\varepsilon_{1,t-j}^2)^2,\qquad
\sup_{t,K}\mathbb{E}\|\kappa_{t,h,K}\|^4<\infty.
\]
Since $\kappa_{t,h,K}$ is $\mathcal A_{t-1}$-measurable, the iid
current shock is independent of it. Thus
$\mathbb{E}[d_{t,K}\mid\mathcal A_{t-1}]=0$ and
$\sup_{t,K}\mathbb{E}\|d_{t,K}\|^4\leq C$. Only the existing $q>4$
moment requirement is used. Orthogonality across periods and stationarity
give the asserted second-moment identity.

Put $H_{n,K}=\sum_{t=1}^n d_{t,K}$ and
$f_n=\mathbb E\|H_{n,K}\|^4$.
Expanding the fourth power, using the martingale property for the
linear term, and H\"older's inequality for the remaining terms gives
\[
f_n\leq f_{n-1}
 +6C^{1/2}f_{n-1}^{1/2}
 +4C^{3/4}f_{n-1}^{1/4}+C.
\]
Induction, starting at $f_0=0$, gives $f_n\leq C_0 n^2$ with
$C_0$ independent of $K$. Dividing by $L^2$ proves the uniform
fourth-moment bound for $Z_{L,K}$.

The exact reindexing of the observed score uses
\[
\kappa_{t,h,K,I_h}
 =\sum_{j=1}^h1\{t-j\in I_h\}g_{j,t-j,h,K},\qquad
\sqrt L\,S_KG_{h,K}
 =L^{-1/2}\sum_{t=2}^{L+h}
 \varepsilon_{1t}S_K\kappa_{t,h,K,I_h}.
\]
Compared with $Z_{L,K}$, only a number of summands bounded in terms of
the fixed $h$ differ, at the two endpoints. Each omitted or added
component has uniformly bounded fourth moment by the same timing
argument. The $L^4$ triangle inequality therefore gives
\[
\left(\mathbb{E}\|\sqrt L\,S_KG_{h,K}-Z_{L,K}\|^4\right)^{1/4}
 \leq C_h L^{-1/2}.
\]

Under Assumption~\ref{np:ass:floor}, choose $c_0>0$ such that
$\mathbb{E}\|Z_{L,K}\|^2\geq c_0$ for all sufficiently large $K$.
Paley--Zygmund and $\mathbb E\|Z_{L,K}\|^4\leq C_0$ give
\[
\mathbb P(\|Z_{L,K}\|^2>c_0/2)\geq c_0^2/(4C_0).
\]
The boundary remainder is $o_p(1)$, so the same positive-probability
lower bound, with a smaller threshold, holds for $\sqrt L\,S_KG_{h,K}$.

Finally, let $Q_{xx,N,K}$ be the series block of $Q_{N,K}$.
Lemma~\ref{np:lem:gram}, the lower Riesz bound on $M_K$, and the
upper bound on $Q_{xx,N,K}$ give the pathwise inequalities
\[
\sqrt L\,\|III_h^{LP}\|_{L^2(\mathcal Y\times\mathcal Y)}
 =\sqrt2\|M_K^{1/2}Q_{xx,N,K}^{-1}\sqrt L\,S_KG_{h,K}\|
 \geq \sqrt{2c}\,C^{-1}\|\sqrt L\,S_KG_{h,K}\|.
\]
Since $T/L\to1$, the claimed lower bound follows.
\end{proof}

\section{Proofs for the Hausman-type test}
\label{app:hausman-proofs}

\subsection{Proof of Theorem~\ref{thm:hausman}}
\label{app:hausman-proof}
\begin{proof}
Under the null, write $\beta_h^*=\beta_h$. The vector expansion in
Appendix~\ref{pa:sec:proposition-proof} gives
\[
\sqrt T(\widehat\beta_h^{LP}-\beta_h^*)
 \xrightarrow{d}N(0,V_h^{LP}).
\]
Lemma~\ref{pa:lem:var} and the smooth fixed-horizon response map imply
\[
\sqrt T(\widehat\beta_h^{VAR}-\beta_h^*)
 =O_p(N^{-1/2}+T^{-1})=o_p(1).
\]
Thus
\begin{align}
\sqrt T(\widehat\beta_h^{LP}-\widehat\beta_h^{VAR})
&=\sqrt T(\widehat\beta_h^{LP}-\beta_h^*)
 -\sqrt T(\widehat\beta_h^{VAR}-\beta_h^*)\nonumber\\
&=\sqrt T(\widehat\beta_h^{LP}-\beta_h^*)+O_p(N^{-1/2}+T^{-1}).
\label{eq:hausman-difference}
\end{align}
No independence between the two estimators is needed.
Assumption~\ref{pa:ass:aggregate} and within identification imply
$V_h^{LP}\succ0$. Lemma~\ref{lem:lp-hac} gives consistency of $\widehat V_h^{LP}$;
Slutsky's theorem therefore gives the $\chi_m^2$ limit. Continuity of its
distribution at $\chi^2_{m,1-\alpha}$ proves the size statement.

Under the fixed alternative,
$\widehat\beta_h^{LP}-\widehat\beta_h^{VAR}\xrightarrow{p}\Delta_h$.
Continuity of matrix inversion at $V_{h,\mathrm{alt}}\succ0$ gives
\eqref{eq:hausman-fixed-alternative}. Its limit is strictly positive
when $\Delta_h\ne0$, so $\mathcal H_h$ exceeds any fixed critical value
with probability tending to one.
\end{proof}

\paragraph{Regularity under the two fixed alternatives.}
For Corollaries~\ref{cor:dynamic-misspecification}
and~\ref{cor:finite-memory}, maintain
Assumption~\ref{pa:ass:sampling} for the respective true system
(with the starred effects and innovations in
\eqref{eq:dynamic-misspecification-dgp}) and a causal stationary
solution with finite fourth moments of the states and innovations.
The dimensions and the tested horizon are fixed, and $N,T\to\infty$.
Write $\mathcal A$ for the full aggregate path of that system and
$\alpha_i^\circ$ for its permanent unit effect. For a stationary
panel array $a$, define its population additive residual by
\[
a^\circ_{it}
 =a_{it}-\mathbb E[a_{it}\mid\mathcal A]
       -\mathbb E[a_{it}\mid\alpha_i^\circ]+\mathbb E a_{it}.
\]
For the practitioner regressor
$R_{it}=(\varepsilon_{1t}y_{i,t-1}',Y_{i,t-1}')'$,
assume the full-sample and half-sample within moments satisfy
\begin{align*}
\frac1{N|I|}\sum_{i,t\in I}\ddot R_{it,I}\ddot R_{it,I}'
 &\xrightarrow{p}Q_R:=\mathbb E[R^\circ_{it}R_{it}^{\circ\prime}]
 \succ0,\\
\frac1{N|I|}\sum_{i,t\in I}\ddot a_{it,I}\ddot R_{it,I}'
 &\xrightarrow{p}\mathbb E[a^\circ_{it}R_{it}^{\circ\prime}],
 \qquad a_{it}=y_{it}\ \text{or}\ y_{1,i,t+h}.
\end{align*}
The LP moments use its available origin periods $I_h$.
For the feasible variance in Section~\ref{subsec:lp-variance},
also impose Assumption~\ref{ass:lp-hac}, the bandwidth conditions of
Lemma~\ref{lem:lp-hac}, and $\Lambda_h\succ0$, where $Q_h$ and
$\Lambda_h$ are defined under the respective DGP in
Appendix~\ref{app:lp-variance-proof}.
Lemma~\ref{lem:lp-hac} then gives the finite positive-definite
limit $V_{h,\mathrm{alt}}=Q_h^{-1}\Lambda_hQ_h^{-1}$.
These conditions do not impose the correctly specified one-step
CLT under the alternative.

The moment conditions imply consistency of within OLS for its
population projection coefficients. All three one-step fits have the
same probability limit, so their SPJ combination has that limit too,
since $2-\tfrac12-\tfrac12=1$. The fixed-horizon recursion is a
polynomial in the one-step coefficients and hence is continuous.

For each true system, let $\mathcal A_t$ be its aggregate history
through $t$, and let $\mathcal F_t^\circ$ be generated by all permanent
effects, aggregate shocks, time effects, and micro innovations through $t$.
All population projections below are evaluated under that system.
Write $x_{it}=\varepsilon_{1t}y_{i,t-1}$,
$Y_{i,t-1}^\circ=(Y_{i,t-1})^\circ$, and
\[
z_{it}=x_{it}^\circ-
\mathbb E[x_{it}^\circ Y_{i,t-1}^{\circ\prime}]
\{\mathbb E[Y_{i,t-1}^\circ Y_{i,t-1}^{\circ\prime}]\}^{-1}
Y_{i,t-1}^\circ,\qquad Q_h=\mathbb E[z_{it}z_{it}'].
\]
These are the population FWL objects of \eqref{eq:Q_CLC block}.

\subsection{Misspecification: panel VAR with omitted variables}
\label{app:dynamic-misspecification-proof}

\begin{lemma}[LP identification under misspecified propagation]
\label{lem:dynamic-lp-identification}
Under \eqref{eq:dynamic-misspecification-dgp} and the sampling,
causal-stationarity, identification, and within moment convergence
conditions in Appendix~\ref{app:hausman-proof}, the candidate error
decomposes as in \eqref{eq:dynamic-lp-error}. For each fixed $h\geq0$,
\begin{gather*}
z_{it}=\varepsilon_{1t}
\{y_{i,t-1}-\mathbb E[y_{i,t-1}\mid\mathcal A]\},\\
\mathbb E[z_{it}P_{i,t,h}]
=\mathbb E[z_{it}G_{i,t,h}]
=\mathbb E[z_{it}U_{i,t,h}]=0.
\end{gather*}
Consequently $\beta_h^{LP}=\beta_h^*$ and
$\widehat\beta_h^{LP}\xrightarrow{p}\beta_h^*$.
\end{lemma}

\begin{proof}
Iteration of \eqref{eq:dynamic-misspecification-dgp} gives
\begin{equation}
\begin{aligned}
y_{1,i,t+h}
={}&e_1^{*\,\prime}\sum_{k=0}^{h}J^*(\mathbf A^*)^kJ^{*\,\prime}
(\alpha_i^*+\lambda_{t+h-k}^*)\\
&+\beta_h^{*\,\prime}x_{it}+c_h'Y_{i,t-1}
+P_{i,t,h}+G_{i,t,h}+U_{i,t,h},
\end{aligned}
\label{eq:dynamic-misspecification-horizon}
\end{equation}
with $P,G,U$ given by \eqref{eq:dynamic-lp-error}.
The first line is additive in unit and date.

Causality and independence of the unit inputs from the aggregate path
give $\mathbb E[y_{i,t-1}\mid\mathcal A]
=\mathbb E[y_{i,t-1}\mid\mathcal A_{t-1}]$.
Since $\mathbb E[\varepsilon_{1t}\mid\mathcal F_{t-1}^\circ]=0$,
$\mathbb E[x_{it}\mid\alpha_i^*]=\mathbb E x_{it}=0$, and thus
\[
x_{it}^\circ=\varepsilon_{1t}
\{y_{i,t-1}-\mathbb E[y_{i,t-1}\mid\mathcal A]\}.
\]
The centered state and $Y_{i,t-1}^\circ$ are
$\mathcal F_{t-1}^\circ$-measurable, so
$\mathbb E[x_{it}^\circ Y_{i,t-1}^{\circ\prime}]=0$ by conditioning
on that field. Hence $z_{it}=x_{it}^\circ$.
The same conditioning gives
\[
\mathbb E[z_{it}P_{i,t,h}]
=\mathbb E\!\left[
\{y_{i,t-1}-\mathbb E[y_{i,t-1}\mid\mathcal A]\}P_{i,t,h}
\,\mathbb E[\varepsilon_{1t}\mid\mathcal F_{t-1}^\circ]\right]=0.
\]
For the future-shock terms,
\[
\mathbb E[z_{it}G_{i,t,h}]
=\sum_{j=1}^h\mathbb E\!\left[
z_{it}\beta_{h-j}^{*\,\prime}y_{i,t+j-1}
\,\mathbb E[\varepsilon_{1,t+j}\mid\mathcal F_{t+j-1}^\circ]
\right]=0.
\]
For $r=t,\ldots,t+h$, $z_{it}$ is measurable before $u_{ir}^*$
is drawn, allowing the current aggregate innovation. Independence gives
\[
\mathbb E[z_{it}u_{ir}^{*\,\prime}]
=\mathbb E\!\left[z_{it}\,
\mathbb E[u_{ir}^{*\,\prime}\mid
\mathcal F_{r-1}^\circ\vee\sigma(\varepsilon_{1r},\lambda_r^*)]\right]=0,
\]
which proves $\mathbb E[z_{it}U_{i,t,h}]=0$.

Finally,
$\mathbb E[z_{it}\mid\mathcal A]
=\mathbb E[z_{it}\mid\alpha_i^*]=0$.
For each $a=P,G,U$, this implies
$\mathbb E[z_{it}a_{it,h}^\circ]=\mathbb E[z_{it}a_{it,h}]=0$.
Applying the population FWL normal equation to
\eqref{eq:dynamic-misspecification-horizon} gives
$Q_h(\beta_h^{LP}-\beta_h^*)=0$.
Since $Q_h\succ0$,
\begin{equation}
\theta_h^{LP}(y,y')=\beta_h^{*\,\prime}(y-y')=\theta_h^*(y,y').
\label{eq:dynamic-misspecification-lp-response}
\end{equation}
The assumed within moment convergence gives consistency.
At $h=0$, the reported SPJ fit has the same limit because its
weights sum to one.
\end{proof}

\begin{proof}[Proof of Corollary~\ref{cor:dynamic-misspecification}]
Lemma~\ref{lem:dynamic-lp-identification} gives the LP probability
limit. Its impact argument applied to each observed equation identifies
$B$ as the observed rows of $B^*$.
The within moment conditions give the population projection limit
for each one-step fit and hence for their SPJ combination.
Continuity of the recursion yields
$\widehat\beta_h^{VAR}\xrightarrow{p}\beta_h^{VAR}$.
The variance conditions in Appendix~\ref{app:hausman-proof} and
Theorem~\ref{thm:hausman} then give power tending to one whenever
$\Delta_h\ne0$.
\end{proof}

\subsection{Misspecification: panel VARMA}
\label{app:finite-memory-proof}
In this lemma, $P_{i,t,h}$, $G_{i,t,h}$, and $U_{i,t,h}$ denote
the components in \eqref{eq:finite-memory-lp-error}, reusing the
notation of Lemma~\ref{lem:dynamic-lp-identification}.
\begin{lemma}[LP identification under misspecified shock transmission]
\label{lem:finite-memory-lp-identification}
Under \eqref{eq:finite-memory-dgp} and the sampling,
causal-stationarity, identification, and within moment convergence
conditions in Appendix~\ref{app:hausman-proof}, the candidate error
decomposes as in \eqref{eq:finite-memory-lp-error}. For each fixed $h\geq0$,
\begin{gather*}
z_{it}=\varepsilon_{1t}
\{y_{i,t-1}-\mathbb E[y_{i,t-1}\mid\mathcal A]\},\\
\mathbb E[z_{it}P_{i,t,h}]
=\mathbb E[z_{it}G_{i,t,h}]
=\mathbb E[z_{it}U_{i,t,h}]=0.
\end{gather*}
Consequently $\beta_h^{LP}=\beta_h^*$ and
$\widehat\beta_h^{LP}\xrightarrow{p}\beta_h^*$.
\end{lemma}

\begin{proof}
Iterating \eqref{eq:finite-memory-dgp} and grouping shocks dated
before, at, and after $t$ gives
\begin{equation}
\begin{aligned}
y_{1,i,t+h}
={}&e_1'\sum_{k=0}^{h}J\mathbf A^kJ'(\alpha_i+\lambda_{t+h-k})\\
&+\beta_h^{*\,\prime}x_{it}+c_h'Y_{i,t-1}
+P_{i,t,h}+G_{i,t,h}+U_{i,t,h},
\end{aligned}
\label{eq:finite-memory-horizon}
\end{equation}
where $P,G,U$ are those in \eqref{eq:finite-memory-lp-error}.
Terms dated $t$ have $j=h-k$, yielding
$\beta_h^{*\,\prime}=e_1'\sum_{j=0}^{\min\{h,q\}}
J\mathbf A^{h-j}J'B_j$.

The derivation of $z_{it}=x_{it}^\circ$ in
Lemma~\ref{lem:dynamic-lp-identification} uses only causality and
fresh-shock independence, which also hold here.
Every shock in $P_{i,t,h}$ is dated before $t$, so
\[
\mathbb E[z_{it}P_{i,t,h}]
=\mathbb E\!\left[
\{y_{i,t-1}-\mathbb E[y_{i,t-1}\mid\mathcal A]\}P_{i,t,h}
\,\mathbb E[\varepsilon_{1t}\mid\mathcal F_{t-1}^\circ]\right]=0.
\]
Each summand of $G_{i,t,h}$ is of the form
$a'y_{i,s-1}\varepsilon_{1s}$ with deterministic $a$ and $s>t$.
Consequently
\[
\mathbb E[z_{it}a'y_{i,s-1}\varepsilon_{1s}]
=\mathbb E\!\left[z_{it}a'y_{i,s-1}
\,\mathbb E[\varepsilon_{1s}\mid\mathcal F_{s-1}^\circ]\right]=0,
\]
and summing gives $\mathbb E[z_{it}G_{i,t,h}]=0$.
For $r=t,\ldots,t+h$, independence of $u_{ir}$ gives
\[
\mathbb E[z_{it}u_{ir}']
=\mathbb E\!\left[z_{it}\,
\mathbb E[u_{ir}'\mid
\mathcal F_{r-1}^\circ\vee\sigma(\varepsilon_{1r},\lambda_r)]\right]=0,
\]
hence $\mathbb E[z_{it}U_{i,t,h}]=0$.
As $\mathbb E[z_{it}\mid\mathcal A]
=\mathbb E[z_{it}\mid\alpha_i]=0$, removing additive components
preserves these moments. The FWL normal equation and $Q_h\succ0$
give $\beta_h^{LP}=\beta_h^*$ and
\begin{equation}
\theta_h^{LP}(y,y')=\beta_h^{*\,\prime}(y-y')=\theta_h^*(y,y').
\label{eq:finite-memory-lp-response}
\end{equation}
Within moment convergence proves consistency; at $h=0$ the SPJ
combination preserves the common probability limit.
\end{proof}

\begin{proof}[Proof of Corollary~\ref{cor:finite-memory}]
Lemma~\ref{lem:finite-memory-lp-identification} gives the LP limit.
At impact, the omitted lagged shock terms are predetermined, so
the one-step projection identifies $B_0$.
Within moment convergence, the SPJ weights, and continuity of the
recursion give
\[
\widehat\beta_h^{VAR\,\prime}
\xrightarrow{p}e_1'J(\mathbf A^\dagger)^hJ'B_0
=\beta_h^{VAR\,\prime}.
\]
The variance conditions in Appendix~\ref{app:hausman-proof} and
Theorem~\ref{thm:hausman} establish the power claim when $\Delta_h\ne0$.
\end{proof}

\subsection{Proof of Lemma~\ref{lem:lp-hac}}
\label{app:lp-variance-proof}

All population objects in this subsection are defined under the
DGP being considered. Use the additive residual operator in
Appendix~\ref{app:hausman-proof} and set
\begin{gather*}
R^\circ_{it}=(x_{it}^{\circ\prime},Y_{i,t-1}^{\circ\prime})',
\qquad w^\circ_{it,h}=(y_{1,i,t+h})^\circ,\\
\xi_h=\{\mathbb E[R^\circ_{it}R_{it}^{\circ\prime}]\}^{-1}
       \mathbb E[R^\circ_{it}w^\circ_{it,h}],
\qquad \eta_{it,h}=w^\circ_{it,h}-\xi_h'R^\circ_{it}.
\end{gather*}
Let $C_h=\mathbb E[x^\circ_{it}Y_{i,t-1}^{\circ\prime}]
\{\mathbb E[Y^\circ_{i,t-1}Y_{i,t-1}^{\circ\prime}]\}^{-1}$,
$z_{it}=x^\circ_{it}-C_hY^\circ_{i,t-1}$, and
$Q_h=\mathbb E[z_{it}z_{it}']$. Define
\[
g_{t,h}=\mathbb E[z_{it}\eta_{it,h}\mid\mathcal A],
\qquad \Gamma_{k,h}=\mathbb E[g_{t,h}g_{t-k,h}'],
\qquad
\Lambda_h=\sum_{k\in\mathbb Z}\Gamma_{k,h}.
\]
The population normal equations give $\mathbb E g_{t,h}=0$.
The following sufficient conditions concern population arrays;
they are additional to the coefficient CLT conditions.

\begin{assumption}[Moments and dependence for feasible variance]
\label{ass:lp-hac}
\leavevmode
\begin{enumerate}
\item[(a)] The respective sampling and causal-stationarity conditions
hold, the dimension and $h$ are fixed, and
$Q_R=\mathbb E[R^\circ_{it}R_{it}^{\circ\prime}]\succ0$.
Moreover
$\mathbb E\|R^\circ_{it}\|^4+\mathbb E|\eta_{it,h}|^4<\infty$.
\item[(b)] Put
$F_t=\mathbb E[(\mathrm{vec}(R^\circ_{it}R_{it}^{\circ\prime})',
 (R^\circ_{it}\eta_{it,h})')'\mid\mathcal A]$.
Its covariance kernel satisfies
$\sum_{k\in\mathbb Z}\|\mathrm{Cov}(F_0,F_k)\|<\infty$.
\item[(c)] The stationary date score has
$\mathbb E\|g_{0,h}\|^4<\infty$,
$\sum_{k\in\mathbb Z}\|\Gamma_{k,h}\|<\infty$, and, for every
coordinate quadruple $(a,b,c,d)$,
\[
\sum_{r,s,v\in\mathbb Z}
\left|\mathrm{cum}(g_{0,h,a},g_{r,h,b},g_{s,h,c},g_{v,h,d})\right|
<\infty.
\]
\item[(d)] Uniformly over contiguous intervals $I$ of length
$|I|\asymp T$, for $A_{it}=R^\circ_{it}$ and $A_{it}=\eta_{it,h}$,
\[
\mathbb E\left\||I|^{-1}\sum_{t\in I}A_{it}\right\|^4
 \leq C|I|^{-2}.
\]
\end{enumerate}
\end{assumption}

\begin{proof}
Write $a_{NT}=N^{-1/2}+T^{-1/2}$ and $L=T_h$.
Conditional iid sampling of unit paths and their conditional
centering imply, for $A=R^\circ,\eta$,
$\mathbb E\|N^{-1}\sum_i A_{it}\|^4\leq C/N^2$.
Together with part (d), this gives
\begin{equation}
\left\{\frac1{NL}\sum_{i,t\in I_h}
 \|\ddot A_{it}-A_{it}\|^4\right\}^{1/4}=O_p(a_{NT}),
\qquad
\frac1{NL}\sum_{i,t\in I_h}\|A_{it}\|^4=O_p(1).
\label{eq:hac-within-error}
\end{equation}
Population additive components disappear under sample double
demeaning, so $\ddot R^\circ=\ddot R$ and
$\ddot w^\circ=\ddot w$.

For either product $R^\circ R^{\circ\prime}$ or
$R^\circ\eta$, its cross-sectional average differs from its
conditional mean given $\mathcal A$ by $O_p(N^{-1/2})$ after
time averaging. This follows from conditional independence, the
fourth moments in part (a), and Jensen's inequality over periods.
Part (b) bounds the time average of the conditional mean minus its
expectation by $O_p(T^{-1/2})$.
Equation~\eqref{eq:hac-within-error} and Cauchy--Schwarz give the
same $O_p(a_{NT})$ order for replacing population residuals by
sample within residuals. Matrix inversion around $Q_R\succ0$
therefore yields
\[
\widehat\xi_h-\xi_h=O_p(a_{NT}),\qquad
\widehat C_h-C_h=O_p(a_{NT}),\qquad
\widehat Q_h-Q_h=O_p(a_{NT}).
\]
These bounds also hold for the proportional-length half samples.
In particular,
$\widehat v_{it,h}=\ddot\eta_{it,h}
-(\widehat\xi_h-\xi_h)'\ddot R_{it}$.
Using \eqref{eq:hac-within-error} once more, the empirical fourth
norms of $\widehat v-\eta$ and $\widehat z-z$ are $O_p(a_{NT})$,
while those of $\widehat v,\eta,\widehat z,z$ are $O_p(1)$.

Let $\widetilde g_{t,h}=N^{-1}\sum_i z_{it}\eta_{it,h}$.
H\"older's inequality for the products and Jensen's inequality
across units give
\[
\left\{\frac1L\sum_t
 \|\widehat g_{t,h}-\widetilde g_{t,h}\|^2\right\}^{1/2}
 =O_p(a_{NT}).
\]
Conditional iid sampling and the finite second moment of
$z_{it}\eta_{it,h}$ also give
$\mathbb E[L^{-1}\sum_t\|\widetilde g_{t,h}-g_{t,h}\|^2]\leq C/N$.
Consequently,
\begin{equation}
\left\{\frac1L\sum_t\|\widehat g_{t,h}-g_{t,h}\|^2\right\}^{1/2}
 =O_p(a_{NT}).
\label{eq:hac-score-replacement}
\end{equation}

Let $\Lambda^{o}_{h,L,b}$ be the Bartlett estimator constructed
from $g_{t,h}$ instead of $\widehat g_{t,h}$. For each lag, expand
the difference of the two score products and use Cauchy--Schwarz.
Since $L^{-1}\sum_t\|g_{t,h}\|^2=O_p(1)$, summing over $b+1$
lags yields
\[
\|\widehat\Lambda_h-\Lambda^{o}_{h,L,b}\|
 =O_p\{b\,a_{NT}+b\,a_{NT}^2\}=o_p(1).
\]

Under part (c), expanding covariances of score products into fourth
cumulants and covariance pairs bounds each sample autocovariance's
variance by $C/L$, uniformly in the lag. The $L^2$ triangle inequality gives
\[
\|\Lambda^{o}_{h,L,b}-\mathbb E\Lambda^{o}_{h,L,b}\|
=O_p(b/\sqrt L)=o_p(1).
\]
Its expectation weights $\Gamma_{k,h}+\Gamma_{k,h}'$ by
$(1-k/(b+1))(1-k/L)$ for $1\leq k\leq b$.
Absolute covariance summability and dominated convergence imply
$\mathbb E\Lambda^{o}_{h,L,b}\to\Lambda_h$.
Thus $\widehat\Lambda_h\xrightarrow{p}\Lambda_h$.
Since $\widehat Q_h\xrightarrow{p}Q_h\succ0$ and $T/L\to1$,
\eqref{eq:lp-feasible-variance} converges to
$Q_h^{-1}\Lambda_hQ_h^{-1}$.

Under the null, the LP error has zero conditional mean given the
permanent effect, and $\mathbb E[z_{it}\mid\mathcal A]=0$.
Its additive date component and micro innovations contribute zero to
the conditional score, so $g_{t,h}=s_{t,h}$.
Lemma~\ref{pa:lem:aggregate} identifies its long-run variance as
$\Lambda_h=\Omega_h$.
Under either alternative the same consistency argument applies to its
own population projections; $\Lambda_h\succ0$ gives the required
positive-definite sandwich limit.
\end{proof}

\section{Estimated aggregate shocks}
\label{app:generated}

The baseline theorems treat $\varepsilon_{1t}$ as observed. This appendix records
an algebraic observation for an augmented regression, not a generated-shock
extension of those theorems. Suppose instead the shock comes from a
first-stage time-series regression $W_t=a'Z_{t-1}+\varepsilon_{1t}$ where
$\mathbb{E}(\varepsilon_{1t}\mid Z_{t-1})=0$, and let $\widehat\varepsilon_{1t}=W_t-\widehat a'Z_{t-1}$,
so that $\widehat\varepsilon_{1t}=\varepsilon_{1t}-Z_{t-1}'(\widehat a-a)$.

Replacing $y_{i,t-1}\varepsilon_{1t}$ by $y_{i,t-1}\widehat\varepsilon_{1t}$ perturbs the
interaction regressor by a linear combination of $y_{i,t-1}\otimes Z_{t-1}$ with
coefficient $\widehat a-a$. Once $y_{i,t-1}\otimes Z_{t-1}$ is included as a nuisance
control, the feasible and oracle regressors differ by an element of the span of that
control, so residualizing either on the controls leaves the same residual. By the
Frisch--Waugh--Lovell theorem the coefficient on $y_{i,t-1}\widehat\varepsilon_{1t}$ in the
feasible regression is numerically identical to the coefficient on
$y_{i,t-1}\varepsilon_{1t}$ in the oracle regression; only the nuisance coefficients differ.
The identity is algebraic and holds for any value of $\widehat a$, so it is exact in
finite samples and requires no conditioning argument. Both regressions must
include the same nuisance interactions and satisfy the relevant rank conditions. Extending the rate and recursive-response
claims requires verifying the augmented design and the mapping to the target;
the identity alone does not establish those results.
Without the nuisance interactions, the one-step error inherits the term
$By_{i,t-1}\{Z_{t-1}'(\widehat a-a)\}$. The first-stage error factor is
common across units, while its loading varies with $y_{i,t-1}$.
If $\widehat a-a=O_p(T^{-1/2})$, this can create a time-series-order
component when its projection on the identifying score does not vanish.

For the positive-horizon projection, generated-shock uncertainty adds to the
date-level uncertainty already present and need not change the baseline order if that additional term is
$O_p(T^{-1/2})$; this requires a separate joint expansion. The appropriate variance calculation depends on the first-stage
construction and should be carried out jointly when the shock cannot be treated as
external data. More general generated shocks, including those from a first stage
that is nonlinear or estimated at a different frequency, require a joint-estimation
or orthogonal-score argument.

\end{document}